\documentclass[12pt,journal,draftcls,letterpaper,onecolumn]{IEEEtran}
\usepackage[dvips]{color}
\usepackage{epsf}
\usepackage{times}
\usepackage{epsfig}
\usepackage{graphicx}

\usepackage{amsmath}
\usepackage{amssymb}
\usepackage{amsxtra}

\usepackage[square, comma, sort&compress]{natbib}

\usepackage{algorithm}
\usepackage{algorithmic}
\begin{document}
%
\title{A Key 6G Challenge and Opportunity - Connecting the Remaining 4 Billions: A Survey on Rural Connectivity}
%
%

\author{Elias Yaacoub$^{1}$,~\IEEEmembership{Senior Member,~IEEE},
        and Mohamed-Slim Alouini$^{2}$,~\IEEEmembership{Fellow,~IEEE}
\thanks{$^1$Department
of Electrical and Computer Engineering, American University of Beirut, P.O. Box 11-0236 / ECE Department, Riad El-Solh / Beirut 1107 2020,
Lebanon. E-mail: eliasy@ieee.org.}
\thanks{$^2$Computer, Electrical and Mathematical Sciences and Engineering (CEMSE) Division, King Abdullah University of Science and Technology (KAUST), Thuwal, Makkah Province, Saudi Arabia. E-mail: slim.alouini@kaust.edu.sa.}
}

\maketitle

\begin{abstract}
Providing connectivity to around half of the World population living in rural or underprivileged areas is a tremendous challenge, but also a unique opportunity. In this paper, a survey of technologies for providing connectivity to rural areas, and that can help address this challenge, is provided. Fronthaul and backhaul techniques are discussed. In addition, energy and cost efficiency of the studied technologies are analyzed. Typical application scenarios in rural areas are discussed, and several country-specific use cases are surveyed. Directions for future evolution of rural connectivity are outlined.
\end{abstract}


\section*{Abbreviations}
\begin{small}
\begin{description}
       \item[1D]\quad  One Dimensional
       \item[2D]\quad  Two Dimensional
       \item[3D]\quad  Three Dimensional
       \item[1G]\quad  First Generation Cellular
       \item[2G]\quad  Second Generation Cellular
       \item[3G]\quad  Third Generation Cellular
       \item[4G]\quad  Fourth Generation Cellular
       \item[5G]\quad  Fifth Generation Cellular
       \item[6G]\quad  Sixth Generation Cellular
       \item[3GPP]\quad  Third Generation Partnership Project
       \item[6LoWPAN]\quad\quad  IPv6 over Low-Power Wireless Personal Area Networks
       \item[AANET]\quad  Aeronautical Ad-Hoc Network
       \item[ADSL]\quad  Asymmetric Digital Subscriber Line
       \item[AP]\quad  Access Point   
       \item[AR]\quad  Augmented Reality
       \item[ARPU]\quad  Average Revenue per User
       \item[BAN]\quad  Body Area Network
       \item[BS]\quad  Base Station
       \item[CAPEX]\quad  Capital Expenditures
       \item[CDMA]\quad  Code Division Multiple Access
       \item[CPE]\quad  Customer Premises Equipment
       \item[CR]\quad  Cognitive Radio 
       \item[CSI]\quad  Channel State Information
       \item[CSMA]\quad  Carrier Sense Multiple Access
       \item[CSMA/CA]\quad\quad  Carrier Sense Multiple Access with Collision Avoidance
       \item[D2D]\quad  Device-to-Device
       \item[DNS]\quad  Domain Name System
       \item[DSL]\quad  Digital Subscriber Line
       \item[DTN]\quad  Delay Tolerant Network
       \item[DWDM]\quad  Dense Wavelength Division Multiplexing
       \item[eMBB]\quad  Enhanced Mobile Broadband
       \item[EU]\quad  European Union
       \item[FSO]\quad  Free Space Optics
       \item[FTTH]\quad  Fiber to the Home
       \item[FTTN]\quad  Fiber to the Neighborhood
       \item[GAIA]\quad  Global Access to the Internet for All
       \item[GEO]\quad  Geostationary Orbit
       \item[GPON]\quad  Gigabit Passive Optical Network
       \item[GPRS]\quad  General Packet Radio Service
       \item[GS]\quad  Ground Station
       \item[GSM]\quad  Global System for Mobile Communications
       \item[HAP]\quad  High Altitude Platform
       \item[HTS]\quad  High Throughput Satellite
       \item[HTTP]\quad Hyper Text Transfer Protocol
       \item[ICT]\quad  Information and Communication Technology
       \item[IoT]\quad  Internet of Things
       \item[IP]\quad  Internet Protocol
       \item[ISP]\quad  Internet Service Provider
       \item[ITU]\quad  International Telecommunication Union
       \item[IVR]\quad  Interactive Voice Response
       \item[KPI]\quad  Key Performance Indicator
       \item[LAN]\quad  Local Area Network
       \item[LED]\quad  Light Emitting Diode
       \item[LEO]\quad  Low Earth Orbit
       \item[LiFi]\quad  Light Fidelity
       \item[LoRa]\quad  Long Range Radio
       \item[LPWAN]\quad Low Power Wide Area Network
       \item[LTE]\quad  Long-Term Evolution
       \item[LTE-A]\quad  Long-Term Evolution Advanced
       \item[M2M]\quad  Machine-to-Machine
       \item[MAC]\quad  Medium Access Control
       \item[MAN]\quad  Metropolitan Area Network
       \item[MANET]\quad  Mobile Ad Hoc Network
       \item[MCS]\quad  Modulation and Coding Scheme  
       \item[MEO]\quad  Medium Earth Orbit
       \item[mHealth]\quad  Mobile Health  
       \item[MIMO]\quad  Multiple-Input Multiple-Output
       \item[mMTC]\quad  Massive Machine-Type Communications
       \item[mmWave]\quad\quad  Millimeter Wave
       \item[MNO]\quad  Mobile Network Operator
       \item[mVAS]\quad  Mobile Value Added Services
       \item[NAT]\quad  Network Address Translation
       \item[NFV]\quad  Network Function Virtualization
       \item[OFDM]\quad  Orthogonal Frequency Division Multiplexing
       \item[OFDMA]\quad  Orthogonal Frequency Division Multiple Access
       \item[OPEX]\quad  Operational Expenditures
       \item[PHY]\quad  Physical Layer
       \item[PON]\quad  Passive Optical Network
       \item[PoP]\quad  Point of Presence
       \item[PSTN]\quad  Public Switched Telephone Network
       \item[PV]\quad  Photovoltaic
       \item[QoE]\quad  Quality of Experience
       \item[QoS]\quad  Quality of Service
       \item[RAN]\quad  Radio Access Network
       \item[RF]\quad  Radio Frequency
       \item[RFID]\quad  Radio Frequency Identification
       \item[RoF]\quad  Radio over Fiber
       \item[ROI]\quad  Return On Investment
       \item[RRH]\quad  Remote Radio Head
       \item[RSSI]\quad Received Signal Strength Indicator
       \item[RSU]\quad  Roadside Unit
       \item[RTP]\quad  Real-Time Transport Protocol
       \item[SCBS]\quad Small Cell Base Station
       \item[SDG]\quad  Sustainable Development Goal
       \item[SDN]\quad  Software Defined Network
       \item[SIP]\quad  Session Initiation Protocol
       \item[SLA]\quad  Service Level Agreement
       \item[SMS]\quad  Short Message Service
       \item[TCP]\quad  Transmission Control Protocol
       \item[TDD]\quad  Time Division Duplexing
       \item[TDMA]\quad  Time Division Multiple Access
       \item[TTL]\quad  Time-To-Live
       \item[TVWS]\quad  Television White Space
       \item[UAV]\quad  Unmanned Aerial Vehicle
       \item[UDN]\quad  Ultra-Dense Network
       \item[UE]\quad  User Equipment
       \item[UHF]\quad  Ultra High Frequency
       \item[UK]\quad  United Kingdom
       \item[URLLC]\quad  Ultra-Reliability and Low-Latency Communications
       \item[USA]\quad  United States of America
       \item[VANET]\quad  Vehicular Ad-Hoc Network
       \item[VHF]\quad  Very High Frequency
       \item[VoIP]\quad  Voice over Internet Protocol
       \item[VR]\quad  Virtual Reality
       \item[VSAT]\quad  Very Small Aperture Terminal
       \item[WAN]\quad  Wide Area Network
       \item[WiFi]\quad  Wireless Fidelity
       \item[WiLD]\quad  WiFi Long Distance
       \item[WiMAX]\quad  Wireless Interoperability for Microwave Access
       \item[WLAN]\quad  Wireless Local Area Network
       \item[WMAN]\quad  Wireless Metropolitan Area Network
       \item[WRAN]\quad  Wireless Regional Area Network
       \item[WSN]\quad  Wireless Sensor Network
\end{description}
\end{small}

\section{Introduction}
\label{sec:Introduction}

By the end of 2018, there were around 3.9 Billion unconnected people~\cite{TheNational_UAE_Article_2018}, out of 4.4 Billion that were unconnected in 2014~\cite{McKinsey_Report_2014}, although the subject of providing connectivity to rural and remote areas has been on the agenda of the ITU long before that~\cite{nonIEEE_ITU1}, and also on that of the World Bank~\cite{nonIEEE_WorldBank1} (In fact, the debate to cover rural areas was active in the late nineteenth century for telegraph and the early twentieth century for telephone~\cite{nonIEEE_BC1_River_WWRF}!). It was noted by McKinsey that 75\% of the unconnected population reside in 20 countries, are mostly concentrated in rural areas, have low income and low literacy rates~\cite{McKinsey_Report_2014}. In fact, a model developed in~\cite{Predictors_Dig_Divide} showed that internet penetration depends on the income per capita and the country's risk (a measure of political, security, economic, legal, tax, and operational risk rating).  A study by the ITU has shown that in terms of absolute numbers, the majority of offline individuals reside in Asia-Pacific, whereas in terms of percentages, the highest numbers correspond to Africa. Furthermore, the study showed that 85\% of the offline population live in Least Developed Countries, whereas 22\% live in Developed Countries~\cite{nonIEEE_ITU_Davos2017}. 

\begin{figure}[t!]
 \begin{center}\hspace{-7.0cm}
  \includegraphics[scale=0.5]{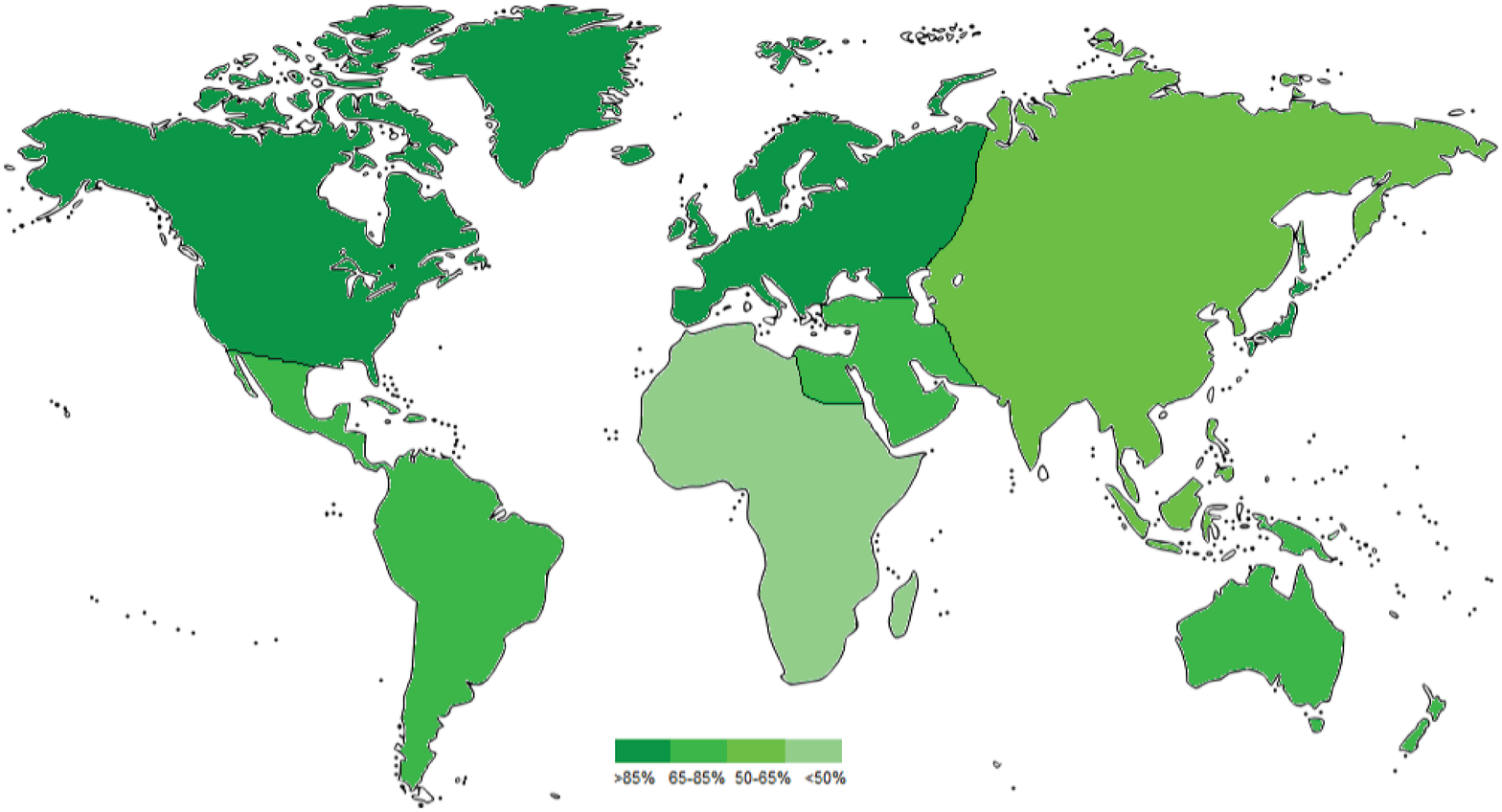}
 \end{center}
 \vspace{+3.0cm}
\caption{Internet penetration percentages across continents.}
\label{fig:Internet_Penetration_Map}
\end{figure}

Indeed, Fig.~\ref{fig:Internet_Penetration_Map} shows the internet penetration rates in percent (\%) of the population of each continent or World region. The results as of May 2019 are: North America (89.4\%), Europe (86.8\%), Oceania/Australia (68.4\%), Latin America / Caribbean (67.5\%), Middle East (67.2\%), Asia (51.8\%), and Africa (37.3\%)~\citep{Internet_World_Stats}.

The barriers to connectivity listed in~\citep{McKinsey_Report_2014, Strategy_Business_Article_2016} include: (i) affordability and low income, (ii) incentives or relevance, and (iii) user capabilities and illiteracy. In~\cite{McKinsey_Report_2014}, a fourth important barrier, infrastructure, is also discussed. Indeed, the infrastructure investment needed to connect the next 1.5 Billion is estimated at 450 Billion USD~\cite{nonIEEE_ITU_Davos2017}.

Certain rural areas in developing countries lack not only communication infrastructures, but also water, electricity, and transportation infrastructures~\cite{nonIEEE_Brewer_RuralTech}. The difficulty of transportation networks can damage the electronic equipment transported to establish rural connectivity~\cite{IEEE_C50}. An algorithm for optimizing road layout and planning in rural areas is investigated in~\cite{IEEE_C14}, and could possibly be expanded to plan the utility networks in conjunction with the transportation infrastructure. In addition, due to the various types of terrain that can be encountered in rural areas (mountains, jungle, desert, etc.), different propagation models apply in different scenarios when wireless communications are used~\citep{IEEE_C26, IEEE_C190, Prop_ITUp1546_Aus}, which further complicates wireless network planning.

Challenges faced in deploying telecommunication networks include (i) the absence of a viable business case due to the sparse and poor population, (ii) increased CAPEX, e.g., due to the scarcity of buildings and the need to build towers to install the BSs, in addition to the high backhaul costs, (iii) limited or absent electricity supply, which increases OPEX due to the need for deploying diesel generators for BS sites, and supplying them them with diesel over difficult transportation routes, and (iv) difficulty of maintenance due to the limited supply of skilled personnel in rural areas~\citep{IEEE_C113, IEEE_C103}. These factors lead to low ARPU, and a long ROI, thus heightening the barriers for building rural networks~\cite{nonIEEE_WTL1}. In certain cases, the barriers faced by some marginalized cases are also due to political, social, or cultural exclusion~\cite{Maitland_IEEE_Computer2018}.

Nevertheless, ICTs are an essential tool to achieve the United Nation's SDGs~\cite{IEEE_J_SDG}. The SDGs include targets related to environment, health, education, gender equality, eliminating poverty, among others. Although the research focus has been on technical aspects of ICTs, there is a lack of a holistic view that aims to achieve social good by overcoming the barriers to connectivity, increasing awareness, and helping all populations to achieve the SDGs~\cite{IEEE_J_SDG}. Thus, providing connectivity to rural areas should not be seen only as a burden and a challenge but also as a great opportunity from a humanitarian perspective. Furthermore, it will also be an important opportunity from a business perspective once technology becomes available and its adoption increases with user awareness.

\subsection{Connectivity Definitions}
\label{subsec:Connectivity_Definitions}
The levels of connectivity vary between different areas. Some areas have some basic level of connectivity, whereas others can be completely disconnected. We define the following four levels of connectivity:
\subsubsection{Not Connected} This scenario corresponds to people without internet connectivity. This typically corresponds to remote rural areas with difficult access, thus leading to limited infrastructure in terms of road, power, transportation, and telecommunications.
\subsubsection{Under Connected} This scenario corresponds to people with limited and/or intermittent connectivity. For example, they can have 2G voice and SMS services, along with possibly intermittent WiFi connectivity from a local network. This could be a local mesh connected to a gateway via a VSAT with limited speed, or a completely local network with partial connectivity to the internet, e.g., through delay tolerant networking.
\subsubsection{Connected} This scenario corresponds to people having GPRS or 3G connectivity~\citep{IEEE_C103}. They can access the internet with medium download speeds. They can also have WiFi connectivity attached to a backhaul network with reasonable speed at some gateway point.
\subsubsection{Hyper Connected} This scenario corresponds to people enjoying state-of-the-art connectivity, e.g., at least 4G  and possibly 5G cellular connectivity, and/or with high speed fixed DSL or FTTH or FTTN connectivity. These subscribers typically live in urban areas, and it is extremely difficult to find them in rural areas, due to the numerous barriers discussed previously.

This paper surveys the literature aiming to provide basic connectivity to the unconnected population, or aiming to increase the connectivity level of those with basic connectivity (\lq\lq Under Connected\rq\rq), or even those that are considered to have reasonable connectivity (\lq\lq Connected\rq\rq) in rural areas. In the following subsection, the position of these people with respect to the 5G connectivity use cases is discussed.

\subsection{Connectivity Use Cases}
\label{subsec:Connectivity_Use_Cases}
\begin{figure}[t!]
 \begin{center}
  \includegraphics[scale=0.3]{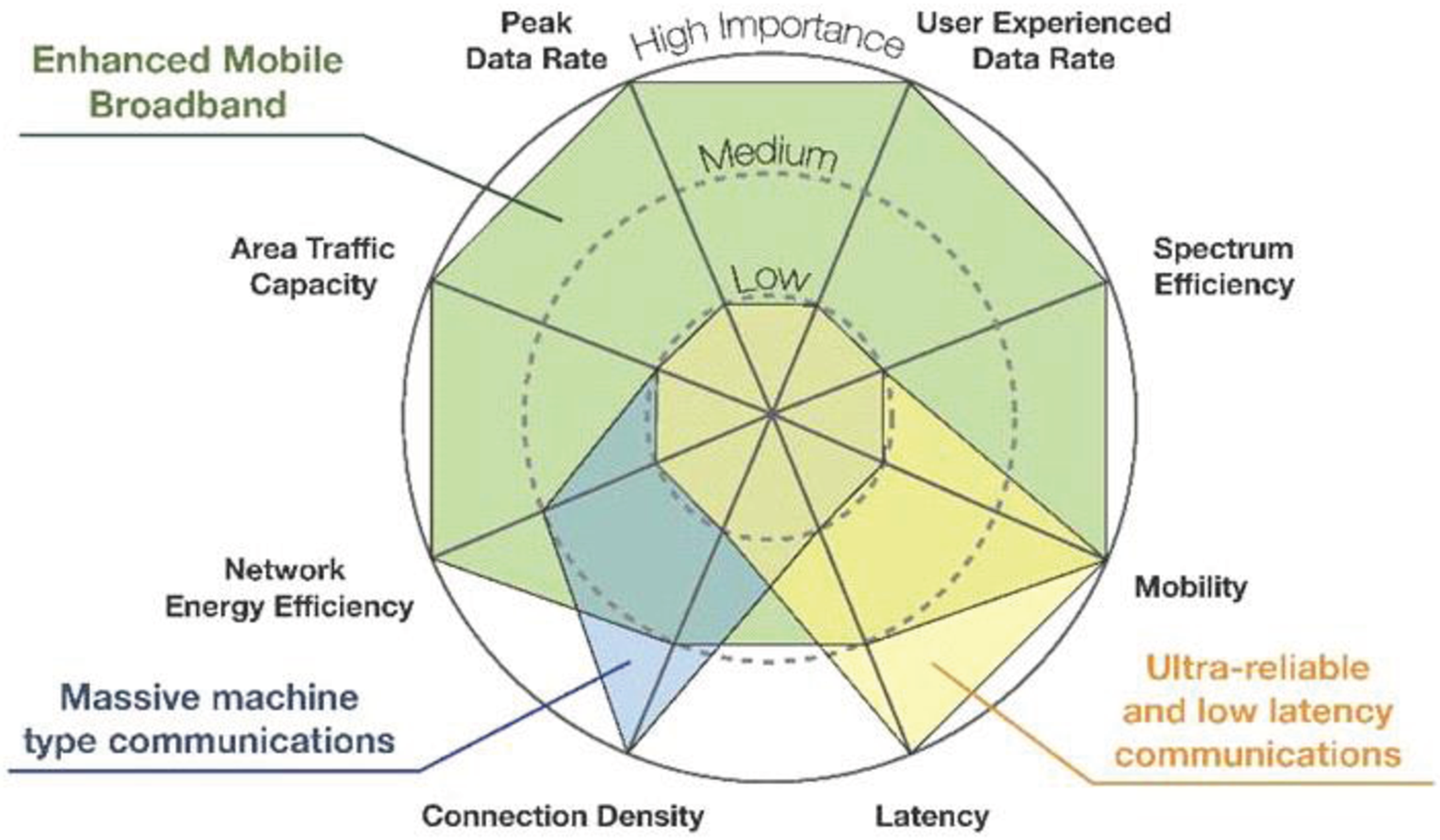}
 \end{center}
  \vspace{-2.0cm}
\caption{5G use cases.}
\label{fig:5G_UseCases}
\end{figure}
\begin{figure}[h!]
 \begin{center}
  \includegraphics[scale=0.3]{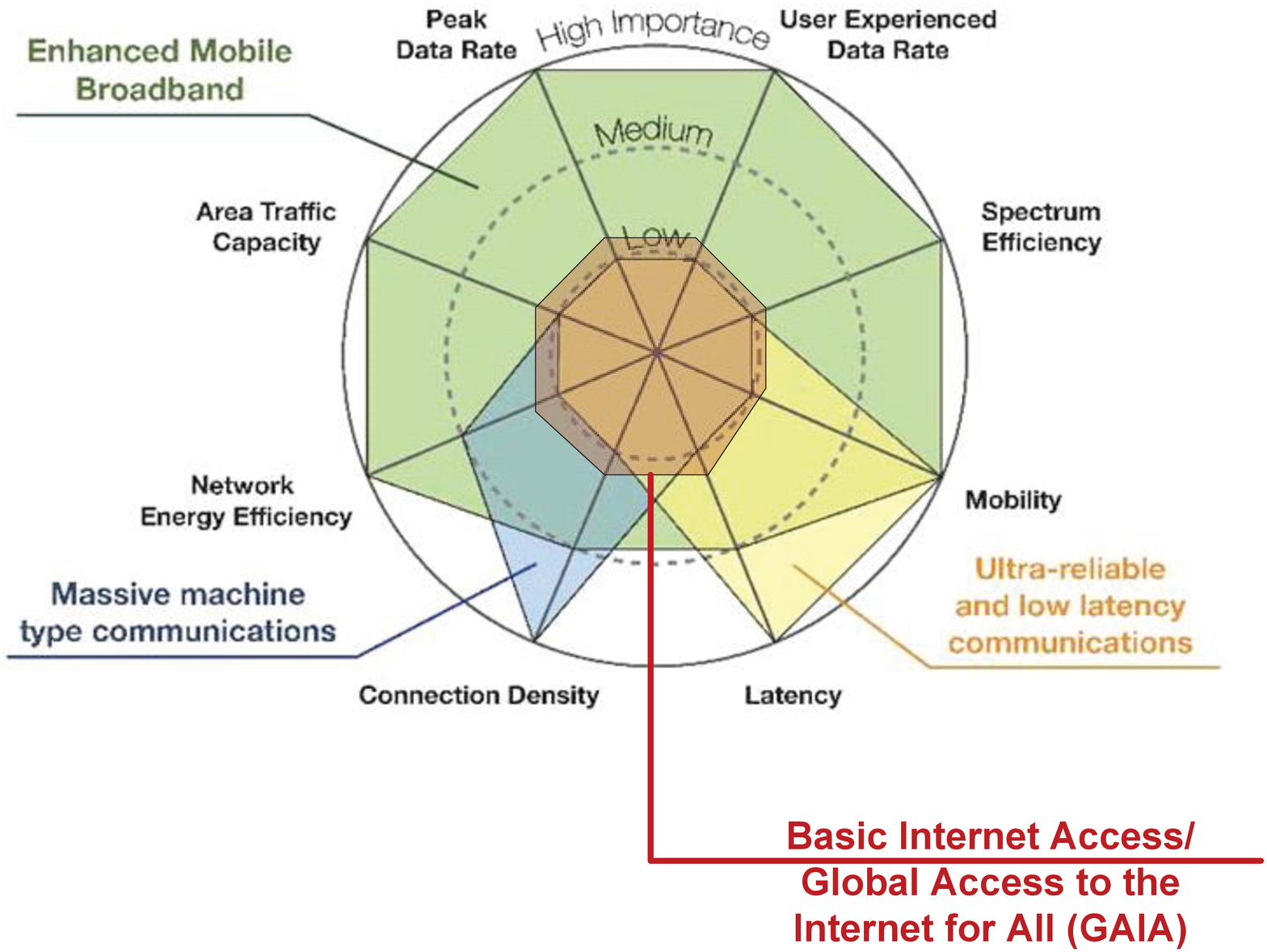}
 \end{center}
  \vspace{-1.0cm}
\caption{5G use cases with ubiquitous basic internet connectivity.}
\label{fig:5G_UseCases_GAIA}
\end{figure}

5G defines three main use cases, representing the pillars of 5G and shown in Fig.~\ref{fig:5G_UseCases}~\cite{ITU_R_M2083}:
\begin{itemize}
  \item Enhanced mobile broadband (eMBB): this use case deals with the increasing load on cellular systems due to the explosion in rich media content, such as audio, video and gaming. It should cope with the tremendous increase in demand for high data rates due to the use of real-time video streaming, social media, large downloads, etc.
  \item Massive machine-type communications (mMTC): This use case is dedicated to accommodate a large number of sensing devices. A major challenge for 5G cellular systems is to handle the dense M2M traffic emanating from IoT devices. Many of these devices will access the network frequently and periodically to transmit relatively short amounts of data.
  \item Ultra-reliability and low-latency communications (URLLC): This involves 5G mission critical services and tactile internet. Tactile internet can be used for example to perform a remote surgery operation by a physician located hundreds of kilometers away using VR/AR techniques.
\end{itemize}

However, these use cases require extensive deployment of infrastructure that can support the high rate low latency communications, which is mostly available in urban areas~\cite{5G_BlindSide}. A major challenge is to provide internet connectivity to the unconnected population of the world, located mostly in rural areas of developing countries~\cite{McKinsey_Report_2014}. Thus, there is a need for a fourth pillar or use case, corresponding to \lq\lq basic internet connectivity\rq\rq or \lq\lq global access to the internet for all (GAIA)\rq\rq~\cite{5G_BlindSide}, as shown in Fig.~\ref{fig:5G_UseCases_GAIA}. This \lq\lq affordable broadband" pillar is considered as the fourth pillar of 5G, leading to an Enhanced-5G standard. It is based mainly on: (i) using unlicensed spectrum, including white space, (ii) energy efficiency, and (iii) using SDN and NFV to reduce infrastructure costs~\cite{IEEE_C215}. As shown in Fig.~\ref{fig:5G_UseCases_GAIA}, the requirements of this fourth pillar do not exceed those of the other three in terms of data rates, device density, or latency. However, the main requirement corresponds to ubiquitous geographical coverage leading to basic connectivity anywhere anytime, with the other 5G use cases providing advanced connectivity in hotspots, in the hope that their coverage growing to cover gradually the areas with little or no connectivity. This scenario corresponds to what we call in this paper as the \lq\lq Beyond 5G or 6G challenge\rq\rq, aiming to connect the remaining unconnected population.

\subsection{Paper Contributions}
\label{subsec:Paper_Contributions}
To the best of the authors' knowledge, there are no detailed survey papers dedicated to the latest solutions for rural connectivity. There are however survey papers addressing fronthaul and backhaul solutions. The most relevant to this work are~\citep{Backhaul_Paper, 5G_Fronthaul_Paper, 5G_Backhaul_Paper}.

The authors of~\citep{Backhaul_Paper} survey the technologies used for backhaul connectivity. Since~\citep{Backhaul_Paper} was published in 2011, a significant part of the paper discusses circuit-switched networks, in addition to packet-switched backhaul. However, rural connectivity is not within the main scope of the paper and is only briefly tackled. In this paper, (i) we focus more on connectivity for rural and remote areas, (ii) we provide a discussion of more recent technologies such as 5G and beyond, (iii) we present the latest advances related to backhauling with UAVs and balloons, and (iv) we present the latest trends and breakthroughs in satellite connectivity.

In~\citep{5G_Fronthaul_Paper}, 5G RAN fronthaul solutions are surveyed in order to meet the 5G performance requirements. In addition,~\citep{5G_Backhaul_Paper} discusses challenges for 5G backhaul connectivity in order to meet the stringent requirements of data rate and latency at the 5G RAN. Thus, it focuses mostly on UDNs and does not consider rural connectivity.
In this paper, we complement the work of~\citep{Backhaul_Paper, 5G_Fronthaul_Paper, 5G_Backhaul_Paper} by focusing on providing connectivity for rural areas. In addition, although we discuss 5G as a possible solution, several other technologies are analyzed and numerous case studies are presented.

\begin{figure}[t!]
 \begin{center}
  \includegraphics[scale=0.8]{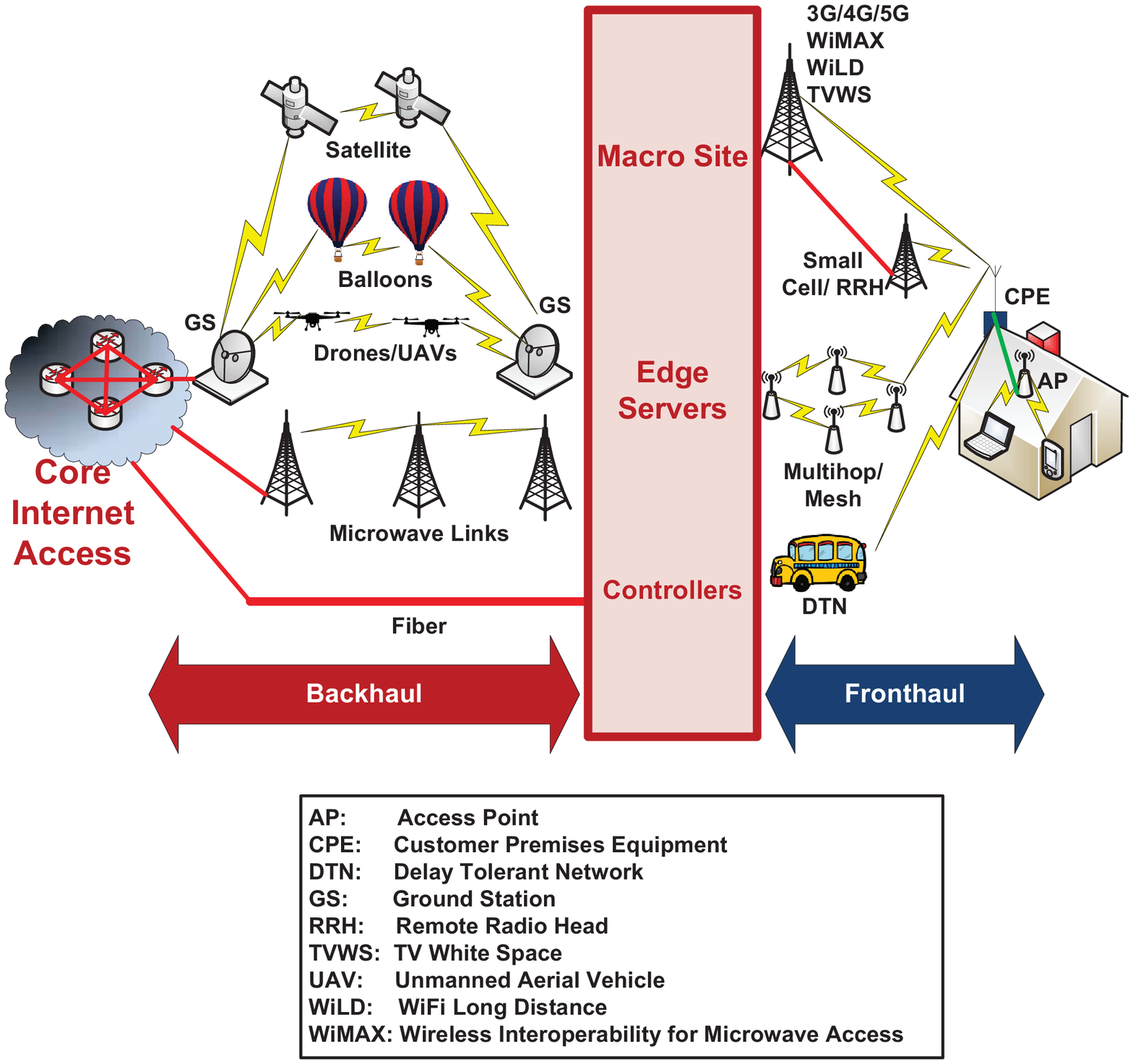}
 \end{center}
  \vspace{-6.0cm}
\caption{Fronthaul and Backhaul examples.}
\label{fig:Fronthaul_Backhaul_Types}
\end{figure}

Different papers in the literature use the terminology referring to fronthaul, backhaul, and midhaul in different ways.
For example, consider a WiFi AP providing access inside a home is connected to a wireless mesh or TVWS network before reaching a site connected to a VSAT terminal or to a fiber PoP. Several papers would treat this scenario by considering WiFi as fronthaul and the mesh or TVWS network as backhaul. In this paper, we consider the more general approach by considering that the satellite or fiber form the backhaul, whereas the multihop or TVWS network is part of the fronthaul, in conjunction with the WiFi network used for direct access. Thus, we follow the following definitions~\citep{FrontMidBackHaul_Website} with illustrative examples shown in Fig.~\ref{fig:Fronthaul_Backhaul_Types}:
\begin{itemize}
  \item Backhaul is the connection to the core network or to the internet, e.g., through a macro BS, a fiber PoP, or a VSAT satellite terminal.
  \item Fronthaul is the connection between the AP and the small cell BS or RRH. It can also denote the access connection between the UE and the AP or RRH.
  \item Sometimes, the connection between the RRH and macro BS could take several hops. Midhaul is referred to the connection that feeds the next link between RRH and BS. In this paper, it is mostly treated as part of the fronthaul section.
\end{itemize}

The rest of the paper is organized as follows: Section~\ref{sec:Backhaul} surveys the technologies used for rural backhaul connectivity, whereas Section~\ref{subsec:Backhaul_Cost_Tradeoffs} analyzes their cost and CAPEX/OPEX tradeoffs. Section~\ref{sec:Fronthaul} surveys the technologies used for rural fronthaul connectivity, whereas Section~\ref{subsec:Fronthaul_Tradeoffs} analyzes relevant issues like electricity provision, spectrum issues, and economical aspects. Section~\ref{sec:Services} surveys the main applications or services that can be provided by connectivity in rural areas, and discusses the barriers that need to be overcome during initial deployment, like increasing user awareness, starting with low-cost simple technologies, and using local content. Section~\ref{sec:Case_Studies} presents an overview of some cases studies for providing connectivity to rural areas, grouped by country, whereas Section~\ref{subsec:Foundations} describes the main foundations or companies launching initiatives for ubiquitous connectivity and for providing internet access to rural and under-privileged areas.  Then, Section~\ref{sec:Future_Directions} discusses future trends and how to build on the efforts providing basic connectivity to rural areas in order to provide more advanced connectivity levels. Finally, Section~\ref{sec:Conclusions} concludes the paper.

\section{Technologies for Backhaul Connectivity}
\label{sec:Backhaul}
In this section, we present an overview of the main backhaul technologies used for providing connectivity to rural areas.

\subsection{Fiber Optics}
This solution consists of laying fiber optic cables throughout the long backhaul distance. Although in some situations it is not feasible (or extremely costly) due to the geographical/terrain considerations (e.g. mountains, lakes, etc.), it remains a possible solution in other scenarios (e.g., plains, desert areas). The main limitation is the cost of civil works, which are generally much more expensive than the cable cost. Although the installation costs can reach 200 USD/meter in dense urban areas~\citep{Zeeshan_FSO_ComMag_2018}, they go down to 30 USD/meter in rural areas, as noted in \citep{Ceragon_WhitePaper_2009}. This clearly plays in favor of fiber deployment when severe geographical constraints do not exist.

To minimize the deployed fiber lengths in rural areas, an optimization approach for point to multipoint communications in GPON networks is proposed in~\cite{IEEE_C91}. The approach finds the minimum spanning tree connecting the customer premises equipment (or the optical equipment that are closest to the customers) to the equipment at the nearest central office, while using a weighted approach taking into account the nature of the terrain and soil composition in the optimization, as these are important factors for fiber deployment in rural areas~\cite{IEEE_C91}. In~\cite{IEEE_C201}, to avoid having under-utilized trees in rural PON networks, a chain of amplifier nodes is used. In both rural and urban cases, the access PON network is connected to a DWDM backbone ring in the approach of\cite{IEEE_C201}. Although PON networks were initially devised for access/fronthaul networks, they are proposed in~\cite{IEEE_C166} as a backhaul solution for 4G/5G wireless networks in rural areas, and a cost analysis is performed for connecting Greek islands with backhaul PON-based optical rings.

In~\cite{IEEE_C66}, RoF is considered to provide backhaul connectivity in rural areas. The objective is to minimize the cost of fiber deployment. First, neighboring villages are clustered, and the Voronoi tessellation is implemented. Then, the positions of the BSs in each area are determined by starting from a certain position, implementing a technique solving the Traveling Salesman Problem to minimize the length of fiber that needs to be deployed between BSs, and then varying the BS positions using Genetic Algorithm. The process is repeated until a suitable placement and its corresponding minimum length fiber deployment are achieved~\cite{IEEE_C66}.

When fiber deployment is too costly, especially when there is no adequate transportation infrastructure (e.g., fiber can be deployed along railroad or power lines~\cite{IEEE_C91}), wireless solutions are more adequate~\cite{nonIEEE_Brewer_RuralTech}, as investigated in the following subsections.

\subsection{Microwave}
This solution consists of placing RF equipment on towers, such that the backhaul transmissions occur over licensed frequencies. In urban areas, the microwave equipment can be placed on poles or the rooftops of existing buildings. In this case, the pole leasing costs should be taken into account in OPEX calculations~\citep{Zeeshan_FSO_ComMag_2018}. However, in long backhaul stretches traversing non-populated areas in order to reach remote rural agglomerations, appropriate towers need to be built. The tower costs should be included in CAPEX, and are estimated at around $50,000$ USD/tower~\citep{Ceragon_WhitePaper_2009, SAWC_2012}. Since microwave frequencies are licensed, appropriate fees should be regularly paid. Instead of considering spectrum costs per capita as in~\citep{Zeeshan_FSO_ComMag_2018}, which is more suitable for urban areas, a cost per link seems more appropriate for low density rural areas, in line with the cost presented in~\citep{Diaz_FiberBackhaul_2015}, since the investigated rural scenario corresponds to a sparsely populated area. In Section~\ref{subsec:Backhaul_Cost_Tradeoffs}, we the costs for different microwave separation distances between towers: $3$, $5$, and $10$ km. This reflects different deployment conditions depending on geographical and weather constraints. It should be noted that, in practical scenarios, in order to increase cost efficiency and adopt a sustainable business model, a wholesale provider can build towers covering a rural area of interest, and multiple mobile network operators could use these towers to route their traffic. Each network operator would then share his profits with the wholesale provider~\cite{Net_Coverage_Maps}.

\subsection{Free Space Optics (FSO)}
FSO can be used either in terrestrial deployments, by placing FSO equipment over transmission towers as in the case of microwave links, or in \lq\lq vertical" deployments, where FSO is used for communications between HAPs, UAVs, balloons, satellites, and/or between these entities and ground stations.

\subsubsection{Terrestrial FSO}
This solution assumes the deployment of FSO towers to ensure backhaul connectivity. We assume tower costs similar to those of microwave towers. This solution does not involve spectrum licenses since it is based on light transmission. However, it is sensitive to certain weather conditions such as fog, and to alignment errors~\citep{Alheadary_Thesis_KAUST_2018, HT_FSO_Weather_RadioEng_2016}.  Therefore, in Section~\ref{subsec:Backhaul_Cost_Tradeoffs}, we investigate different separation between terrestrial FSO towers within the ranges discussed in~\citep{HT_FSO_Weather_RadioEng_2016}: $0.5$, $3$, and $5$~km.

In~\cite{IEEE_C177}, the concept of RoF, where RF signals are modulated over optical carriers, is extended to FSO systems to build a radio over FSO (RoFSO) system. Signals over a fiber backbone are transferred to FSO in order to reach rural or hard to reach areas where fiber optic cables are not available. The system designed in~\cite{IEEE_C177} was tested and achieved almost error free transmission in good weather conditions, but was sensitive to rainfall and to scintillation effects.

\subsubsection{Vertical FSO}
In situations where the erection of towers is not practical, FSO backhaul communications can be performed using HAPs such as drones or UAVs. Since these devices hover at higher altitude, they are less sensitive to weather conditions (e.g., they fly above fog), they can be separated by longer distances than terrestrial FSO, e.g., distances of $5$, $10$, and $20$~km can be considered between two flying platforms. However, they cost around $50,000$ USD per platform, and their operational costs are estimated to $859$~USD per flying hour~\citep{Zeeshan_FSO_ComMag_2018}. Nevertheless, when such devices can be fully solar powered, their operational costs can be significantly reduced. In the cost analysis of Section~\ref{subsec:Backhaul_Cost_Tradeoffs}, we assume a maintenance cost comparable to that of microwave links (considered to be $375$~USD/year in~\citep{Zeeshan_FSO_ComMag_2018}) and add an increase of $33$\% to account for the additional complexity of the equipment, such that the total becomes $500$~USD/year.

FSO links for satellite communication have also received attention in the literature. In~\cite{FSO_Feeder_VHTS}, optical feeder links based on DWDM FSO were proposed for very high throughput satellite systems. The objective is to increase the throughput between ground stations and GEO satellites, which would significantly reduce the number of needed stations with traditional Ku or Ka bands. A 15-year roadmap for the development of the proposed system is presented in~\cite{FSO_Feeder_VHTS}. In~\cite{FSO_Feeder_VHTS2}, the performance of such a system is analyzed in the presence of atmospheric turbulence and enhancements were obtained by using a zero-forcing precoder proposed in~\cite{FSO_Feeder_VHTS2}. The communication between satellites and users was considered in~\cite{FSO_Feeder_VHTS2} using Ka band RF multibeams.

Concerning the use of FSO for satellite communications, it should be noted that most satellite networks use RF/mmWave frequencies (C/Ku/Ka) for communication between satellites and ground stations. These same frequencies can be used for inter-satellite communication. However, FSO use is becoming increasingly popular for inter-satellite communication in space, due to providing high bandwidth, having high directivity, and requiring less power and mass for the transceivers~\citep{KK_ComST2017}. It can be (and is sometimes) used for satellite-ground station communication, but faces significant challenges compared to its use for satellite-satellite communications~\citep{KK_ComST2017, SGH_FSODL_2018}.
Challenges for using FSO for satellite-to-ground communications include~\citep{KK_ComST2017}:
\begin{itemize}
  \item Absorption and scattering loss (by the gas molecules and aerosols particles present in the atmosphere),
  \item Attenuation due mostly to fog, but also rain and snow,
  \item Atmospheric turbulence, due to the variation of the temperature and pressure along the propagation path in the atmosphere,
  \item Beam divergence loss, due to diffraction close to the receiver's aperture,
  \item Background noise from the sun, or from diffracted light collected by the receiver, and
  \item Cloud blockage.
\end{itemize}

In~\citep{SGH_FSODL_2018}, the authors indicate that these challenges cannot allow a guaranteed data rate between satellites and ground stations using FSO, but operation is still possible by adaptively varying the achievable data rate.
The challenges in inter-satellite FSO communications are mainly due to the relative movement of the satellites, leading to Doppler shift, and to satellite vibration and tracking~\citep{KK_ComST2017}. These challenges are also partially existing in satellite-to-ground communications.
Hence, the challenges for FSO in space are less critical, and indeed significantly better performance can be achieved, especially due to the fact that the optical signal is not traversing the atmosphere. In 2008, an optical inter-satellite link achieved a data rate transmission of $5.6$~Gbps for distances up to $5,000$~km in the European Data Relay satellite System (EDRS). Newer terminals can transmit at a data rate of 1.8 Gbps for distances up to $45,000$~km. The inter-satellite Ka band link achieved $300$~Mbps~\citep{EDRS_Overview}. In EDRS, LEO satellites relay the data from ground stations to GEO satellites using FSO for the GEO-LEO satellite communication, and RF bands for communication with the ground stations (similarly to the satellite-satellite Ka link, the satellite-ground Ka link achieved $300$~Mbps)~\citep{EDRS_Overview}. Some of the recent satellite constellations discussed in Section~\ref{subsubsec:Backhaul_Sat}, e.g., Starlink, are also expected to use FSO for its inter-satellite communication ~\citep{UT_article_Starlink_2018, WP_Article_2019, SpaceX}.

\subsection{HAPs/Drones/UAVs/Balloons}\label{subsubsec:Backhaul_UAV}
To solve the coverage problem in large rural areas without relying on the deployment of costly wired infrastructure for backhaul, HAPs and UAVs are generally recommended~\cite{IEEE_BC1}.

UAVs are proposed in~\cite{IEEE_J11} to provide backhaul connectivity for ground 5G base stations. A steering algorithm to allow the antennas of the UAV and the ground BS to be dynamically steered towards each other is proposed in~\cite{IEEE_J11}, and tested via an experimental setup. In\cite{TVWS_HAP}, a multihop network of HAPs is proposed for providing backhaul connectivity to TVWS deployments in rural areas. The HAPs communicate with each other using mmWave frequencies, and with the ground stations using FSO.

An interference alignment scheme to maximize the sum-rate capacity of HAPs communicating with GSs is proposed in~\cite{IEEE_J28}. The scheme assumes $M$ HAP drones and $N$ GSs, with no CSI at the HAPs. A tethered balloon with $(M-1)\times (N-1)$ antennas is used as a decode and forward (DF) relay with half-duplex operation: The HAPs transmit their data to the GSs first, then the tethered balloon relays the data after performing pre-coding~\cite{IEEE_J28}.

Aerostats filled with lighter than air gas are tethered to the ground and used to provide connectivity to rural areas in~\cite{IEEE_C17}. Although these tethered aerostats can be used to provide backhaul connectivity by communicating with each other using directive antennas, they were also used in~\cite{IEEE_C17} to provide fronthaul WiFi access by using high gain omnidirectional antennas. The performance of a balloon tethered at $200$~m altitude and providing WiMAX connectivity was evaluated in~\cite{IEEE_C187} via simulations and shown to perform effectively in terms of delay, throughput, and traffic load.

Balloons can currently provide cellular connectivity to rural areas~\cite{IEEE_Spectrum_Loon2}, and are planned to provide backhaul connectivity to 5G networks~\cite{IEEE_Spectrum_Loon1}. Multiple balloons communicate in a multihop fashion using mmWave frequencies, before reaching a ground station. For example, seven balloons were able to relay a signal over a distance of $1,000$~km~\cite{IEEE_Spectrum_Loon2}. Furthermore, significant progress has been made in controlling their navigation paths, and they can stay longer above a target area. For example, balloons launched from Australia were used to cover areas in New Zealand or Argentina~\cite{IEEE_Spectrum_Loon2}.

\subsection{Satellite}\label{subsubsec:Backhaul_Sat}
Even before the internet, satellite backhaul connectivity was considered to provide basic telephony services to rural areas~\cite{IEEE_C19}.

Recently, satellites are being considered for providing 5G connectivity, with plans to reduce latency and use virtualized network functions~\cite{Sat5G_Proj_Website}. They can also provide a parallel backhaul link for optimizing resources whenever a terrestrial backhaul is available~\cite{Sat_5G_CTN}. Indeed, an SDN based approach is proposed in~\cite{IEEE_C117}, where satellite and terrestrial backhaul are integrated in 5G networks. Traffic can be routed dynamically on either the satellite or terrestrial backhaul, or split over both, depending on quality of service requirements and traffic engineering policies~\cite{IEEE_C117}.

Satellite communications are expected to provide backhaul connectivity for the IoT~\citep{Sat_5Trends_ITU, IEEE_J9}, where for example sensors in a remote agricultural area or those used to monitor the environment in the Amazonian jungle can benefit from even low bandwidth small satellites (SmallSats) that collect the data periodically~\cite{Sat_5Trends_ITU}.

Satellites can operate at~\citep{IEEE_J9}:
\begin{itemize}
  \item Geostationary orbit (GEO) at an altitude of around $36,000$ km, where the satellite remains pointed towards the same location on earth,
  \item Medium earth orbit (MEO), where the satellite operates at altitudes between $2,000$ and $35,000$ km, and
  \item Low earth orbit (LEO), where the satellite operates at altitudes between $160$ and $2,000$ km.
\end{itemize}

GEO satellites can be used with cheap user equipment at the ground stations, since the satellite is at a fixed location with respect to the ground. LEO and MEO satellites have the advantage of lower latency due to lower altitude, but require more complex antennas at the satellite and more importantly at ground stations in order to track the satellite in orbit~\citep{IEEE_J9}. Some satellites use elliptical orbits, e.g., the Molniya constellation consisting of five satellites, in order to solve this problem since they need 1D scanning instead of 2D antenna scanning by ground station antennas~\citep{IEEE_J9}.

Satellite communications are overcoming the traditional challenges that prevented them previously from being a competitive backhaul solution. Indeed, there are several reasons that allow satellite to compete for backhaul connectivity. Mainly, the satellite operator can act as a service provider for the MNO~\citep{ESOA_WhitePaper_2016}, and thus the satellite launch and management costs do not affect the MNO directly, as long as the costs of bandwidth leasing and/or of SLA between MNO and satellite operator allow satellite to compete with other backhaul technologies~\citep{ESOA_WhitePaper_2016}.
Therefore, the following advancements led to addressing most of the challenges traditionally faced by satellite connectivity and especially the use of satellite for backhaul:
\begin{itemize}
  \item Increase in capacity and decrease in cost per Mbps: The main driver for the decreased cost per Mbps over satellite links is the increasing use of HTSs~\citep{Hughes_SatBackhaul}. HTS can achieve $20$~times more throughput than the traditional fixed satellite service~\citep{ADLittle_HTS, MWEE_Sat_Trends}. They allow for narrower beams, with frequency reuse across multiple beams, thus reusing multiple spot beams to cover a service area, as opposed to the traditional wide beam approach~\citep{ADLittle_HTS}. Furthermore, they are launched in GEO, LEO, and MEO orbits~\citep{MWEE_Sat_Trends}. This increase in capacity and the abundance of offered HTS bandwidth led to a decrease in costs per Mbps~\citep{Hughes_SatBackhaul, Newtec_SatBackhaul}. The price decrease has followed an exponential trend in the last few years, and it is expected to continue, albeit with an almost linear slope, in the coming years~\citep{Gilat_WhitePaper_2018}.
  \item Decrease in delay: Latency in satellite links has always been a problem that cannot be overcome due to the distance traveled by the signal. However, latency is being reduced through innovative techniques in other parts of the protocol stack, since not much can be made about the signal travel time. For example, TCP acceleration is used to decrease latency~\citep{Newtec_SatBackhaul, Gilat_WhitePaper_2018}. In addition, the increased use of LEO and MEO satellites can partially reduce the delay problem, since they operate at lower altitudes than GEO satellites.  Furthermore, another method that can reduce delay is the operation of satellite backhaul networks at Layer~2, since traditionally satellite communications are operated at Layer~3 of the protocol stack. This Layer~2 operation not only reduces latency but also makes integration of satellite backhaul with the traditional MNO backhaul easier~\citep{Gilat_WhitePaper_2018}.
  \item Decrease in complexity: MNOs do not have to worry about the management of their satellite backhaul connectivity, since there are several flexible business models that govern their relationship with the satellite operator. In fact, the satellite operator can offer a leasing agreement to MNO, or they can sign a SLA where the MNO sets the KPIs and target values, whereas the satellite operator ensures that the required targets are met~\citep{ESOA_WhitePaper_2016, Gilat_WhitePaper_2018}.
\end{itemize}

In addition, flexible business models allow MNOs to use satellite backhaul as a black box and focus on running their network and managing their business. The following business models were described in~\citep{ESOA_WhitePaper_2016}:
\begin{itemize}
  \item \lq\lq {\em Directly contracting with satellite operators for raw capacity. In this case, the MNO leases satellite capacity, buys a hub, and runs its own satellite network.
  \item Directly contracting with satellite operators for an end-to-end managed service solution (\lq\lq one stop shopping\rq\rq). In this case, the satellite operator provides and manages the ground equipment, bandwidth, and support, based on a Service Level Agreement (SLA).
  \item Entering into a service agreement with a service provider or who provides end-to-end connectivity solutions and operates the satellite network.}\rq\rq
\end{itemize}

Next, we provide an overview of HTS satellite networks aiming to provide global connectivity, in addition to CubeSat networks that are gaining increasing popularity, although their current role in providing ubiquitous coverage seems to be limited.

\subsubsection{HTS LEO Satellite Networks in Space}
In this section, we describe some of the new satellite networks that are launching large numbers of small satellites in order to provide large bandwidth capacity with the aim of providing internet connectivity to every corner of the world~\citep{WP_Article_2019}. They are contributing to the proliferation of HTS satellites discussed in the previous sections, and more specifically at LEO orbits.
OneWeb plans to launch a network of $2,000$ LEO satellites operating at an altitude $1,200$ Km in collaboration with Airbus. The first six satellites of this constellation were launched in February 2019. Satellites are around $150$ Kg in mass, and operate in the Ku band~\citep{WP_Article_2019, OneWeb}.
But OneWeb is not alone. SpaceX, founded by Elon Musk, plans to launch a constellation of $12,000$ satellites, named \lq\lq Starlink\rq\rq. It has the same objective of providing internet to the underserved areas of the planet. Furthermore, SpaceX plans to use parts of the profits earned from this project to fund its Mars colonization plans! The first satellites to be launched are also in the low weight category ($100$-$500$ Kgs) and will operate from a LEO orbit between $1,100$-$1,300$ Km (although a couple of test satellites are orbiting at $500$ Km). They will operate in the Ku/Ka bands in their communication with ground stations, but can use FSO for their inter-satellite communication in space~\citep{UT_article_Starlink_2018, WP_Article_2019, SpaceX}. Amazon has also joined the race with its Kuiper project, with plans to deploy $3,236$ LEO satellites~\cite{Sat_Amazon_Kuiper}.

The deployment of Starlink and OneWeb satellites at these altitudes reduces significantly the GEO satellite delay from $600$ ms by around an order of magnitude. However, these large numbers of satellites put in orbit pose the problem of end of life issues and adding to the space debris problem. The operators of these constellations (OneWeb and Starlink) have put in place mechanisms for end of life recovery of satellites and for managing them after their service lifecycle is completed.

These large constellations of satellites could rely on satellite images in order to detect the population concentration zones in rural areas in order to direct their beams where they are most needed. For example, an approach based on convolutional neural networks and deep learning is used in~\cite{IEEE_C144} in order to detect building in rural areas from satellite images. A software system is proposed in~\cite{IEEE_C162} to detect buildings from aerial photographs. When coupled with other methods to determine network coverage gaps in these areas, e.g.~\cite{Net_Coverage_Maps}, the planning could target the uncovered areas more efficiently.

\subsubsection{CubeSat Networks}
CubeSats are small satellites with cubic sizes denoted as 1U, 2U, and so on, with \lq\lq 1U\rq\rq corresponding to a $10\times 10\times 10$ cm$^3$ cube~\cite{AK_CubeSat_2019}. Despite the increasing popularity of CubeSats and the increase in their number of launches, CubeSat networks are not currently used for backhaul connectivity. In fact, their satellite to ground communications require high energy consumption to achieve data rates only in the order of kilobytes per second~\citep{Ennis_CubeSat_Thesis}. For example, a Swiss company named ELSE, plans to launch a constellation of $64$ CubeSats, called Astrocast~\citep{ELSE_Astrocast}. The Astrocast platform aims to serve users with satellite phone calls in fixed areas allowing the transmission of only $1$ KB of data per day~\citep{ELSE_Astrocast, AK_CubeSat_2019}.

Recent advances in research are however leading to increasing data rates~\citep{AJN_CubeSat_2019}. Kepler Communications in Canada plans to launch $140$ CubeSats in order to develop a satellite backhaul~\citep{Kepler_CubeSat}. However, it was mentioned in~\citep{AJN_CubeSat_2019} that $140$ CubeSats are not sufficient to provide continuous global coverage. At $500$ m altitude, it was shown in~\citep{AJN_CubeSat_2019} that the number of CubeSat satellites needed for continuous coverage is $71$ satellites per orbital plane and $36$ orbital planes are needed, thus $36\times71=2,556$ CubeSats are required. In addition, despite this \lq\lq global coverage\rq\rq, access time is intermittent. For example, with the Fernbank Observatory in Atlanta, Georgia,  considered as a point of interest, it is covered by a first CubeSat for $500$~s, followed by a $100$-second period of no coverage, and then by a period of coverage by a second CubeSat for a duration of $700$~s~\citep{AJN_CubeSat_2019}.

Nevertheless, in~\citep{AJN_CubeSat_2019}, CubeSats with multiband transmission are investigated, where the CubeSat can use the following bands: radio frequencies ($2$–$30$ GHz), millimeter wave ($30$–$300$ GHz), Terahertz band (up to 10 THz), and optical frequencies (with typical bands of $850$ nm/$350$ THz, $1300$ nm/$230$ THz, and $1550$ nm/$193$ THz). Link budget and constellation planning are described, and the results are evaluated via simulations (no actual deployment is yet performed). With these advances, CubeSats can provide an infrastructure extending the IoT to an internet of space things (IoST)~\citep{AK_CubeSat_2019}. They can be used themselves for imaging/sensing and sending the observed data, or for relaying the measurement data of ground sensors in remote locations to a control center, thus providing (possibly multihop) backhaul connectivity in space~\citep{AK_CubeSat_2019} (more accurately, this is similar to fronthaul access described in Section~\ref{subsubsec:Fronthaul_Sat}, but for machine type communications: The satellite is the actual BS, although backhaul is still provided by a network of satellites allowing the data to reach its final destination). As stated in~\citep{AK_CubeSat_2019}, this operation can be envisaged to provide wireless connectivity to remote areas in the same way used for IoST, benefiting from the increased data rates obtained by the multiband approach of~\citep{AJN_CubeSat_2019}.

Since the advanced CubeSat performance shown in~\citep{AJN_CubeSat_2019, AK_CubeSat_2019} is demonstrated via simulations (without actual deployments), then the current main driver for the currently reduced costs and the increased competitiveness of satellite backhaul is the business deployment of HTS satellites through networks like OneWeb and SpaceX, among other networks belonging to other (older) satellite operators. The anticipated role of CubeSats in backhaul connectivity, when the theoretical results reach the deployment phase, is expected to increase competition and reduce prices even further.

\section{Backhaul Cost Issues and Tradeoffs}
\label{subsec:Backhaul_Cost_Tradeoffs}
In this section, the previous solutions for providing backhaul connectivity to rural areas will be investigated in terms of their cost, consisting of CAPEX and OPEX. These include fiber, microwave, and FSO (terrestrial and \lq\lq vertical"), in addition to satellite.

The closest study to this work is in~\citep{Zeeshan_FSO_ComMag_2018}, where the costs of fiber optics, microwave links, in addition to terrestrial and vertical FSO are analyzed and compared. However,~\citep{Zeeshan_FSO_ComMag_2018} considers a fronthaul/backhaul scenario, with a dense deployment of BSs, such that $100$ macro BSs and $1000$ small cell BSs are deployed in an area of $5\times$5 km. Clearly, the results of such a scenario cannot be generalized to a pure backhaul scenario where the objective is to transport the traffic over hundreds of kilometers.
In this paper, we consider remote rural areas where the backhaul link needs to traverse long distances, without necessarily having any access/fronthaul BSs along the way, before reaching the remote area. Such an area is considered in~\citep{IEEE_C34} for example, where $12$ hours are needed to reach the area from the nearest urban town. Thus, stretches of hundreds of kilometers could be traversed by the backhaul link before reaching the target area. Therefore, we base our subsequent cost analysis on a distance of $100$~km, and we analyze the costs of laying fiber, erecting microwave or terrestrial FSO towers, or using HAPs for vertical FSO.  Nevertheless, we use the cost parameters defined in~\citep{Zeeshan_FSO_ComMag_2018}, and make the necessary adjustments whenever these parameters do not apply to the considered rural scenario, while providing appropriate justifications or references.

\begin{table}[h!]
\begin{tiny}
\caption{Parameter Values used in the CAPEX/OPEX Calculations}
\label{tab:Backhaul_Costs}
\begin{center}
\vspace{-1.0cm}
\begin{tabular}{|c|}
 \hline
 {}\\
 \hline
  \begin{tabular}{|p{8.0cm}|p{1.5cm}|p{1.5cm}|p{3.5cm}|}
  \hline
   {\bf Parameter} & {\bf Value} & {\bf Type (CAPEX/ OPEX)}  & {\bf Reference}\\
  \hline
   Total Backhaul Distance (m)  & $100,000$ & - & - \\
  \hline
  \end{tabular}
 \\
 \hline
 {}\\
 \hline
   {\bf Fiber}\\
   \hline
 \begin{tabular}{|p{8.0cm}|p{1.5cm}|p{1.5cm}|p{3.5cm}|}
  \hline
   Fiber Cable Cost (USD/m)  & $10$ & CAPEX & \citep{Zeeshan_FSO_ComMag_2018} \\
  \hline
   Fiber Installation Cost (USD/m)  & $30$ & CAPEX & \citep{Ceragon_WhitePaper_2009} \\
  \hline
   Fiber Equipment Cost (USD)  & $3,500$ & CAPEX & \citep{SAWC_2012} \\
  \hline
   Fiber Power and Maintenance Cost per Link per Year (USD)  & $200$ & OPEX & \citep{Zeeshan_FSO_ComMag_2018} \\
  \hline
  \end{tabular}
  \\
 \hline
 {}\\
 \hline
   {\bf Microwave}\\
   \hline
 \begin{tabular}{|p{8.0cm}|p{1.5cm}|p{1.5cm}|p{3.5cm}|}
  \hline
   Microwave Tower Cost (USD)  & $50,000$ & CAPEX & \citep{Ceragon_WhitePaper_2009} \\
  \hline
   Microwave Equipment Cost (USD)  & $40,000$ & CAPEX & \citep{Ceragon_WhitePaper_2009} \\
  \hline
   Spectrum License Costs per Year per Link (USD)  & $3,000$ & OPEX & \citep{Diaz_FiberBackhaul_2015} \\
  \hline
   Microwave Power and Maintenance Cost per Year (USD)  & $375$ & OPEX & \citep{Zeeshan_FSO_ComMag_2018} \\
  \hline
  \end{tabular}
   \\
 \hline
 {}\\
 \hline
   {\bf Terrestrial FSO}\\
   \hline
  \begin{tabular}{|p{8.0cm}|p{1.5cm}|p{1.5cm}|p{3.5cm}|}
  \hline
   Terrestrial FSO Tower Cost (USD)  & $50,000$ & CAPEX & \citep{Ceragon_WhitePaper_2009} \\
  \hline
   Terrestrial FSO Equipment Cost (USD)  & $20,000$ & CAPEX & \citep{Zeeshan_FSO_ComMag_2018} \\
  \hline
   Terrestrial FSO Link Maintenance Cost per Year (USD)  & $8,000$ & OPEX & \citep{Zeeshan_FSO_ComMag_2018} \\
  \hline
  \end{tabular}
   \\
 \hline
 {}\\
 \hline
   {\bf Vertical FSO}\\
   \hline
  \begin{tabular}{|p{8.0cm}|p{1.5cm}|p{1.5cm}|p{3.5cm}|}
  \hline
   Vertical FSO Equipment Cost (USD)  & $50,000$ & CAPEX & \citep{Zeeshan_FSO_ComMag_2018} \\
  \hline
   Vertical FSO Operation Cost per Hour (USD)  & $859$ & OPEX & \citep{Zeeshan_FSO_ComMag_2018} \\
  \hline
   Vertical FSO Operation Cost per Year (USD)  & $7,524,840$ & OPEX & Calculated from previous row \\
  \hline
   Vertical FSO Solar Powered Operation Cost per Year (USD)  & $500$ & OPEX & Assumed Comparable to microwave maintenance costs (with 33\% increase) \\
  \hline
  \end{tabular}
   \\
 \hline
 {}\\
 \hline
   {\bf Satellite}\\
   \hline
  \begin{tabular}{|p{8.0cm}|p{1.5cm}|p{1.5cm}|p{3.5cm}|}
  \hline
   VSAT Equipment Cost (USD)  & $4,000$ & CAPEX & \citep{Gilat_WhitePaper_2015} \\
  \hline
  Satellite Hub $+$ Installation Cost (USD)  & $500,000$ & CAPEX & \citep{Gilat_WhitePaper_2015} \\
  \hline
   HTS Capacity Cost (USD/Mbps/Month)  & $250$ & OPEX & \citep{Gilat_WhitePaper_2018} \\
  \hline
   Annual Maintenance Costs (\% of CAPEX)  & 15\% of CAPEX & OPEX & \citep{Gilat_WhitePaper_2015} \\
  \hline
  \end{tabular}
    \\
 \hline
 {}\\
 \hline
\end{tabular}
\end{center}
\end{tiny}
\end{table}

\begin{figure}[t!]
 \begin{center}
  \includegraphics[scale=0.6]{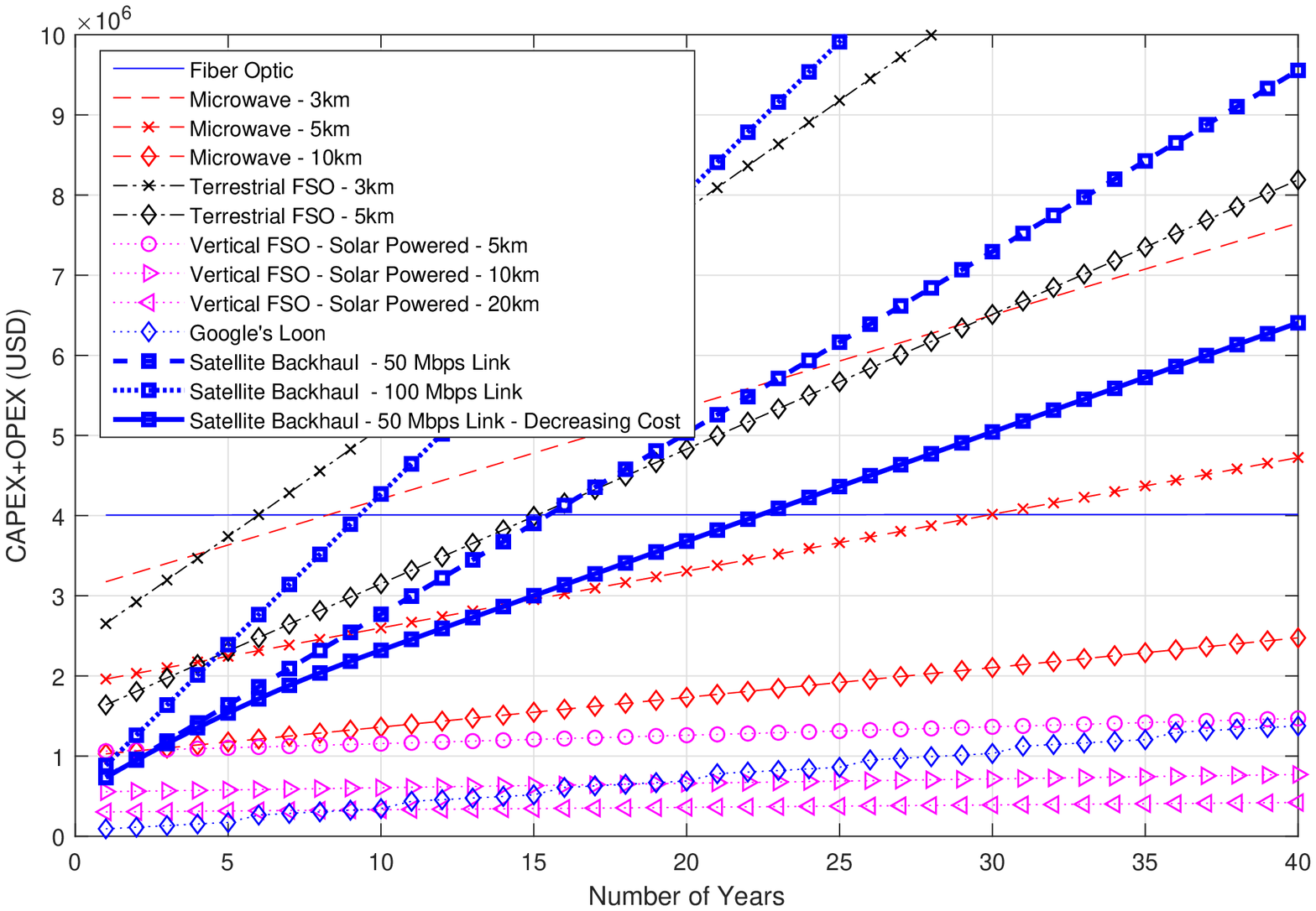}
 \end{center}
\caption{Long-term Backhaul CAPEX and OPEX costs.}
\label{fig:Backhaul_Costs}
\end{figure}

In~\citep{Zeeshan_FSO_ComMag_2018}, the costs shown are for CAPEX in addition to 1-Year OPEX costs. However, to understand the cost tradeoffs between the various technologies, a longer time visibility is needed. In fact, a relatively high CAPEX technology can be more cost efficient than another technology with lower CAPEX, if the first technology has a lower OPEX, with savings that accumulate over the years. The Vertical FSO scenarios are an order of magnitude more costly than the other scenarios, which is consistent with the 1-Year results of~\citep{Zeeshan_FSO_ComMag_2018}. Hence, they are not shown in the comparisons to provide more clarity in the figure. The results without Vertical FSO ar shown in Fig.~\ref{fig:Backhaul_Costs}. Although the costs of fiber deployment appear to be constant, they are increasing slowly. The reason is that we have only one link in the studied scenario of $100$~km, and thus no repeater maintenance costs. In fact, the distance between repeaters can be shown to be in the order of $350$-$400$~km~\citep{Tomar_FiberRepeater_IRJET2017}. This low OPEX ($200$ USD per link per year as in Table~\ref{tab:Backhaul_Costs}) allows fiber to be competitive in the long run. In fact, Fig.~\ref{fig:Backhaul_Costs} shows that Terrestrial FSO is more cost effective than fiber in the first $6$ and $15$ years for a tower separation of $3$~km and $5$~km, respectively. Afterwards, fiber becomes more cost efficient.

Similarly, Fig.~\ref{fig:Backhaul_Costs} shows that Microwave is more cost effective than fiber in the first eight and $30$ years for a tower separation of  $3$~km and $5$~km, respectively. Afterwards, fiber becomes less costly. Interestingly, solar powered Vertical FSO seems to be very cost efficient on the long run, as it is less costly than fiber even for separation distances of $5$~km.
However, solar powered Vertical FSO can be achieved through different techniques, e.g. solar powered drones or balloons. To consider a concrete example with sufficient details in order to investigate further this scenario, we consider the special case of Google's Loon project~\citep{Loon}.

The Loon project consists of launching balloons in the stratosphere to provide connectivity to rural area~\citep{Loon}. A detailed feasibility study for this project is presented in~\citep{Burr_LoonFeasibility_2015}, where the following information is provided:
\begin{itemize}
  \item Each balloon costs around $17,870$ USD, which we will round to $18,000$ USD in our calculations (CAPEX).
  \item Maintenance costs per balloon amount to $1,230$ USD each $100$ days, which we will round to $5,000$ USD/year.
  \item A balloon covers a diameter of $40$~km, which we will reduce to a $33.3$~km distance, thus requiring three balloons to cover $100$~km. To be on the conservative side, we will assume that four balloons are needed for covering a $100$~km backhaul link.
  \item A balloon's service life is five years. Thus, every five years, we will regenerate CAPEX costs.
\end{itemize}

The cost results are shown in Fig.~\ref{fig:Backhaul_Costs}. Clearly, the Loon project costs are within the range obtained in Fig.~\ref{fig:Backhaul_Costs} for solar powered Vertical FSO. The cost jumps every five years are due to deployment of new balloons.

In addition, Fig.~\ref{fig:Backhaul_Costs} shows that Satellite is more cost effective than fiber in the first $8$ and $15$ years for a link bandwidth of $100$~Mbps and $50$~Mbps, respectively. Afterwards, fiber becomes more cost efficient. Furthermore, Satellite with $100$~Mbps and $50$~Mbps is more cost efficient than Microwave with $3$~km separation for the first $10$ and $23$ years, respectively.

The previous results assumed a cost of $250$~USD/Mbps/Month for a satellite backhaul link. However, according to the forecast in~\citep{Gilat_WhitePaper_2018}, the bandwidth prices are decreasing rapidly due to the increased deployment of HTS satellites. Therefore, we also consider the following scenario:
\begin{itemize}
  \item We assume a $50$~Mbps link.
  \item Prices start from $250$~USD/Mbps/Month.
  \item The price decreases to $100$~USD/Mbps/Month within  ten years following a linear slope (These numbers are inline the estimates provided in~\citep{Gilat_WhitePaper_2018}).
  \item The price stabilizes at $100$~USD/Mbps/Month afterwards.
  \item Other costs remain as indicated in Table~\ref{tab:Backhaul_Costs}.
\end{itemize}

The case corresponding to the above assumption is compared to the previous scenarios as shown in Fig.~\ref{fig:Backhaul_Costs}. The updated Satellite scenario now outperforms Terrestrial FSO throughout the investigated time interval. It also outperforms the scenario with Microwave Links separated by $3$~km for the whole duration, and Microwave Links separated by $5$~km for the first $15$ years.
However, according to the results of Fig.~\ref{fig:Backhaul_Costs} using the assumptions of Table~\ref{tab:Backhaul_Costs}, Satellite Backhaul is still far from competing with Solar-Powered Vertical FSO. It should be noted that the cumulative costs in Fig.~\ref{fig:Backhaul_Costs}  are intentionally calculated over an exaggerated period of $40$ years, in order to show the CAPEX-OPEX tradeoffs between the various solutions on the long run.

\section{Technologies for Fronthaul Connectivity}
\label{sec:Fronthaul}

This section describes fronthaul technologies. Different types of rural regions require different solutions, as there is no single technology that can meet the requirements for any rural setup. In fact, adopting a given solution would depend on the population density, geographic/terrain characteristics, and distance to the nearest gateway/exchange point~\cite{IEEE_C41}, among other parameters. A summary of the main literature investigating backhaul and fronthaul technologies is presented in Table~\ref{tab:Fronthaul_Backhaul_Summary}.

\subsection{5G Networks}
The advanced performance features of 5G namely very high data rates, ultra low latency, enhanced QoS and QoE, correspond mostly to an urban setup where a complex architecture is deployed and high speed backhaul is available~\citep{IEEE_C45, IEEE_J10}. Indeed, some of the main techniques used in 5G such as cell densification, mmWave frequencies, and massive MIMO are mainly concerned with increasing the data rate and less related to providing ubiquitous coverage~\cite{5G_BlindSide}.

Bringing 5G to rural areas and providing global access to the Internet for all~\cite{5G_BlindSide} requires some additional modifications to meet the challenges inherent in rural areas. A network would typically start with reduced features and gradually evolve with time in order to reach the 5G performance targets. For this reason, we mentioned in Section~\ref{subsec:Connectivity_Use_Cases} that rural connectivity might be the fourth use case of 5G, or possibly the combination of the four use cases could form the foundation of 6G.

Thus, in this section, we present an overview of the literature attempting to extend 5G access to rural scenarios. In~\cite{IEEE_C69}, 5G small cells are proposed to provide connectivity to small villages in rural areas. The backhaul is provided wirelessly via massive MIMO. Each SCBS provides coverage to the local village, and uses a small cell backhaul radio based on MIMO to communicate with a central BS equipped with massive MIMO in a nearby town. The central BS equipped with massive MIMO communicates with several SCBSs, and is connected to the backbone network via fiber. The SCBSs can be powered by renewable energy sources to avoid the limited availability of electricity in certain rural areas~\cite{IEEE_C69}. In~\cite{Khaturia2019ConnectingTU}, a frugal 5G network is proposed for rural ares, where the access in villages is provided via WLAN, e.g., WiFi, connected to the 5G fiber backbone via a middle mile network, that could be a multihop network (see Section~\ref{subsubsec:Fronthaul_Multihop}) reaching a macro 5G BS connected to the fiber point of presence. The novelty in~\cite{Khaturia2019ConnectingTU} is in making use of SDN and NFV to propose a fog based architecture where network slicing can be performed closer to the network edge at the access part. Since a significant portion of the rural traffic is local, in addition to the fact that content of common interest to the rural population can be cached closer to the network edge, this solution can lead to faster service without routing unnecessary traffic to the network core.

In~\citep{IEEE_C211, IEEE_C212}, UAVs are used to provide access to rural areas, rather than being used for backhaul transport. Each UAV is assumed to be a moving 5G BS, covering a given area. It communicates with a fixed BS site for backhaul. The fixed BS sites are connected to each other and to the core network via fiber optic cables. In addition, the fixed BSs are powered by solar panels, and the UAVs are battery powered and can be recharged periodically at the BS sites. When a UAV is recharging, another UAV provides 5G access to its target area. Thus, the number of UAVs is twice the number of covered areas. This approach was found in~\citep{IEEE_C211, IEEE_C212} to be more cost effective than deploying a network consisting solely of fixed ground BS sites. It was also shown to lead to economically feasible deployments with subscription fees of around $20$~Euros/month in certain rural areas~\cite{IEEE_C57}. However, it was shown in~\citep{IEEE_C57, IEEE_J10} that the scenario of 5G large cells with very large coverage areas is more cost effective than UAVs, pushing down the subscription fees to round $2$~Euros/month. These large cells need to be accompanied by smaller cells to meet the capacity demand, similarly to the scenario of~\cite{IEEE_C69}. In~\cite{IEEE_C56}, UAV mission planning is performed, with the objective of ensuring coverage while minimizing energy consumption. UAVs stay at neighboring ground sites to recharge their batteries as in~\citep{IEEE_C211, IEEE_C212}, and then move to target areas to provide 5G coverage. The energy minimization problem is solved using Genetic Algorithms, and results are shown to compare favorably to other works in the literature~\cite{IEEE_C56}.

The UAV-based approach of~\citep{IEEE_C211, IEEE_C212} is investigated in~\citep{IEEE_C163} for scenarios with challenged networks, e.g., in case of a disaster area with many ground BSs destroyed. In that case, UAVs can act as RRHs connected to the remaining BSs. The UAV placement can be optimized to enhance coverage and increase the 5th percentile spectral efficiency~\citep{IEEE_C163}.

\subsection{Technologies for IoT Connectivity}
With the proliferation of IoT, several papers in the literature have investigated the possibility of implementing IoT in rural areas and providing access to the sensors to send their measured data over a communications network. In~\cite{IEEE_C205}, to facilitate the deployment of M2M devices in rural areas, SMS over 2G GSM networks is used to transmit the M2M traffic. M2M devices communicate wirelessly with a nearby gateway, and the gateway translates their data into SMS format and sends it over the network. The constrained application protocol (CoAP) is used as a replacement to HTTP in M2M environments.

In~\cite{IEEE_C94}, a mesh network using 6LoWPAN is proposed in order to transmit IoT data from hard to reach areas or areas with limited connectivity. IoT devices would relay their data in a multihop fashion until reaching a gateway that is connected to the internet, e.g., via GPRS or 3G. In~\cite{IEEE_C61}, range extension in rural areas was performed by using IEEE 802.15.4 with multihop transmission in the $868$~MHz band to extend the hop range, along with time slotted channel hopping to increase robustness. In~\cite{IEEE_C87}, to ensure $k$-th order receive diversity of IoT measurements in an LPWAN, the placement of gateways is investigated, such that the transmission of each sensor is received by $k$ gateways. It was also mentioned that although a certain number of sensors can reach a given gateway, the gateway would in practice have a maximum number of devices that it can connect to. This might affect the results in urban and suburban areas, but not in rural areas where the device density would be much less~\cite{IEEE_C87}.

In~\cite{IEEE_C121}, LoRa radio, an LPWAN technology, was used to provide connectivity for an IoT network used for monitoring water quality in a remote rural area. The LTE-based narrowband IoT (NB-IoT) could have been used for the same purpose. However, LoRa LPWAN was preferred due to the use of unlicensed spectrum, and the low data rates required~\cite{IEEE_C121}. In the approach of~\cite{IEEE_C121}, a LoRa module collects the sensor measurements, and sends them to a LoRa gateway. The measurements are collected from the village water tank and the water locks around the village. The gateway can then forward the data to the cloud using GSM connectivity. Data can be stored, processed, and analyzed in the cloud servers and alerts can be sent if needed to the relevant authorities to take appropriate action. In~\cite{IEEE_C33}, experimental measurements showed that LoRa transmission range can exceed $5$~km in rural areas. In~\cite{IEEE_C4}, measurements from IoT devices were collected by a moving gateway (vehicle or UAV) using LoRa, after being forwarded by the gateway using LTE to an LTE BS and then to the cloud. It was noted in that a UAV provided larger coverage than a terrestrial vehicle, especially when the altitude of the UAV increased. UAVs were also used in~\cite{IEEE_C79} to collect 5G measurements from IoT sensors in farming applications, and relay them to 5G BSs with mobile edge computing (MEC) to perform local processing in the absence of connectivity to an internet cloud.

In~\cite{IEEE_C192}, IoT is used to monitor biogas plants. Biogas represents an environmentally friendly fuel that can be used for cooking in rural areas, instead of firewood and crop residue, which emit hazardous smoke~\cite{IEEE_C192}. An Arduino device is used to monitor the consumption of biogas. The results are sent via SMS to an android mobile application on the user's phone. The application then updates a remote database whenever internet connectivity is available. The data stored on the database server can then be analyzed to detect consumption trends and usage statistics~\cite{IEEE_C192}.

\subsection{FSO}
An extension to FSO in indoor scenarios consists of using visible light communications to transmit data. For example, LiFi uses communication through LEDs and photodiode receivers to provide high speed internet connectivity up to $10$~Gbits/s in indoor scenarios~\citep{LiFi1, LiFi2}. With cheap off-the-shelf LEDs, speeds of $10$~Mbits/s can be reached~\citep{LiFi3, LiFi4}. The LiFi LEDs can be connected to an access point or router providing backhaul connectivity, e.g., to fiber optic cable.

However, to be able to reach these high data rates at the network access part in a rural setting, the backhaul needs to be able to support them. Thus, it is hard to implement the system in rural areas without fiber backhaul, and impossible to implement it in areas without electricity. However, in~\citep{LiFi3, LiFi4, LiFi5}, it was mentioned that research is ongoing on solar panels so that they can be used to provide a backhaul for LiFi using light communications, in addition to their initial role in providing energy from solar radiation. Thus, they can provide both electricity to power the LEDs and a backhaul channel to carry the traffic of the high speed fronthaul LiFi communications.

In~\cite{Nature_Photonics_Africa}, a joint fronthaul-backhaul design is proposed where terrestrial FSO links can be used transmit the access traffic (e.g., from LiFi) in rural areas until reaching PON networks connected to the fiber backhaul.

\subsection{Direct Satellite Access}\label{subsubsec:Fronthaul_Sat}
Although satellites can be used to provide backhaul connectivity, they are also used to provide fronthaul access to rural areas where population is sparse, such that the deployment of terrestrial backhaul is not justified, while at the same time the population in these areas can afford satellite access.  For example, internet access fees via satellite can range between $50$-$150$ USD/month~\cite{Satellite_Access_2019}, which is acceptable for developed countries, but is generally far beyond the reach of the population in rural areas of developing countries. A better solution in these countries would be for an MNO to provide local access at reduced (potentially State subsidized) prices, and have its local access BSs connected to its core network via satellite backhaul. This corresponds to the scenario described in Section~\ref{subsubsec:Backhaul_Sat}. In\cite{IEEE_C119}, it was proposed that the user ground equipment used to gain satellite access be provide by Governments using subsidized funds. This allows providing coverage quickly to rural areas. In fact, it was suggested in~\cite{IEEE_C186} that if wired fiber and DSL deployments start from high density areas and move outwards, then satellite connectivity can provide access starting from the outside inwards, until wired deployments catch up. Similarly, in~\citep{IEEE_C169, IEEE_C138}, it was proposed that satellites provide direct cellular access to users in rural areas, whereas terrestrial cellular networks provide coverage to urban areas. Techniques for avoiding co-channel interference between the two systems, based on using adaptive beamforming at the satellites, were also proposed~\citep{IEEE_C169, IEEE_C138}. In~\cite{Lin_5G_Sat_2019}, challenges facing direct satellite access for 5G UEs were discussed. UEs within the coverage of the same spot beam might have varying delays to the satellite depending on their location, and hence timing advance should be properly implemented. However, the delays incurred in traditional techniques might necessitate the use of other measures, e.g., relying on global navigation satellite system (GNSS) based techniques, where the UE can perform timing advance based on its position with respect to the satellite. Another challenge is to handle Doppler effects for LEO satellites (they are considered negligible with GEO satellites). In addition, coverage issues are a challenge with LEO satellites, since even if beam pointing is performed to cover the same area as the satellites move, the UEs have to be handed over between beams and/or satellites every few seconds~\cite{Lin_5G_Sat_2019}.

In addition to providing access for rural areas, satellites can provide access in disaster areas until wireless networks gradually become active again, in which case the satellite network can provide backhaul connectivity to isolated islands of WiFi, WiMAX, or cellular connectivity~\citep{IEEE_C72, IEEE_C98}. A direct satellite control channel can be used to monitor the network activity, even if the satellite acts as backhaul for data traffic~\citep{IEEE_C72, IEEE_C98}.

\subsection{WiFi/WiMAX/Multihop/Mesh Networks}\label{subsubsec:Fronthaul_Multihop}

In~\cite{IEEE_C164}, in order to provide affordable access for rural areas, WiFi is proposed as a fronthaul last mile access solution, and WiMAX is proposed as a backhaul solution. The interoperability between the two networks is discussed and analyzed. In~\cite{WiMAX_3G_IJRRCS}, WiMAX is proposed as an overlay network over 3G cellular network, and the $900$~MHz band is proposed for 3G in order to expand the coverage in rural areas.

Multihop transmissions allow nodes providing access to rural villages to communicate with each other in a multihop fashion until reaching a gateway node that is connected to the internet. This way, the multihop links provide a backhaul allowing internet connectivity to extended to rural areas that were initially unconnected~\citep{IEEE_C70, IEEE_C208}. Multihop and mesh network are mostly based on WiFi and/or WiMAX in the literature. Long range WiFi, or WiLD, is considered attractive for rural areas from a cost perspective, since it relies on unlicensed spectrum and on low-cost widely available WiFi equipment~\cite{IEEE_C15}.

Long distance evaluation of IEEE 802.11g links was performed in~\cite{IEEE_C102} in a flat desert area. Basic connectivity was reached up to a nine km distance, but the transfer of larger files (larger than $10$~MB) was possible only up to seven km. Unlike pure rural areas, long distance IEEE 802.11g WiFi links implemented in a semi-urban area, where nodes in distant farms were connected to a central node in an urban area, were shown in~\cite{IEEE_C183} to be subjected to interference from other WiFi deployments in the surroundings. In rural areas, the density of deployments is significantly lower.

In~\cite{IEEE_C34}, multihop long range WiFi connectivity was used to provide access to a group of rural villages, located within a $10$~km radius from a central village. The \lq\lq long-range" was achieved by resorting to directive antennas that can extend the range of WiFi to make it suitable for rural coverage~\cite{IEEE_C158}. The WiFi stations at the center of each village are connected in a multihop fashion~\cite{IEEE_C34, IEEE_C158}. The central village can be connected to the network via a VSAT terminal for example~\cite{IEEE_C34}. In~\cite{IEEE_C20}, instead of resorting to multihop, the nodes in a WiFi mesh within the range of each other perform collaborative transmission by transmitting simultaneously the same signal and adjusting the phase of their transmissions such that the signals add constructively at the destination. Thus, collaborating nodes form a sort of distributed antenna array which enhances the performance at the receiver. To reduce the deployment costs of a rural WiFi mesh network, the antenna tower heights need to be kept to the minimum required to obtain line of sight connectivity~\cite{IEEE_C54}. In~\cite{IEEE_C54}, an algorithm is also provided to protect the survivability of the network, by having each node connected to at least two other nodes.

A system for managing a WiFi mesh network in rural areas is proposed in~\cite{IEEE_C131}, where users are authenticated by logging in to a portal. In~\cite{IEEE_C120}, an approach using mobile devices for extending the WiFi multihop connectivity to provide last-mile coverage in rural areas is proposed. Virtualization is used such that a mobile device can be split into two virtual devices: (i) a traditional mobile device using WiFi connectivity, and (ii) a virtual access point (VAP) providing connectivity to other mobile devices. A tree based structure is adopted, with a traditional WiFi AP connected to the internet positioned at the root of the tree. Then, devices connected to the AP download a network auto-configuration software (NAS) to alow them to act as VAP. The approach of~\cite{IEEE_C120} caters for practical challenges such as the selection of IP addresses, DNS resolution, and NAT. The method of~\cite{IEEE_C120} is not only applicable for range extension in rural areas, but also for providing connectivity in disaster scenarios. To perform a low cost deployment in rural areas, single channel single radio WiFi devices are typically deployed in a mesh network~\cite{IEEE_C209}. To avoid interference from potentially other deployments in single channel single radio IEEE 802.11s devices, a channel switching approach is implemented in Linux in~\cite{IEEE_C209}, in order to allow the devices in a mesh network to dynamically switch their channel. Within the same single channel single radio long distance WiFi multihop network, TDMA is generally used to avoid concurrent transmissions on neighboring links, thus avoiding interference~\citep{IEEE_C174, IEEE_J23}. In~\cite{IEEE_C174}, angular separation between links is considered in the TDMA allocation problem in order to increase the efficiency, by allowing simultaneous transmission on the same time slot for links with sufficient angular separation, which is denoted as spatial TDMA (STDMA) in~\citep{IEEE_J23}, where link scheduling algorithms for STDMA mesh networks are presented and analyzed. In~\cite{IEEE_C165}, changes were made to the initial CSMA/CA protocol in order to outperform TDMA in IEEE 802.11n mesh networks in rural areas. The changes are implemented by software and do not need hardware upgrade. They consist of adapting the protocol to the special conditions of long range transmission in rural areas with low deployments of interfering networks and mostly line of sight connectivity between communicating nodes. They can be summarized by: (i) using a coarse/fine grained approach for rate adaptation, by selecting the subset of MCSs suitable for a given RSSI in coarse-grained part, then performing probing for the MCS to be used only among those in the previously selected subset, (ii) using efficient retransmission by adopting the most reliable MCS in the selected set if a frame is not received after one retransmission, (iii) performing more frame aggregation, and (iv) reducing the size of the contention window. The improvements due to TCP packet aggregation in wireless mesh networks were also demonstrated in~\cite{IEEE_C182}.

A system based on WiFi mesh, named WiBACK and developed by Fraunhofer FOKUS, was used to provide connectivity to rural areas, e.g.,~\citep{IEEE_C62, IEEE_C59, IEEE_C35, IEEE_C81}. It implements multiprotocol label switching (MPLS) in order to differentiate traffic and maintain the QoS of multimedia services. Its control plane implements the IEEE 802.21 standard. Details of its topology management function and capacity management function are presented in~\cite{IEEE_C139}. In~\citep{IEEE_C62}, WiBACK was used to provide cellular connectivity to rural areas with limited or absent cellular coverage: Nanocell GSM base stations functionality is implemented at the access side of the WiBACK devices, such that the GSM traffic is terminated at the access point. The GSM voice traffic is then transformed into VoIP and carried along the web and video traffic over the WiBACK system, using SIP and RTP to reach the backhaul network, where a SIP gateway allows the interconnection with the GSM and PSTN networks. In~\citep{IEEE_C59}, WiBACK was used to relay IEEE 802.11a traffic provided to a remote farm. In~\citep{IEEE_C35, IEEE_C81}, \lq\lq eKiosks" were deployed in a rural area, equipped with WiFi connectivity to provide hotspot internet access to the local population, with WiBACK providing backhaul connectivity to reach a gateway connected to the internet.

In~\cite{IEEE_C173}, a multihop WiMAX mesh network is considered for providing broadband connectivity to rural areas. An algorithm is proposed to build a routing tree between WiMAX BSs, with the tree rooted at the BS connected to the backhaul network (thus acting as gateway). The performance of such a WiMAX system can be enhanced by using adaptive smart antennas for WiMAX~\cite{IEEE_C146}.In~\cite{IEEE_C208}, the problem of joint routing and scheduling in a multihop network with directional antenna was formulated as a linear program, and a scheduling algorithm is proposed based on directed edge coloring in a multi-graph. In~\cite{IEEE_C179}, an energy aware routing approach is presented for wireless mesh networks, such that the routing path selected is the one that maximizes network lifetime. The purpose is to perform efficient routing in rural areas where permanent energy availability is not guaranteed. Similarly, in~\cite{IEEE_C136}, concepts to allow energy awareness while routing traffic in WiBACK networks were proposed. In~\cite{IEEE_C101}, it was shown that routing algorithms taking into account the PHY and MAC layer parameters, such as the Hybrid Wireless Mesh Protocol (HWMP), outperform other algorithms like Dynamic Source Routing (DSR) and Optimized Link State Routing (OLSR) in IEEE 802.11s mesh networks in rural areas.

In multihop networks where multiple gateways connected to the backhaul are available, the selection of the appropriate gateway, along with the path to that gateway, can be optimized, especially when the gateways have different capabilities. In~\cite{IEEE_C130}, a gateway-aware routing approach is proposed for multihop networks in rural areas, and shown to achieve enhanced performance. The approach of~\cite{IEEE_C130} finds the route from each node to the internet, taking into account the capacity of each gateway, and the bottleneck capacity of the multihop path from the source node to that gateway.

In~\citep{IEEE_C173, IEEE_C146, IEEE_C208, IEEE_C179, IEEE_C136, IEEE_C130}, routing was optimized through access points or nodes that were installed in the rural area. However, the initial placement of these access points can be optimized depending on the characteristics of the rural area and the type of population agglomeration. Thus, optimizing the locations can complement the optimization of route selection. Indeed, in~\cite{IEEE_C82}, the optimized placement of mesh access points (MAPs) in rural areas was studied. Several optimization algorithms (Hill Climbing, Virtual Force, Time-Efficient Local Search, and Random) were implemented for different rural settlement models (Dispersed, Linear, Nucleated, and Isolated), and the one leading to the best performance for each settlement type was selected. After deploying the MAPs, at least one gateway needs to be selected to provide connectivity to the internet. In certain rural areas, this is governed by the availability of backhaul infrastructure, e.g., fiber optic cable reaching a certain village. In other scenarios, e.g., where wired connectivity is absent and backhaul will be provided by a VSAT terminal for example, the placement of the gateway can also be optimized. This was investigated in~\cite{IEEE_C55}, where the same settlement models of~\cite{IEEE_C82} were used, and several gateway placement algorithms (Grid based, Incremental Clustering, Multihop Traffic-flow Weight, and Random) were compared.

\subsection{mmWave}\label{subsubsec:Fronthaul_mmWave}
Channel models for mmWave have mainly been investigated for urban areas, with sufficient investigations for rural areas still lacking~\citep{nonIEEE_BC1_River_WWRF}. An attempt was made in~\cite{mmWave_PropModel}, where it was shown that the 3GPP rural macrocell propagation model suffered from some flaws, and the authors proposed a more accurate model based on measurements performed in a rural area in Virginia.

As an example of a mmWave technology, Terragraph is a 60~GHz multi-hop multi-point wireless distribution network~\citep{Terragraph1, Terragraph2}. It is part of the facebook connectivity project~\cite{Terragraph3}. It is based on WiGig standards IEEE 802.11ad and IEEE 802.11ay. It was initially proposed for urban areas with complicated wired infrastructure, e.g., where it is hard to deploy fiber to each home/neighborhood. In this case, terragraph provides fiber-like speeds by deploying devices on lamp posts, building rooftops, etc. It can provide access to users and/or relay their data in a multihop approach until reaching a fiber point of presence~\citep{Terragraph1, Terragraph2, Terragraph3}. However, it can be used in rural areas similarly to the multihop techniques discussed in Section~\ref{subsubsec:Fronthaul_Multihop} for more traditional technologies like WiFi and WiMAX, especially in villages where a relatively large population exists in order to benefit from street infrastructure and rooftops to deploy the Terragraph nodes. Furthermore, although it can be connected to different types of backhaul, the high Terragraph speeds can best be reached with a fiber backhaul. Hence, it is mostly useful in areas where a fiber backbone is deployed up to a large rural town (e.g., parallel to railroad tracks or power lines), and then Terragraph can provide connectivity to rural villages in the area surrounding this town.

\subsection{Delay Tolerant Networks (DTNs)}\label{subsubsec:Fronthaul_DTN}
DTNs are suitable for scenarios with limited connectivity and no infrastructure to carry the communication data. Mobile vehicles (e.g., cars, buses) can collect the data from the source and carry it to the destination, which makes them a good candidate to provide connectivity to rural areas with no communication infrastructure~\cite{IEEE_C198}. In fact, relay nodes can also come into play, where the data of the rural users is stored in these aggregation/relay nodes, until a DTN vehicle passes and collects it in bulk. Pioneering work using buses as \lq\lq data mules" to carry data in rural areas of developing countries was performed in~\cite{IEEE_C28_Comp1}. Two well-known routing protocols, ad hoc on-demand distance vector (AODV) and optimal relay path (ORP) were compared in~\cite{IEEE_C28} for a DTN scenario. AODV worked better with multihop communication between mules, when the number of data mules increases. ORP performed better in the opposite case, since it was initially conceived to relay data between disconnected MANETs. The authors of~\cite{IEEE_C28} proposed a new algorithm that considers an adaptive approach between AODV and ORP to dynamically optimize performance, depending on the situation. These results were further validated by the work in~\cite{IEEE_C90}, where a detailed simulation comparing different algorithms was performed. It was found that each routing algorithm performs best under certain conditions, which outlines the need for an environment-aware dynamic routing approach.

To be a able to efficiently deliver data under intermittent connectivity, a new layer for DTN, the bundle layer, is inserted in the protocol stack above the transport layer and below the application layer\cite{IEEE_C159}. A modification to the traditional TCP/IP networking architecture is proposed in~\cite{IEEE_C198} to suite vehicular DTNs, where: (i) the bundle layer was added above the MAC and below the networking layer (instead of being above the transport layer as in~\cite{IEEE_C159}), to aggregate and de-aggregate traffic faster in a DTN setup instead of using small IP packets, and (ii) the data and control plane are separated. In~\cite{IEEE_C142}, multihop transmissions in vehicular DTNs were investigated. An efficient approach to transmit the data to other vehicles going in the same direction to speed up delivery is proposed. However, such an approach might not always be applicable in rural areas, due to the state of the transportation networks and the potentially limited number of vehicles that can carry the traffic at a given time. For these reasons, the buffers might become full before successful delivery. Therefore, a buffer management approach is proposed in~\cite{IEEE_C122}, where data bundles are allocated a TTL. When the buffer becomes full, those with the least TTL are deleted to make room for newer data. This avoids having older data that is occupying resources for too long from preventing the delivery of newer data. In~\cite{IEEE_C191}, the impact of packet lifetime is studied in vehicular DTNs, where it can be adapted to enhance performance based on vehicle speed, vehicle density, and quality of service requirements. DTNs can be extended to a scenario with D2D communications, where the user devices relay the traffic in a multihop fashion. However, in a rural area, the density per square kilometer is much less than in an urban scenario. In~\cite{IEEE_C154}, it was shown that multihop communications is feasible via D2D, with a density of at least $12$ devices per square km sufficient to successfully transmit the data if the communication range between devices is $300$~m. Indeed, dissemination of information of common interest (IoCI) was investigated in~\cite{IEEE_J21} by considering opportunistic social networks in an integrated operation with cellular networks. Users share information with other users in their social network thus extending the reachability of cellular BSs. It was shown in~\cite{IEEE_J21} that the delivery ratio of IoCI before it expires exceeds $90$\% when opportunistic networks of more than $20$ devices are assisting in the information dissemination.

A testbed showing the transfer of news information and email over DTNs is described in~\cite{IEEE_C141}. In~\cite{IEEE_C168}, WSNs and DTNs are used jointly for carrying environmental sensor measurements from rural areas through public transport buses. The DTN is also used to carry health data from the remote health posts in the rural villages to the city hospital. In~\cite{IEEE_C88}, an incentives-based DTN approach was presented, in order to avoid having selfish nodes preventing the delivery of other nodes data. The approach of~\cite{IEEE_C88} also incorporates security and privacy measures, since the purpose is to transmit the mHealth data of patients in a rural area to a health center in the city, using a vehicular DTNs from the vehicles in the rural area, those heading from the rural area to the city, and those in the city. The approach~\cite{IEEE_C88} protects the privacy of the patients, and allows to measure the behavior of relay nodes through a reputation metric maintained by a trusted authority. In~\cite{IEEE_C167}, anonymity and security in DTNs are maintained by resorting to identity based cryptography (IBC). Two entities, each knowing its own private key and the identity of the other entity, can independently compute a shared secret key for their communication. However, this approach requires the existence of a trusted entity called the public key generator (PKG). When entities are not under the same PKG, other methods are proposed in~\cite{IEEE_C167}, including hierarchical IBC. Some limitations of IBC were outlined in~\cite{IEEE_C126}, e.g., the distribution of PKG parameters and revocation issues with intermittent connectivity in DTNs. Instead, the authors of~\cite{IEEE_C126} propose an approach based on social contacts within a given region, where each source knows one or affiliated entities (AEs) of the destination. The public keys of AEs are assumed known, or an exchange of a symmetric key with the source is assumed to be performed. For inter-region communications, the gateways connected to the internet can handle more traditional cryptographic techniques. Other security measures leading to pseudonymity, based mainly of grouping the nodes into groups to confuse attackers, are presented in~\citep{IEEE_C151, IEEE_C116}.

\subsection{Power Line Communications (PLC)}
In~\cite{IEEE_C80, IEEE_C13}, broadband communications over power lines are proposed as a solution for providing connectivity to rural areas. It is argued that electric utility companies cannot compete in areas where DSL or fiber broadband networks are deployed. However, in rural areas, they might have a niche area to provide network connectivity. This is particularly true for rural areas having access to the electricity grid, but not yet to broadband communication networks. Using PLC for broadband connectivity also allows utility companies to easily deploy smart meters and move into the smart grids era in rural areas~\cite{IEEE_C80}. Potential markets for PLC include Brazil~\cite{IEEE_C80} and South Asia~\cite{IEEE_C13}. PLC can be used for backbone using the medium voltage (MV) lines, for fronthaul/last mile access using low voltage (LV) lines, and for in-building or in-house wiring~\cite{IEEE_C13}. Hybrid models, where, for example, a fiber backbone is connected to a PLC LV fronthaul, or where a PLC HV backhaul is connected to a wireless fronthaul, are also possible\cite{IEEE_C13}.

\subsection{TV White Space (TVWS)/ Cognitive Radio (CR)}\label{subsubsec:Fronthaul_TVWS}
Spectrum measurements have shown large availability of TV White Space spectrum, especially in rural areas, e.g.,~\citep{IEEE_J18, IEEE_C52}.

The IEEE 802.22 WRAN standard based on using CR technology to transmit on under-utilized TVWS is investigated in several references as a solution to the rural connectivity problem~\citep{IEEE_J18, IEEE_C52, IEEE_C78, IEEE_C196, IEEE_C39, IEEE_C11, nonIEEE_Congo_Africom18, IEEE_C171, IEEE_C49}. It uses unlicensed spectrum, and provides a wide coverage area, typically $30$~Km radius and can be extended to $100$~Km~\cite{IEEE_C11, IEEE_J2}. The IEEE 802.11af standard, known as Super WiFi, is also used for providing access using TVWS frequencies. It uses CSMA at both the BS and the client devices, whereas IEEE 802.22 uses OFDMA~\cite{IEEE_J18}. However, IEEE 802.22 covers significantly larger areas whereas IEEE 802.11af deals with much shorter ranges, comparable to those of traditional WiFi (IEEE 802.11a/b/g). Thus the two standards can be used in a complementary fashion: IEEE 802.22 BSs can be used in rural areas instead of local 802.22 CPE, thus providing a sort of \lq\lq mid-haul" connectivity with the backbone network connected to the TVWS database, and these BSs can be connected (whether in a wired or wireless way) to 802.11af Super WiFi APs providing short range access to user devices~\cite{IEEE_C1}. This way, optimized placement of 802.22 CPEs can be performed in order to maximize capacity as in~\cite{IEEE_C1}, an optimization that cannot be done directly on mobile user devices. In~\cite{IEEE_C2}, Super WiFi is compared to long range WiFi, which relies on traditional WiFi (IEEE 802.11a/b/g) with directional antennas and increased power. Although long range WiFi is cheaper, Super WiFi propagates for longer distances and penetrates through walls better, since it uses lower frequencies ($300$-$700$ MHz) compared to long rang and traditional WiFi (2.4 GHz and 5 GHz). In~\cite{IEEE_C2}, both standards IEEE 802.22 and IEEE 802.11af are discussed under Super WiFi, while noting that the first covers significantly longer distances. In~\cite{IEEE_J15}, a comparison was made between TVWS and LTE in suburban and rural scenarios, and the results showed a significantly higher energy efficiency for TVWS.

In~\cite{IEEE_C194}, a tool that can be used to plan networks with cognitive protocols and dynamic frequency selection is proposed. It caters for CR networking and uses simulated annealing to optimize performance. A throughput maximization tool using Hill Climbing is proposed in~\cite{IEEE_C96} for TVWS mesh networks relaying traffic from rural areas to the neighboring fiber points of presence. The output optimizes route assignment, power allocation, and frequency reuse. In~\cite{IEEE_C135}, measurements showed that the use of white spaces in rural areas can significantly reduce the number of required WiFi APs (up to $1650$\% in rural areas where the population density is below $20$ people per square km), compared to a WiFi only deployment scenario. In urban areas, the benefit of larger coverage areas achieved by TVWS frequencies is outweighed by the need for denser deployments to meet the QoS requirements of the dense population. Furthermore, it was shown in~\citep{IEEE_C199, TVWS_MiddleMile} that additional TVWS APs in urban areas will suffer from increased interference, and have less available TV spectrum, unlike the situation in rural areas where TVWS deployments are more favorable. The use of directional antennas was shown in~\cite{TVWS_MiddleMile} to significantly reduce the interference problem.

In~\cite{IEEE_C30}, a TDMA mesh network exploiting TVWS was implemented. Due to using lower frequencies, larger coverage was achieved compared to a WiFi mesh network based on the 2.4 Ghz spectrum. The mesh nodes were built from commodity hardware in~\cite{IEEE_C30}, with a node costing $330$ USD. In~\cite{IEEE_J6}, a novel approach is proposed for rural broadband access using TVWS: In every village, CPEs form collaborative clusters using slotted Aloha communications. In the uplink, the CPEs of each cluster implement distributed beamforming to send the same signal to the BS, such that the received signals add up constructively at the BS. The BS communicates with clusters from several rural villages. In the donwlink, the BS communicates with each CPE individually. Each CPE is connected to a WiFi AP that provides access inside users' homes, whereas the BS is connected to the internet via appropriate backhaul connectivity. The CPEs make use of the TV antennas deployed on top of rooftops in order to perform efficient transmission at low cost. In fact, the feasibility of multiuser MIMO (MU-MIMO) in rural areas was demonstrated in~\citep{MU_MIMO_ZFBF, MU_MIMO_UCA_Aus, MU_MIMO_UCA_Aus2, MU_MIMO_UCA_Aus3, TVWS_MIMO_OFDM_Aus}, where in~\citep{MU_MIMO_UCA_Aus3, TVWS_MIMO_OFDM_Aus} a model using OFDM over TVWS, and that accurately predicts performance, was proposed (except for closely positioned users, where the model underestimates the actual performance). Another approach is described in~\cite{TVWS_5G_Rural}, where WiFi APs and/or 5G mmWave small cells are used inside villages/homes, and a village connectivity point is connected via TVWS to a macro UHF BS that connects the villages together and has access to a gateway with fiber backhaul connectivity.

Finally, it should be noted that although all the references described in this section discuss TVWS, thus benefiting from the unused spectrum or the freed spectrum due to the transition to digital television, there have been some attempts in the literature proposing the combination of internet traffic with digital video broadcasting (DVB) data in order to provide connectivity for rural areas, e.g.,~\citep{IEEE_J8, IEEE_C133, IEEE_J7}. In addition, the presence of interactivity with digital TV was used in~\cite{IEEE_J22} to provide better accessibility to information for disabled people, who are present in higher proportions in poor and rural areas.

\subsection{Community Networks}\label{subsubsec:Fronthaul_CommunityNetworks}
Community networks consist of deploying a local network to provide connectivity in a given rural area or village. In~\cite{nonIEEE_DIY_acmCompass18}, a local cellular deployment was proposed by using plug-and-play BSs for access, TVWS for backhaul, and by resorting to a virtualized core using a cloud infrastructure. The objective is to provide local broadband access without relying on mobile network operator, to which the business case might not be profitable. Other work have considered similar scenarios, e.g.~\citep{nonIEEE_DIY_Comp1, nonIEEE_DIY_Comp2, nonIEEE_DIY_Comp3}, while focusing on cellular connectivity rather than broadband internet access. This is justified since some works have shown that around 70\% of calls in rural areas take place within the vicinity of the same BS, i.e., most rural users call other users in relatively close proximity~\cite{Rural_CoteDivoire}.  In~\cite{Village_BS}, the village BS is proposed, in order to provide local GSM connectivity in rural areas while being off-grid and off-network. It can also provide local data services in a DTN fashion. In~\cite{IEEE_C85}, community cellular networks based on GSM were proposed, while avoiding the licensed spectrum costs by using GSM white spaces similarly to TVWS. The approach of~\cite{IEEE_C85} works without modification to handsets. The community BSs receive measurement reports from handsets, thus allowing them to determine the available channels. Given the low density in rural areas, they can operate as secondary users over the primary licensed GSM spectrum, even without the collaboration of the primary operator. A related but different concept is adopted in~\cite{IEEE_J13}, where the \lq\lq HybridCell" system is presented. In this system, a local community cellular network is established. Whenever the commercial cellular network is congested or its signal quality degrades, the mobile phone switches automatically to the local community network. Whenever both parties (the caller and the callee) are on the same network (either local or commercial), the call proceeds normally. When each is on a different network, the call will be routed to voice mail and/or an SMS is queued for delivery to the callee to notify him/her of the call. The approach of~\cite{IEEE_J13} is tested in a crowded refugee camp where commercial networks are congested. However, it can also be implemented in a rural scenario where connectivity is intermittent. In~\cite{IEEE_C137}, the large idle periods in rural areas due to low traffic density are exploited to reduce the power consumption of GSM base stations, thus reducing the operational costs in community networks.

These local community networks provide affordable access. However, it remains challenging to inter-operate with other commercial mobile networks, and appropriate gateways are needed~\cite{nonIEEE_DIY_acmCompass18}. Nevertheless, this problem is being addressed. For example, Kuha~\cite{nonIEEE_Kuha1} provides plug-and-play BSs that can be used in community networks to provide 4G/LTE broadband connectivity. They can be installed and maintained by the end user, and can be connected via an existing internet connection (e.g., fiber, microwave, satellite) to the backbone network. Whenever such connectivity is not available, Kuha can provide backhaul connectivity via satellite. Although end users can install and maintain BSs (provide power and internet access), the BSs can then be controlled remotely (automatic upgrades, software downloads, management, automatic operation, etc.)~\cite{nonIEEE_Kuha1}. An example of such a deployment can be found in~\citep{nonIEEE_Kuha2, nonIEEE_Nokia1}, where Kuha BSs are connected to Nokia's operation and support system (OSS) to build community networks providing connectivity to rural areas. In another example~\cite{IEEE_C176}, femtocells, or home BSs (HBSs), were connected to fixed outdoor directive antennas, pointed in the direction of 3G or 4G cellular BS. The purpose is to extend cellular coverage in rural areas where the population density does not justify the additional deployment of BSs. In the approach of~\cite{IEEE_C176}, local \lq\lq infopreneurs" handle user subscription and HBS deployment, according to a business agreement with the mobile network operator owning the macro BS. Community projects can be accompanied by local capacity building, to build a team of local experts who can support and maintain the community network~\cite{IEEE_J5}. They can also train other community members in \lq\lq train the trainer" fashion.

Community networks using WiFi mesh networks are also deployed, e.g.,~\citep{CWMN_Urban, CWMN_Rural}, even in urban areas where the objective is to benefit from the dense availability of WiFi APs to reduce the costs of broadband access~\citep{CWMN_Urban}. In community networks, since in general people within the same rural community have similar interests, pre-fetching of popular content during off-peak hours and content caching would help in avoiding congestion over the backhaul bandwidth that is likely to be limited in rural areas~\citep{CWMN_Rural}.

\begin{table}[h!]
\begin{footnotesize}
\caption{Summary of Key References for the Different Fronthaul/Backhaul Technologies Used}
\label{tab:Fronthaul_Backhaul_Summary}
\begin{center}
\begin{tabular}{|p{3cm}|p{7cm}|p{5cm}|}
  \hline
   {\bf Ref.} & {\bf Technology} & {\bf Fronthaul/Backhaul}  \\
  \hline
   \citep{IEEE_C91, IEEE_C201, IEEE_C166}  & Fiber (GPON) & Backhaul \\
  \hline
   \citep{IEEE_C66}  & Fiber (RoF) & Backhaul \\
  \hline
   \citep{IEEE_C177}  & FSO & Backhaul (Terrestrial)\\
  \hline
   \citep{FSO_Feeder_VHTS, FSO_Feeder_VHTS2}  & FSO & Backhaul (Satellite) \\
  \hline
   \citep{IEEE_C34, IEEE_C158, IEEE_C20, IEEE_C54, IEEE_C131, IEEE_C120}  & WiFi Multihop & Fronthaul \\
  \hline
   \citep{IEEE_C208, IEEE_C173, IEEE_C146}  & WiMAX Multihop & Fronthaul \\
  \hline
   \citep{IEEE_C29}  & WiFi/GPRS & Fronthaul (WiFi)/Backhaul (GPRS)\\
  \hline
   \citep{IEEE_C50}  & WiFi mesh/VSAT & Fronthaul (WiFi)/Backhaul (VSAT)\\
  \hline
   \citep{IEEE_C62, IEEE_C59, IEEE_C35, IEEE_C81, IEEE_C139}  & WiFi mesh (WiBACK) & Backhaul\\
  \hline
   \citep{IEEE_J18, IEEE_C52, IEEE_C78, IEEE_C196, IEEE_C39, IEEE_C11, nonIEEE_Congo_Africom18, IEEE_C171, IEEE_C49, IEEE_J2, IEEE_C1}  & TVWS & Fronthaul \\
  \hline
   \citep{IEEE_C198, IEEE_C28_Comp1, IEEE_C90, IEEE_C159, IEEE_C142, IEEE_C122, IEEE_C191, IEEE_C141, IEEE_C168, IEEE_C88, IEEE_C167}  & Delay Tolerant Networks (mostly using WiFi) & Fronthaul \\
  \hline
   \citep{IEEE_C80}  & Power Line Communications & Fronthaul \\
  \hline
   \citep{IEEE_C205}  & 2G SMS & Fronthaul (for M2M data) \\
  \hline
   \citep{IEEE_C69}  & 5G & Fronthaul (small cells)/Backhaul (massive MIMO) \\
  \hline
   \citep{IEEE_C211, IEEE_C212, IEEE_C57, IEEE_C56, IEEE_C163}  & 5G & Fronthaul (UAVs)/Backhaul (fiber) \\
  \hline
   \citep{IEEE_C17}  & Tethered Balloons & Fronthaul (WiFi)/Backhaul (WiFi with directive antennas) \\
  \hline
   \citep{IEEE_J11, TVWS_HAP}  & UAVs & Backhaul \\
  \hline
   \citep{IEEE_Spectrum_Loon2}  & Balloons & Backhaul (mmWave) \\
  \hline
\end{tabular}
\end{center}
\end{footnotesize}
\end{table}

\section{Fronthaul Considerations and Tradeoffs}
\label{subsec:Fronthaul_Tradeoffs}
Different tradeoffs need to be taken into account when connectivity is provided for the first time to remote and rural areas that were initially unconnected. For example, in certain scenarios, mobility can sacrificed for the purpose of providing connectivity when it does not exist in the first place. For example, since it is hard to provide full wireless coverage to a rural area with an economically viable solution, it might be much more feasible to provide access to certain hotspots, e.g., schools, markets~\cite{nonIEEE_Brewer_RuralTech}. This limited coverage/mobility scenario can constitute a first step in deploying rural connectivity before expanding to solutions supporting larger mobility. When coverage is provided to a hotspot only, high gain directive antennas can be used to provide wireless connectivity from a relatively large distance. Thus, a pole on a hill can provide access to several sparse villages. Another approach is to use the communication devices at hotspots in a multihop fashion from village to village until reaching a village with backhaul connectivity~\cite{nonIEEE_Brewer_RuralTech}.

Another dimension that can be considered is the social dimension, where people can willingly share their connection with those who cannot afford it (or they can together share the cost), as long as geographical reachability is feasible with the used technology. In fact, multiple graphs are considered in~\cite{nonIEEE_RuralLowIncome_SmartGIFT18}, where one corresponds to the geographical availability of network connectivity (although some of the users might not be able to afford it), the second corresponds to affordability (i.e., corresponds to users who can pay for the connectivity, although some of them might be out of coverage), and the third corresponds to social shareability (i.e., showing the willingness of users to share the connection with some other users). Based on these graphs, connectivity can be extended to parts of the rural and low income areas that were otherwise unreachable, due to the joint use of these three dimensions~\cite{nonIEEE_RuralLowIncome_SmartGIFT18}.

This section analyzes these fronthaul considerations, focusing on mobility, provision of electricity, in addition to spectrum aspects.

\subsection{Mobility and Moving Cells}
\label{subsec:Fronthaul_Moving_Cells}
This section describes connectivity with limited mobility in rural areas, in addition to connectivity in hard to reach areas with moving cells, e.g., on board a train, boat, or plane.

\subsubsection{Nomadic Versus Intermittent Mobile Connectivity}
When users go to internet kiosks at fixed locations to get connected, they activate their devices at the kiosk location. This is a kind of nomadic mobility that allows people to have connection ate certain hotspot areas. However, in other scenarios, e.g., as in~\cite{IEEE_J16}, users are mobile, but they receive intermittent connectivity as they encounter \lq\lq infostations" along their trajectory. Thus, mobile devices need to be in \lq\lq hunting " mode (to use the terminology of~\cite{IEEE_J16}) so that they discover the presence of infostations. This process consumes energy, and thus the behavior of the mobile devices need to take it into account in order to optimize performance. The infostations could be connected in a mesh network until reaching a gateway with backhaul connectivity, whereas the mobile devices need to discover the infostations before performing the intended transaction within the contact time~\cite{IEEE_J16}.

\subsubsection{Trains, Boats, Planes}\label{subsubsec:Moving_Cells}
AANETs were extensively reviewed and discussed in~\cite{Hanzo_ProcIEEE_InternetAboveClouds}. The objective of AANETs is to provide ubiquitous connectivity to airplanes, taking into account the harsh propagation conditions and the strict security constraints. This ubiquitous connectivity is ensured by having connections between airplanes (air-to-air), between airplanes and satellites (air-to-satellite), between airplanes and ground stations (air-to-ground), in addition to communications between satellites (satellite-to-satellite), between ground stations (ground-to-ground), and between satellites and ground stations (satellite-to-ground). This complex network of interconnectivity provides coverage to airplanes wherever they are and allows passengers to be continuously connected~\cite{Hanzo_ProcIEEE_InternetAboveClouds}.

Satellite connectivity is an important component of AANETs discussed in~\cite{Hanzo_ProcIEEE_InternetAboveClouds}. Indeed, to meet the increasing passenger demands for ubiquitous connectivity, satellite connectivity is becoming more popular for transport networks, especially for trains, boats, and planes.  In~\cite{ESA_Train_Sat}, a system for satellite connectivity to trains is described. It consists of a VSAT terminal on top of the train connected to the satellite, and an internal train WiFi network to distribute the data to the passengers.

In case an MNO is providing access to train passengers, then the train's VSAT can be considered part of a moving BS, with satellite providing backhaul connectivity as described in Section~\ref{subsubsec:Backhaul_Sat}. In~\cite{BBC_TrainBoatPlane}, a scenario for deploying satellite connectivity to trains in the UK is described. Ofcom, the mobile operator, also planned deployment for boats and planes. This is performed by pointing an earth station positioned on the train, boat or plane towards a geostationary satellite. Advancements in pointing accuracy and reduced pointing errors will allow high data speeds to be achieved (in the order of $50$~Mbps per earth station)~\cite{BBC_TrainBoatPlane}.  In addition, the use of HTS with multiple beams not only increases throughput, but allows handover between multiple beams of the same satellite or between different satellites~\cite{MWEE_Sat_Trends}. The use of high throughput spot beams is convenient for in-flight connectivity, thus allowing bandwidth intensive applications like video streaming to be used inside airplanes~\cite{SES_HTS}. This kind of applications is also popular with cruise passengers in boats, where satellite connectivity allows these passengers to meet their expectations. Thus, in addition to its important use in boats dedicated to business (cargo, freight, etc.), satellite connectivity is also serving touristic cruises~\cite{SES_Caribbean}.
However, connectivity for boats and planes is considered non-terrestrial roaming and faces some administrative and legal challenges in order to be operational. The costs of roaming in these scenarios are expected to be high~\citep{BBC_TrainBoatPlane, Rocco_BoatPlane}. For example, in~\citep{Rocco_BoatPlane}, an example is described for satellite connectivity for boats between Denmark and Germany, where passengers were surprised by the high roaming fees while the boat was in International waters (for a short duration), as they assumed they were connected to either Danish or German operators with acceptable roaming prices.

\subsection{Power Grid Connectivity in Rural Areas}
\label{subsec:Fronthaul_Energy_Efficiency}
Many rural areas do not have access to the power grid~\cite{IEEE_C60}. Even when the grid connectivity exists, the power supply is often partial or intermittent. The use of solar panels in remote rural areas has its problems, e.g., the theft of solar panels used to power the telecommunications equipment~\cite{IEEE_C65}, or the telecommunication equipment itself~\cite{IEEE_C60}. Frequent power outages can lead to equipment malfunction and increased downtime, a problem that can be exacerbated when maintenance teams reside away from the rural area~\cite{IEEE_C60}. Thus, energy efficiency in operating telecom equipment in rural areas is of utmost importance.

In~\cite{IEEE_C22}, cellular coverage is planned for a rural area in India by starting with 2G GSM, with the hope of upgrading to newer technologies later. Microwave was used for the backhaul as it was found more cost effective than fiber or satellite links. Due to the intermittent power availability, the deployed BSs were powered by solar panels. To further reduce CAPEX costs, the telecom infrastructure was installed on the power transmission towers whenever possible, with appropriate safety measures taken to protect the equipment.

In areas completely off-grid, solar panels might be the only available source of power. Thus, they must be properly dimensioned to reduce the probability of downtime of the backhaul devices in periods of overcast or cloudy weather~\cite{IEEE_C140}. In~\cite{IEEE_C_PV_BTS} and~\cite{LTE_SolarBS_Korea}, BSs powered by solar panels were designed for GSM and LTE, respectively, whereas in~\citep{IEEE_C_WindFi, RE_PV_Wind}, BSs powered by both solar panels and wind energy are proposed. In~\citep{nonIEEE_LuleaTechRep_2016, IEEE_C214}, the use of massive MIMO beamforming to provide connectivity in rural areas was shown to lead to significant energy savings, especially when coupled with the use of ultra-lean design, where green networking concepts are implemented by switching-off certain functions of the BSs in the absence of transmission. In fact, beamforming allows concentrating the radiation on areas where the population exists, this avoiding the waste of power in unnecessary directions. The reduced energy consumption then makes it more feasible to power the BSs with solar panels~\citep{nonIEEE_LuleaTechRep_2016, IEEE_C214}.

In~\cite{IEEE_C197}, tethered lighter than air platforms are proposed for electricity generation in rural areas that are not connected to the power grid. In the approach suggested in~\cite{IEEE_C197}, mmWave power is beamed through the large antennas dimensioned for the platforms, and captured through a waveguide integrated into the tether. This is an interesting future area of research, where power can be generated wirelessly from mmWave to complement power generation from solar panels. Appropriate antenna sizing and dimensions, depending on the beaming platform used (high tower, stratospheric balloon, or satellite), are discussed in~\cite{IEEE_C128}. In~\cite{IEEE_J19}, massive MIMO antennas are used for energy harvesting using a TDD approach: In the downlink slot, the MIMO antenna powers the user devices, which then transmit their data in the uplink slot. This approach is suitable for IoT data collection for example. It was shown in~\cite{IEEE_J19}, via simulations, to increase the energy efficiency, especially when hybrid precoding is used to reduce the number of RF chains (a scenario most convenient with a sparse used density, typical in rural areas).

The connectivity to the power grid varies depending on the distance to the rural area from the nearest power station, the population density, the nature of the terrain, etc. Thus, different scenarios require different measures. Next we discuss three situations where the population density increases gradually.

\subsubsection{Single User} In the simplest most extreme scenario, a nomadic or mobile user moving in a remote isolated area might need connectivity in addition to the ability to power the communications equipment. He might have a portable small VSAT in his car, similarly the the scenario for trains and boats discussed in Section~\ref{subsubsec:Moving_Cells}, along with a small solar panel to provide the necessary power. In case the area has basic connectivity through some fronthaul technology, e.g., TVWS, the user can stop whenever connectivity is required and can set up a solar powered CPE to access the nearest TVWS BS.

\subsubsection{Small Population Agglomerations} In the more common scenario of a relatively small or low density population agglomeration, e.g., a small village or a group of neighboring villages, the use of renewable-powered BSs complemented as needed by Diesel generators, or the presence of a micro-operator for power, might be common local solutions.

In~\cite{IEEE_C34}, long-range WiFi APs used to provide internet access to rural villages were powered by solar panels, whereas special user end-user equipment used to communicate with the access points was designed: It had a battery that could be charged by using electricity provided by diesel generators, which the villagers used to turn on for a few hours for their daily activities since power from the mains grid was absent. Even when APs are powered by solar panels, in scenarios where the usage of WiFi is intermittent in rural areas, the APs can be put to sleep in order to save energy. A wake-up mechanism can then be put in place as proposed in~\cite{IEEE_C156}, where a beacon signal is transmitted by the sending device with a predefined signature. Once, this signal is detected by a sensor connected to the sleeping AP, the sensor wakes up the AP and communication can take place.

In~\cite{IEEE_C113}, a \lq\lq micro telecom" approach is proposed to provide 2G GSM connectivity to rural areas. Under this approach, each small village is equipped with a small BS, with each village BS connected to a central BS using a microwave link, where a central BS covers a group of surrounding village BSs. These central BS sites can communicate with each other via multihop until finally reaching a BS site connected to the backbone network. Power needs of such a deployment can be achieved with $100$~Watts for each village site and $400$~Watts for each central site, which can be achieved through the use of solar panels~\cite{IEEE_C113}.

In rural regions that are off-grid, recharging smart phones becomes a challenge. Recharge outlets in these rural regions are provided for a fee, which increases the costs for the rural population. Smart phones with reduced specifications and tailored data bundles consisting of applications draining less battery power can be used to alleviate this problem~\cite{IEEE_C203}.

A DC microgrid model is proposed in~\cite{IEEE_C109} to deal with the absence of electrification in rural areas. Users deploy solar panels in their homes. They use the energy they need, and transfer the surplus to the grid. When their battery levels go down, they consume energy from the grid. Other users, not deploying a solar system at their homes, act only as consumers from the grid. A microgrid controller monitors and controls these activities~\cite{IEEE_C109}. Such a microgrid can be used to power a small village or a small group of houses in a rural area, and to provide electricity for the telecom equipment used for rural connectivity. It can provide the energy needed for other services such as water pumping, e.g., see~\cite{IEEE_J20} where water pumps are designed to work with solar panels in an energy efficient manner. In~\cite{IEEE_C210}, an approach for dimensioning solar panels and wind turbines is proposed, along with a corresponding simulation software, so that users in rural areas can use it meet their energy needs. Similarly, in~\cite{IEEE_J14}, dimensioning of a microgrid consisting of solar panels and wind turbines is performed by taking into account weather conditions (hourly solar irradiation and wind speed data) at each deployment location. The optimization in~\cite{IEEE_J14} is performed using a Genetic Algorithm aiming to find the best tradeoff between cost and reliability. For off-grid rural areas with limited solar energy, a hybrid approach based on wind turbines and diesel generators operating in a complementary manner is proposed in~\cite{IEEE_C180}. The cost of energy generation is minimized using a quadratic programming approach.

Several microgrids in rural villages can collaborate to exchange any surplus of energy and meet the demands of local consumption, especially when energy generation from renewable energy sources is adopted. Therefore, a coalitional game theory approach is adopted in~\cite{IEEE_C63}, in order to exchange energy between rural microgrids that are not connected to the mains power grid, with the objective or reducing the energy costs of the coalition. A similar collaborative approach between $12$ neighboring villages in an off-grid rural area in India is described in~\cite{IEEE_C145}, where linear programming is used to optimize the energy costs. The area contains a mix of energy sources including solar, wind, biomass, micro-hydro, and diesel generator that is used whenever the renewable sources cannot meet the demand~\cite{IEEE_C145}. The presence of multiple microgrids is considered beneficial, even when these microgrids are connected to the main grid~\cite{IEEE_C40}. In fact, they provide resiliency and robustness in case of any intermittent supply or outage of the main grid~\cite{IEEE_C40}. Also, in~\citep{IEEE_J25, Private_Sector_Microgrid}, microgrids without connectivity to the main grid are considered as a distributed solution for the electrification problem in rural areas. In~\citep{Private_Sector_Microgrid}, government intervention (subsidies, loans, etc.) and tariffing methods to encourage the private sector to deploy microgrids in rural areas are discussed. It is suggested in~\citep{IEEE_J25} that they should be owned by the community, which reduces theft, increases awareness and sense of ownership, and allows for job creation by educating local personnel to perform maintenance operations (similar remarks were also presented in~\citep{Private_Sector_Microgrid}). Houses can have power meters to measure consumption from the village's microgrid, which can also be used for water pumping, refrigeration (e.g., for medicine), street lighting, providing power to common facilities (school, local medical center), and powering telecom equipment~\cite{IEEE_J25}. Thus, microgrids can meet local demand, can coordinate their supply to meet the demand of a larger rural area, and can remain helpful after the main power grid reaches the rural area.

\subsubsection{Larger Rural Population Agglomerations}
\begin{figure}[t!]
 \begin{center}
  \includegraphics[scale=0.7]{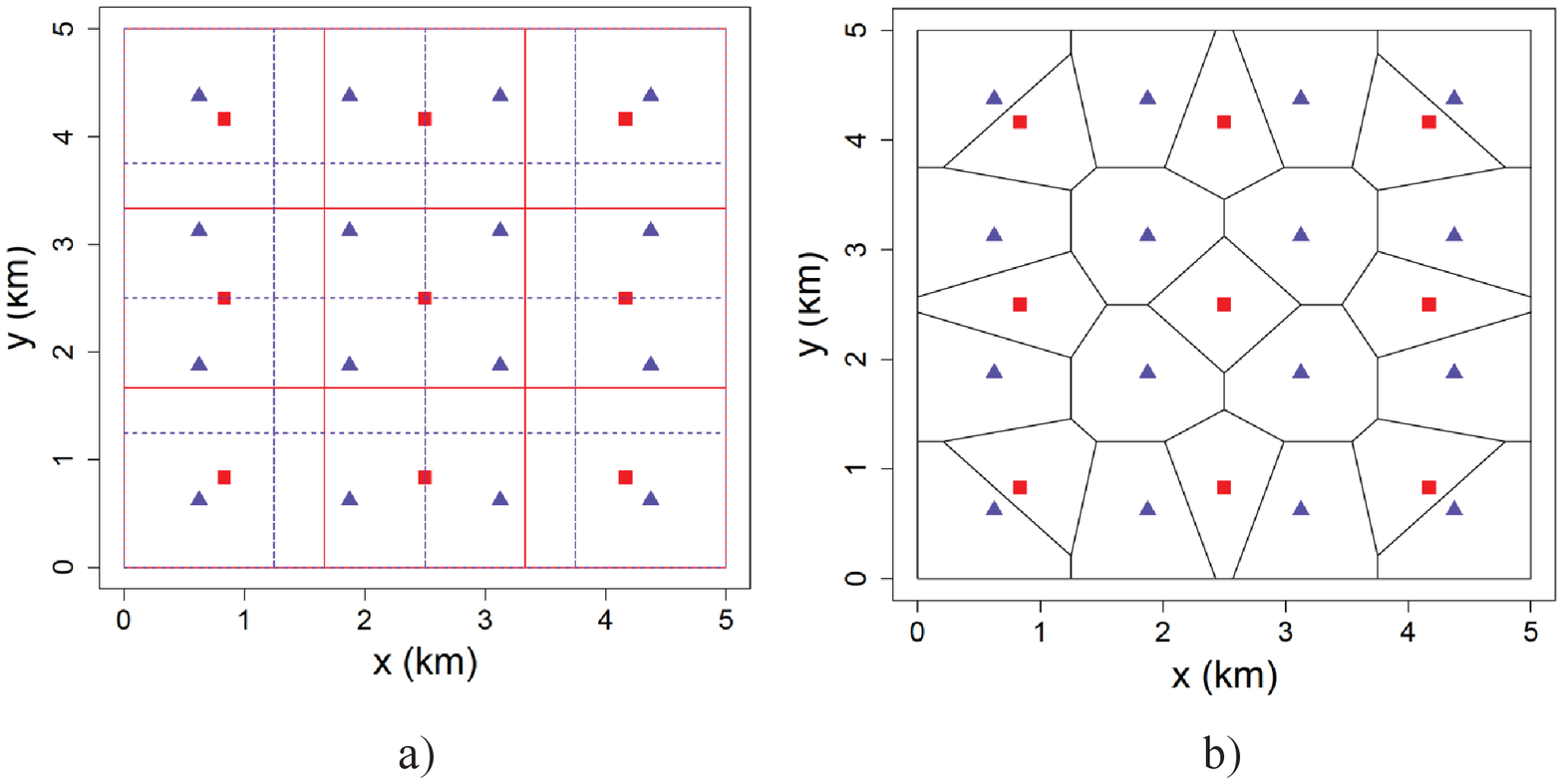}
 \end{center}
  \vspace{-12.0cm}
\caption{Cell coverage with two operators under cooperative and non-cooperative scenarios: (a) Non-collaborative mode where the solid lines grid represents the Voronoi cells for Operator~1 (BSs shown as red squares) while the dotted lines grid represents the Voronoi cells for Operator~2 (BSs shown as blue triangles); (b) Collaborative mode where the BSs of Operator~1 (red squares) and those of Operator~2 (blue triangles) form a single virtual network.}
\label{fig:Operator_Coop_Coverage}
\end{figure}
\begin{figure}[h!]
 \begin{center}
  \includegraphics[scale=0.7]{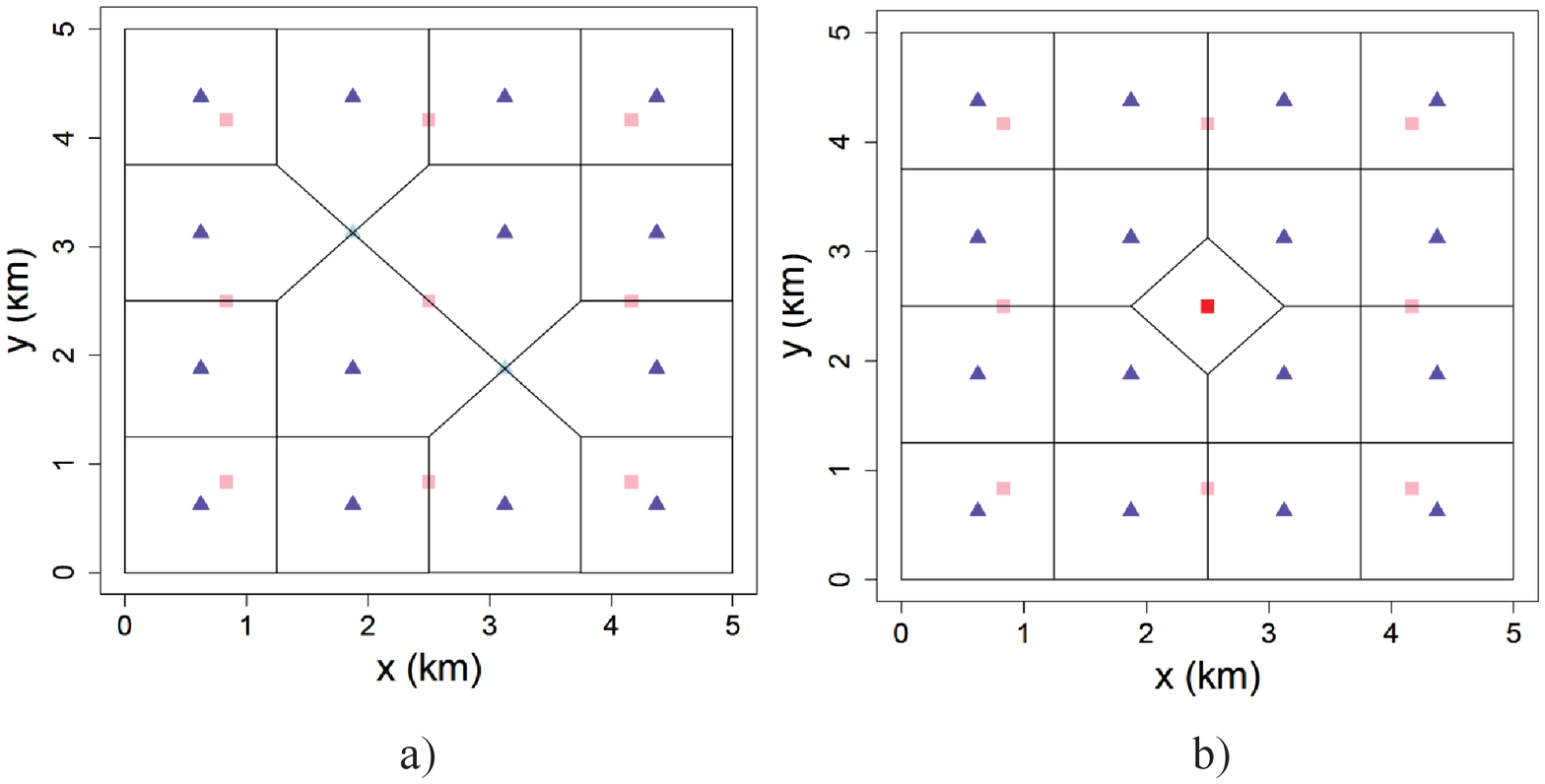}
 \end{center}
  \vspace{-12.0cm}
\caption{BS on/off switching with operator collaboration: (a) Active BSs with uniform user distribution; (b) Active BSs with Gaussian spatial user distribution centered at (2.5,2.5) and having unit variance.}
\label{fig:Operator_Coop_OnOff}
\end{figure}

In larger areas with bigger and denser population agglomerations, ARPU might be higher and lead to more profitability for mobile operators in these dense population islands located in large rural areas. Furthermore the mains power grid might have reached these areas, although power might be intermittent, which allows mobile operators to have more flexibility in powering their network. For example, they can not only rely on renewable energy sources to power the communications network and use the power grid when the generated renewable power is not sufficient, but also sell the excess of renewable energy to the grid whenever the generated power by solar panels exceeds the network needs~\cite{Hakim1}. Furthermore, this approach can be coupled with BS sleeping strategy where certain macro BSs provide coverage whereas smaller BSs need only to be active whenever the traffic demand increases in these areas. Thus, the PV cells powering BSs can be linked together as a sort of microgrid owned by the mobile operator, who can then route the surplus of renewable power to BSs where it is needed most, while others are switched off, and still be able to sell any surplus to the main power grid~\cite{Hakim2}.

Another approach that allows operators to have a sustainable business model in rural areas is the collaboration between operators whenever more than one are covering a certain area. In addition to sharing towers and equipment for example, they can run their networks as a single virtual network. Operator~1 will receive fees when it serves subscribers of Operator~2 and vice-versa. This approach is investigated in~\cite{Hakim3}, where optimal solutions to set this inter-operator roaming price are discussed, depending on the profit margins and the availability of renewable energy at the BSs of the various collaborating operators.  Fig.~\ref{fig:Operator_Coop_Coverage} shows an example of an area covered by two operators, and indicates the coverage areas of each BS whenever they act separately or act in full collaboration such that their networks form a single virtual network. Fig.~\ref{fig:Operator_Coop_OnOff} shows a collaborative scenario with BS on/off switching, corresponding to an extreme scenario where Operator~2 relies mostly on PV panels whereas Operator~1 relies mostly on fossil fuels to power the network. Clearly, most of the BSs of Operator~1 are switched off in order to optimize the costs, and the network is served as much as possible by BSs powered by renewable energy. The energy cost savings of Operator~1 would allow him to pay the roaming fees of his subscribers to Operator~2 while still gaining some profit. These conclusions hold assuming either a uniform distribution of subscribers over the area as in Fig.~\ref{fig:Operator_Coop_OnOff}~(a) or a Gaussian distribution where the population is concentrated in the center of the area and the density reduces gradually as we move towards the boundaries, as in Fig.~\ref{fig:Operator_Coop_OnOff}~(b).

\subsection{Spectrum and Economical Aspects}
\label{subsec:Fronthaul_Spectrum_Economical_Aspects}
This section describes the relevant references discussing the issues related to spectrum regulation, spectrum allocation, and spectrum auctions, taking into account the conditions specific to rural areas.

Spectrum auctions are not justified given the low population density in rural areas since upfront spectrum costs are not justified by the low expected ARPU. Therefore, one of the potential solutions consists of having governments \lq\lq require", in the auctioning process, that the auction winners will use the spectrum to provide coverage not only for urban areas but also for rural areas. Another solution would be to use unlicensed spectrum first, and then after the connectivity is established, licensed spectrum can be used (to avoid the interference problems with unlicensed spectrum when connectivity increases)~\cite{nonIEEE_Brewer_RuralTech}.

When licensed spectrum is used, lower frequencies are more suitable for rural areas due to their better propagation characteristics, and thus a larger area can be covered with less sites. Although a lower carrier frequency entails a lower bandwidth, this is not a problem in most situations due to the sparse population density in rural areas~\cite{IEEE_C172}.


In~\cite{nonIEEE_Spectrum_MIS2017}, the role of national regulation authorities (NRAs) in promoting the use of TVWS is discussed. An approach for multi-criteria decision analysis to evaluate spectrum management frameworks was proposed and used to evaluate the Federal Communications Commission (FCC) in the USA and Ofcom (UK) frameworks for dynamic spectrum management. It was found that it is preferable to keep tight control on the geo-location spectrum database by operating it by the NRA, in order to more efficiently protect the TV users from secondary access, while allowing license-exempt spectrum access to secondary users~\cite{nonIEEE_Spectrum_MIS2017}. Similarly, in~\cite{Spectrum_Regulation_TVWS}, TVWS spectrum regulations in several countries were discussed, and it was noted that many developing countries, notably in Africa, where the TVWS spectrum is the most available, are lagging behind developed countries in terms of spectrum regulation. In~\cite{Montsi_Big_Data_GWS17} a database for real-time smart spectrum sharing was designed. A testbed was implemented, and it was successfully able to serve $35,000$ connection requrests in less than $15$ seconds. Furthermore, a discussion was provided in~\cite{Montsi_Big_Data_GWS17} about the capability of the cloud database to handle spectrum sharing from different regions, and to be able to coordinate dynamic spectrum allocation across neighboring countries in border regions.

In~\cite{IEEE_SpectrumTV_GIIS2013}, a game theoretic scenario was considered for providing TVWS spectrum in rural areas, where a BS aims to sell secondary spectrum to APs in a wireless mesh network. The problem is modeled as a Stackelberg game where the BS attempts to maximize its profit and the APs want to maximize the QoS of their users. In~\cite{nonIEEE_SpectrumTV_jICTs2014}, the same authors build on the work in~\cite{IEEE_SpectrumTV_GIIS2013} to provide a more elaborate game theoretic scenario. A game model based on Bertrand duopoly market was adopted in~\cite{nonIEEE_SpectrumTV_jICTs2014}, where two primary users (PUs) compete for providing services (in this case TVWS spectrum in rural areas) to secondary users (SUs). The objective is to maximize the profits of the PUs while meeting the QoS constraints of the SUs. In~\cite{IEEE_J4}, the TVWS spectrum allocation is modeled as a two-stage Stackelberg game: (1) between the central BS (CBS) connected to fiber backhaul and fixed CPEs, and (2) between CPEs and UEs, where several UEs are connected to each CPE. The CBS distributes the total available data rate to CPEs, and each CPE distributes its share to its connected UEs. It was shown that although network entities behave selfishly, the scenario of~\cite{IEEE_J4} leads to optimal rate distribution to UEs, depending on the UE willingness to pay.

In~\cite{IEEE_C85}, the use of GSM \lq\lq white spaces" in rural areas was proposed. The objective is to build community cellular networks with affordable prices to end-users, by avoiding the payment of spectrum licenses for GSM frequencies. It was argued that this was feasible due to the low spectrum occupancy in rural area. This generally applies also for traditionally license-exempt spectrum, e.g., as in~\cite{IEEE_C89}, where it was shown that WiFi 5~GHz spectrum occupancy is significantly low in rural areas, especially when compared to occupancy in urban areas.

\section{Service to End-Users/Applications}
\label{sec:Services}
To increase internet adoption in rural areas, users need to see the benefits provided for their daily lives, which will increase adoption, and eventually make the business case viable for operators. Relevance to rural users can be demonstrated through, e.g., eEducation, eCommerce, and eGovernment~\cite{Strategy_Business_Article_2016}. Furthermore, users need to adopt the offered services, which will increase demand and encourage operators to enhance the level of connectivity provided. This requires user awareness, simple accessible applications, and the provision of content that is of interest to the local community.

\subsection{Description of Provided Services}
\label{subsec:Service_Description}
\begin{figure}[t!]
 \begin{center}
  \includegraphics[scale=0.6]{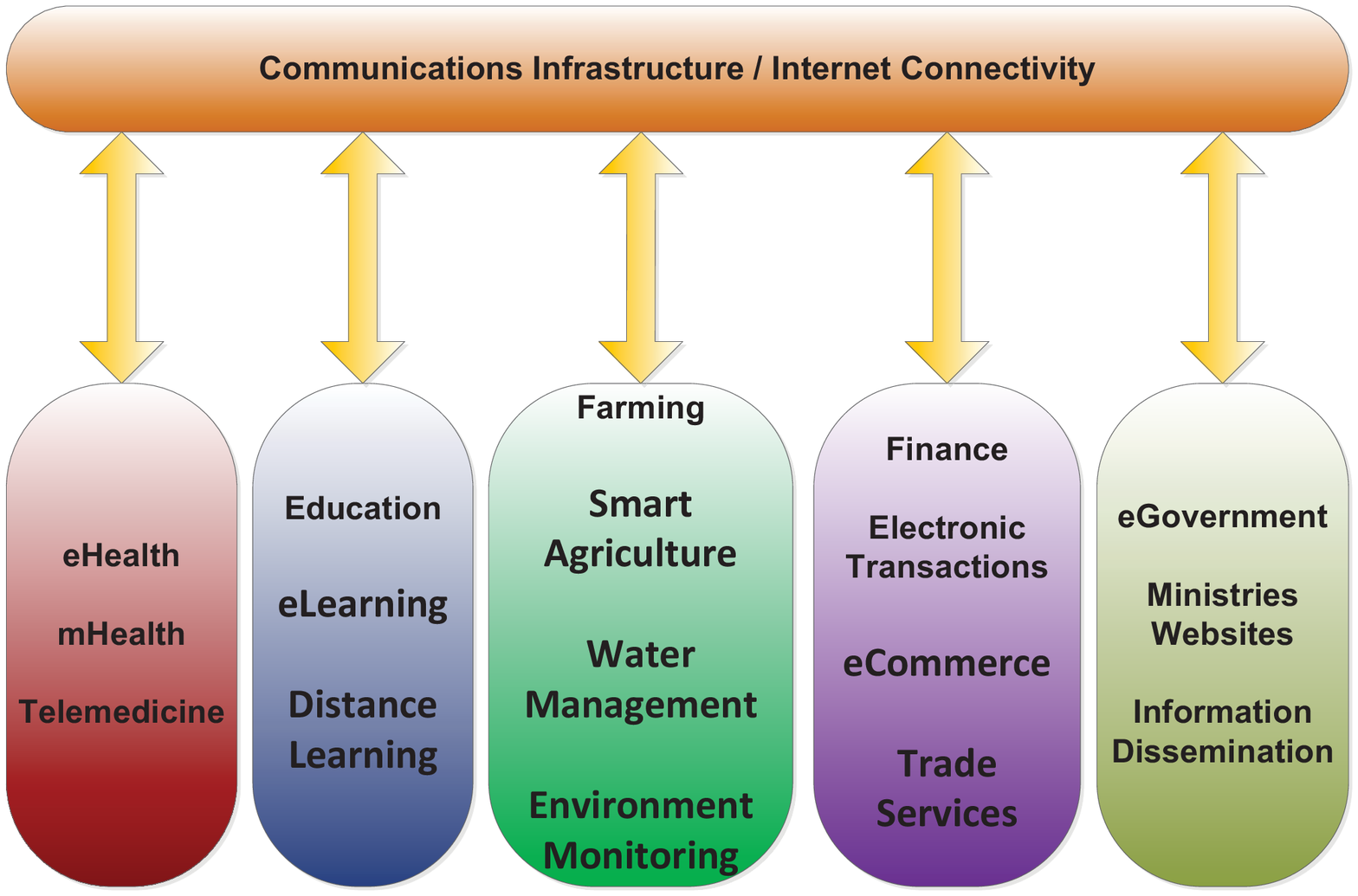}
 \end{center}
  \vspace{-8.0cm}
\caption{Interaction between the communications infrastructure and the various services.}
\label{fig:Interaction_Comm_Services}
\end{figure}

Typical services needed in rural areas and that can be facilitated by telecommunications networks include eHealth, eCommernce, eGovernment, in addition to environment monitoring and farming~\cite{IEEE_C103}, as shown in Fig.~\ref{fig:Interaction_Comm_Services}.

\subsubsection{Health}
In rural areas, patients lack quality healthcare. Telemedicine represents a suitable solution to address this problem~\cite{IEEE_C155}. Primary health centers can be overburdened with daily visits while the referral to secondary and tertiary health centers does not use technology to accurately transfer patent information~\cite{IEEE_C124}. In~\cite{IEEE_C124, IEEE_C83}, two different systems are proposed where in both, a social worker helps patients in a rural area communicate with doctors remotely using multimedia technology relying wired broadband supported by wireless connectivity (WiMAX in~\cite{IEEE_C83}) in hard to reach areas. A simulation of the WiMAX physical layer was performed in~\cite{IEEE_C170}, with the aim of supporting telemedicine applications in rural areas. In~\cite{IEEE_C46}, ultrasound imaging was performed in rural areas by trained non-physician personnel, and the results were transferred to a cloud system where the physicians could perform diagnosis in an urban hospital. The approach of~\cite{IEEE_C46} requires that the portable ultrasound device be connected to a WiFi router, which necessitates some form of basic internet connectivity in the rural area to send the measurements to the cloud storage. In~\cite{IEEE_C160}, image processing techniques were proposed for detecting eye diseases, where a technician can take a picture of the patient and the disease can be diagnosed remotely.

With the advent of the IoT, sensor measurements can be performed by BANs in rural areas; they are transmitted to smart mobile phones~\citep{IEEE_C32, IEEE_C152, IEEE_C110}, e.g., using Bluetooth low energy~\citep{IEEE_C32}, where they can be stored and transmitted to the cloud whenever network connectivity is available, whether through WiFi, WiMAX, or cellular networks. Measured mHealth signals could include electroencephalography~\citep{IEEE_C152} and electrocardiography~\citep{IEEE_C110}. Cloud storage can allow remote diagnosis and patient monitoring~\cite{IEEE_C32}, possibly with the help of a \lq\lq health ontology", as proposed in~\cite{IEEE_C161}. In~\cite{IEEE_C206}, a movable booth, that can be transported by bicycle or motorcycle, is designed to monitor the health parameters of children under five years old in rural areas. Sensors located inside the booth, and connected to a single microcontroller, measure the vital signs and the microcontroller transfers the data to the cloud for processing.  In~\cite{IEEE_C207} the telemedicine process is extended to telepsychiatry, where the images captured by a mobile's camera are processed, and the extracted features are transmitted to make a diagnosis about the individual's wellbeing. The features are transmitted instead of the images in order to preserve the privacy of the patients.

Thus, in rural areas, connectivity for health services can help in telemedicine and remote diagnosis, tracking of disease and epidemic outbreaks, training health workers through eEducation, and increasing the population awareness about certain diseases and best practices to avoid them~\cite{IEEE_C103}. These constitute preliminary steps until the infrastructure allows supporting the revolutionary healthcare aspects promised by 5G, such as remote surgery with haptic feedback using URLLC, diagnostics using robots and machine learning, etc.~\cite{5G_Healthcare}.

\subsubsection{Education} Education is also an important service to be provided to remote rural areas. In~\cite{IEEE_C83}, remote education is provided to rural areas via multimedia based virtual classes. In~\cite{IEEE_C195}, a \lq\lq flipped classroom" model was used in order to support the online education of PhD candidates while overcoming the limited connectivity in rural areas that hindered the implementation of video conferencing based online classroom setup. In~\cite{IEEE_C148}, an interactive education system using satellite networks is tested for rural areas. Scenarios with direct satellite connectivity, or with WiFi for local area connectivity after receiving the signal from the VSAT terminal, were both tested and shown to achieve acceptable packet loss rates. The internet of educational things (IoET) is proposed in~\cite{IEEE_C67} for underprivileged rural areas. Students in the first grade are equipped with tablet computers. Due to the lack or reliable internet connectivity, students can mainly use the tablets for reading eBooks. To make the learning experience more attractive and introduce IoET, the approach of~\cite{IEEE_C67} consists of using several sensors so that the students become acquainted with the environment (temperature, humidity, etc.), since they mainly come from a farming background. The sensors are connected to a Raspberry PI device and packaged within an appealing enclosure that the students can design using a desktop 3D printer. Students can access readings and pictures through a web application on the tablet, using a local WiFi connection. Thus, IoET was introduced in~\cite{IEEE_C67} using local content without relying on a backbone network.

\subsubsection{Farming}
Farming is a typical application area for rural zones that can benefit from internet connectivity, or even from local network connectivity. For example, IoT devices can be used to perform precision irrigation in areas with limited availability of irrigation water. Drip irrigation not only uses the right amount of water, but optimizes the watered area near the plant roots~\cite{IEEE_C18}. RFID sensors can also be used to track and monitor livestock~\cite{IEEE_C18}.

In~\cite{IEEE_C79}, IoT sensors are proposed for farming applications in rural areas without internet or cloud connectivity. UAVs are used to collect the measurements from IoT sensors, and relay them to the nearest 5G BS. The BS is provided with a renewable energy generator, complemented by a diesel generator. UAVs can be recharged at a recharge station located at the BS site before resuming their operation. The site also contains edge computing servers to process the farming-related measurements in the absence of cloud connectivity. Due to the variability of power availability, caused partially by the varying numbers of UAVs being recharged, the number of active servers can vary. Therefore, measurements from arriving drones are queued for processing, and a queuing model is proposed in~\cite{IEEE_C79} to capture the behavior of the system.

An IoT approach to detect leaf diseases in farming scenarios in rural areas was presented in~\cite{IEEE_C127}. In the approach of~\cite{IEEE_C127}, several sensors (temperature, humidity, soil moisture etc.) and a camera are attached to to a Raspberry PI device. Measurements are collected and transmitted to the cloud where they can be stored in a database and retrieved by the farmers through a web application to check the status of their plantations. The camera captures figures of the leaf which are then transmitted and processed with suitable feature extraction techniques in order to detect whether any disease is affecting the plants.

In~\cite{IEEE_C4}, LoRa is used to collect measurements from IoT devices. The gateway collecting the measurements of the devices using LoRa used LTE as a backhaul to forward the measurements an LTE BS and then to the cloud through the LTE core network. The scenario of~\cite{IEEE_C4} applies to agribusiness where the gateway is a moving vehicle used in the agricultural process, e.g., to collect cane.

In~\cite{IEEE_C107}, it was noted that farming applications that require real time coordination such as crop transportation during grain harvesting, might suffer from the intermittent cellular coverage in rural areas when the Message Queue Telemetry Transport (MQTT) protocol is used for transmitting IoT measurement data. MQTT is implemented on top of TCP, which slows down performance due to its guaranteed delivery property, especially when the packet losses are due to the wireless channel conditions, not to network congestion. Adopting a last-in first-out transmission approach instead of a first-in first-out approach was shown in~\cite{IEEE_C107} to be more convenient, since the latest update is the more relevant, which leads to a reduced delay in sharing the most up to date information.

\subsubsection{Financial Services} The deployment of automated teller machines (ATMs) and the use of point of sale (POS) devices pose numerous challenges in rural areas. One of the most important challenges is the lack of reliable connectivity so that the transactions with the users' cards can be performed in real-time. In~\cite{IEEE_C114}, a solution is proposed to this problem by using public key infrastructure (PKI). Private keys are stored in the card and the rural ATM, and the public keys issued by a trusted authority (e.g., Central Bank of a given country) can be exchanged between the two entities. This way, authentication, encryption (for confidentiality), and integrity can be performed locally without the need to have a connection. Enhanced security can be achieved by using a mobile phone along with the smart card according to the approach described in~\cite{IEEE_C114}. However, certain issues remain to be addressed, the most important one being the risk of overdrawing from a certain bank account (since the ATM does not have a permanent connection). A possible solution proposed in~\cite{IEEE_C114} is to store the account balance on the card, and to allow trusted ATMs to modify it in case of withdrawal. The information at the bank's server will then be updated periodically (either when the connection is available with the ATM, or manually where an authorized bank employee collects data from ATMs using a memory stick).

The previous works have tried to customize the banking and payment transactions to rural areas, where it is assumed that network intensive transactions such as blockchain are hard to implement. However, in~\cite{IEEE_J24}, blockchain financial transactions are extended to rural areas under certain conditions. It is assumed that reliable connectivity is provided by local community BSs such as Nokia Kuha~\citep{nonIEEE_Kuha1, nonIEEE_Kuha2, nonIEEE_Nokia1}, but the connectivity between the BS and the internet is intermittent. In that case, transactions can be made locally where the entities involved include users, miners, and proxy nodes. Incentives are provided to local miners, whereas proxy nodes can be co-located with one of the community BSs (one proxy per village). The proxy nodes complete the already performed local transactions when backhaul network connectivity is available. The feasibility of the proposed approach in~\cite{IEEE_J24} was demonstrated using a proof-of-concept testbed using Raspberry Pis and off-the-shelf computers.

\subsubsection{eCommerce/Trade Services}In~\cite{IEEE_C104}, a system for supporting self-help groups in rural areas was proposed. The objective is to support micro entrepreneurs in rural areas to expand their businesses and support the local economy. The system of~\cite{IEEE_C104} uses mVAS and IVR to allow transactions between rural stakeholders in order to support micro entrepreneurs in rural areas and allow them to expand their business activities. The system is mobile based since it assumes the population lacks sufficient education and financial means to own computers and use web based services. However, it caters for the creation of a web portal that can be expanded in the future as the users become more computer aware.

In~\citep{IEEE_C73}, an ePayment method in rural areas was proposed using SMS. Under this approach, users have to register with the system. They can then top up their accounts by purchasing vouchers and sending the code via SMS or by transferring amounts to online systems like~\cite{Dokupay}. Afterwards, they can use SMS for their payment transactions whenever they make any purchase, and they receive confirmation accordingly, also via SMS. The results of~\citep{IEEE_C73} have shown that the average response time with SMS is around $30$-$40$ seconds.

In~\cite{IEEE_C71}, a system for eProcurement was designed over low end smartphones. It allowed small scale retailers in a rural area to replenish their stocks without having to close their shops and move to the nearest urban center $70$~km away. Furthermore, it allowed the providers to update the availability and pricing information of products online, and to schedule bulk delivery to specific delivery points closer to the retailers, where the payments can be made in cash upon delivery (the system does not support online payment since most of the people in that rural area do not have bank accounts and credit cards).

\subsubsection{eGovernment} Before establishing eGovernment in rural areas, a bottom up approach is proposed in~\cite{IEEE_C84}, where the services related to health, agriculture, and education in rural areas should reach a certain benchmark level before embarking on an eGovernment project. Otherwise, the eGovernment project would not achieve the intended benefits as the rural area is not yet ready for its adoption, according to the approach of~\cite{IEEE_C84}.

\subsubsection{Other Services}
In~\cite{IEEE_C8}, kiosks are proposed to provide employment opportunities in rural areas in order to support the local economy. Job seekers can post their information at the kiosk, whereas employers post available opportunities. The same kiosks can be used for buying/selling purposes, where the sellers can advertise their products at the kiosks.

Bus ticketing in rural areas was considered in~\cite{IEEE_C125}. Ticket vending stations in rural areas suffered from slow connectivity to central servers. Therefore, the approach of~\cite{IEEE_C125} was based on machine learning implemented on central servers, and then publishing the resulting models on text files that can be downloaded by the Android-based ticketing stations at the start of each business day to help in predicting departures and destinations and thus speed up the ticketing process.

In~\cite{IEEE_C157}, surveillance of long linear infrastructures, e.g., pipelines, power grid, railroads, involving long stretches of rural areas with limited connectivity is discussed. Typically, surveillance and monitoring activities are performed using UAVs, possibly grouped in flying ad hoc networks, with real-time video transmission sometimes required. Scenarios studied in~\cite{IEEE_C157} include (i) cellular coverage, (ii) the use of towers for line of sight communication with the UAV, and (iii) the use of MIMO transmissions by having several connected antennas along stretches of the monitored infrastructure. In general, the three methods are feasible since there is usually room alongside the monitored infrastructure to place the communication infrastructure. The approach of~\cite{IEEE_C157} allows modeling the wireless coverage in the monitored rural area by using only a limited set of signal strength measurements and then interpolating the results to properly perform mission planning for the UAV trips. In~\cite{IEEE_C99}, the scenario of UAVs used for monitoring with cellular connectivity using LTE was considered, and interference measurements were performed. To reduce interference from ground UEs or BSs, techniques like adaptive beamforming or coordinated multipoint were proposed. A more detailed analysis was performed in~\cite{IEEE_J12} for the same scenario, where beamforming and interference cancellation were proposed at the UAV side, along with intercell interference coordination at the BS side.

In\cite{IEEE_C204}, a crowd-sourcing approach is used to predict the signal strength of cellular networks in rural areas. It is based on using an application that collects location information along with signal strength data, and then uses the collected samples to predict the coverage over the whole area. It can be used for assessing the coverage in order to implement services like eHealth, eCommerce, etc.

\subsection{User Awareness}
\label{subsec:User_Awareness}
Many people in developing countries are not aware of the potential of the internet in changing their daily lives, and some have even not heard of it~\cite{Strategy_Business_Article_2016}. The lack of digital literacy is indeed a barrier to internet adoption~\cite{McKinsey_Report_2014}. Furthermore, if users do not know of the existence of a service, they cannot use it. A survey in~\cite{nonIEEE_BharaNet1_ppt} showed that 70\% of surveyed institutions did not know of the existence of BharatNet, a project for providing connectivity to rural areas in India~\cite{nonIEEE_BharaNet2_pdf}.
Rural communities must determine their needs, see a potential benefit in the technology, and acquire the skills needed to use it in order to meet those needs and achieve the intended benefits. Therefore, public education campaigns are needed in addition to deploying the necessary technology to provide rural connectivity~\cite{IEEE_C74}. For example, in~\cite{IEEE_C37}, training was performed for teachers in rural areas in order to be able to use information and communication systems in their education approach. In~\cite{IEEE_C58}, an SMS-based gateway is developed in order to allow teachers, with limited or no internet connectivity in rural areas, to communicate and exchange experiences by accessing a chat room where their counterparts with internet connectivity are interacting. In~\cite{IEEE_C147}, a mentoring approach is proposed where engineers would train adequate groups in rural areas to use innovative technologies. In~\cite{IEEE_C181}, a helpdesk was established to assist farmers, that are mostly poor and illiterate, in making the best decisions for their farming activities. They can call an assistant and provide their farmer ID, with the assistant having internet access and connected to a central database containing the encountered problems and their solutions.

Even in developed countries, adequate training and awareness initiatives are performed when connectivity is provided to rural areas, e.g., in~\cite{IEEE_C100} where connectivity is provided to clinics in rural areas and the medical personnel was trained to use electronic health records.

A \lq\lq living lab approach", where the users in the rural area take part in the innovative process, which should be user-centric and tailored to their needs, helps increasing technology adoption in rural areas~\cite{IEEE_C71}. Furthermore, using local languages in applications can accelerate deployment and break language barriers to technology adoption, since people would feel the technology more relevant to their needs, especially in poor rural areas where the population is under-educated~\cite{IEEE_J1}.

\begin{figure}[t!]
 \begin{center}
  \includegraphics[scale=0.6]{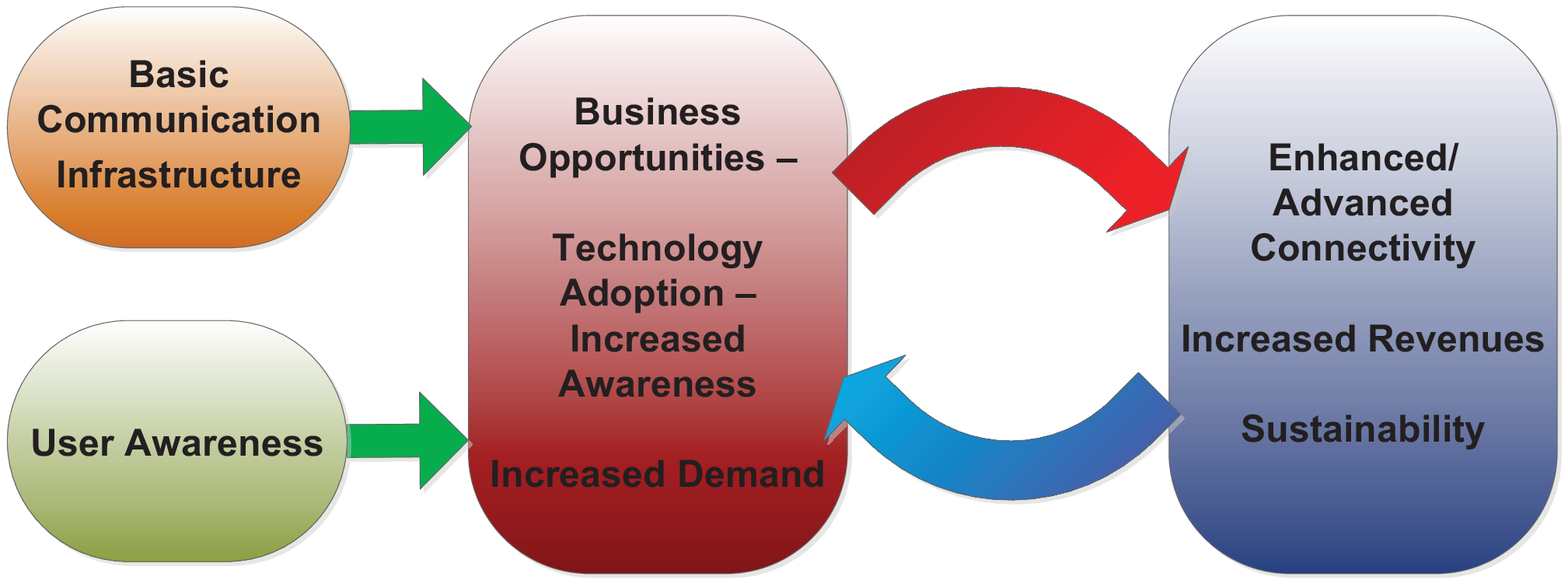}
 \end{center}
  \vspace{-11.0cm}
\caption{Interaction between connectivity and awareness.}
\label{fig:Interaction_Awareness_Connectivity}
\end{figure}

Indeed, providing a basic connectivity in rural areas, coupled with user awareness and basic adoption of the technology, will lead to creation of local business opportunities, which in turn will lead to more adoption and increased demand. The higher demand will lead to more revenues for operators and will require more advanced deployments to cope with the demand, which will lead to gradually enhancing the connectivity as the business model will become gradually more sustainable and profitable. This \lq\lq virtuous circle\rq\rq is shown in Fig.~\ref{fig:Interaction_Awareness_Connectivity}.

\subsection{Making Things Simple}
\label{subsec:Making_Things_Simple}
In rural areas, in order to start building basic infrastructure, cheap off-the-shelf equipment can be used (although it should be robust to support the weather conditions specific to the rural area considered), in addition to open source software whenever possible.

In~\cite{IEEE_C97}, a network management and monitoring platform was built from several open source software programs, in order to allow a local wireless internet service provider to manage a network in a rural area in an affordable way. The open source tools used in~\cite{IEEE_C97} allowed network monitoring, network management, intrusion detection, and firewall functionalities.

The common trend of reduction of antenna sizes in mobile handsets in order to achieve low form factors reduces their efficiency, which limits the performance of handsets, particularly in rural areas where the signal typically has to travel longer distances. Therefore, in~\cite{IEEE_C175}, tunable antennas are proposed to increase the efficiency. In~\cite{IEEE_C92}, tin cans were used to increase the gain of CPE antennas used in rural areas. In~\cite{IEEE_C203}, in order to allow people in low-income rural areas to afford smart phones, low-end smartphones are suggested as a solution. They provide functionality similar to smart phones but possess lower hardware specifications.  In~\cite{IEEE_C71}, an application designed for low-end smartphones was used to allow local retailers to perform eProcurement. In~\cite{IEEE_C200}, GSM voice channels are used for data transmission. The data bits are modulated and transmitted knowing that the voice codec will affect the signal. The forward channel is used for transmission, and the reverse channel is used for feedback. This method aims to transmit low data rates (although higher than SMS rates), in rural areas where voice connectivity is available over GSM but no data connectivity is provided. Typical applications for the method of~\cite{IEEE_C200} include micro-financing and point-of-sale transactions.

To provide connectivity for the elderly, poor, and digitally illiterate, the \lq\lq basic internet\rq\rq initiative is defined~\cite{BasicInternet_Website}, where internet access can be provided free of charge when only static content (text and images) is requested, since this kind of traffic amounts to around 2-3\% of the bandwidth~\cite{IEEE_J30}. Those requesting dynamic content can do that for a fee. Internet neutrality is maintained since the filtering is done based on data type and not on content provider~\cite{IEEE_J30}. In addition, techniques for human computer interaction (HCI) that are suitable for the elderly and illiterate are discussed in~\cite{IEEE_J30}. For example, simple authentication methods based on biometrics can be used, and voice instructions can accompany web navigation. The GAIA initiative also aims to provide global internet coverage~\cite{GAIA_IETF1}.

If reaching every home in a rural area is not currently possible, providing hotspot areas with broadband connectivity can be considered a starting point. These could include the village school for example, or a public library. For example, a discussion of providing public libraries with broadband access in rural areas of the United Sates is discussed in~\cite{IEEE_C36}, as they can serve the unconnected rural community in the surroundings. This would allow the population to access eGovernment services although they do not have internet access in their areas. An attempt for providing internet access to libraries in some Latin American countries is described in~\cite{IEEE_J17}, allowing consulting and downloading digital content, although at very low speed. In~\cite{IEEE_J3}, the GSM network is proposed as a backhaul solution for WiFi or WiMAX traffic in rural areas where backhaul connectivity is not available. The operator would have to deploy a multiplexer at the GSM base transceiver station and another at the base station controller, in order to multiplex the voice and IP traffic. The data rates achieved would be very low, but would be sufficient for email and basic browsing.

\subsection{Using Local Content}
Community Networks, discussed in Section~\ref{subsubsec:Fronthaul_CommunityNetworks}, can be used to provide local content. Hence, they provide a first step towards providing broadband access to local rural communities, and after they succeed and prosper they can be connected to the other commercial networks.

In~\cite{IEEE_C202}, VANETs with RSUs that are not necessarily connected to the internet or a backbone network are proposed for rural areas. RSUs can exchange information between themselves through passing vehicles in a way similar to delay tolerant networks. RSUs can also act as sink nodes for WSNs sending relevant information to vehicles in an integrated WSN-VANET scenario~\cite{IEEE_C112}. Thus, in the approach of~\cite{IEEE_C202}, local safety information on the rural roads and local content can be shared between vehicles and RSUs. As the penetration of smart vehicles and the deployment of backbone infrastructure to rural areas increase, this network can then evolve into a full-fledged VANET with fully connected RSUs. This approach is complemented in~\cite{IEEE_C129} by proposing the use of satellites to transmit downlink traffic to VANETs in rural areas without RSUs, until the number of RSU deployments increases. RSU placement can be optimized using the weighted approach of~\cite{IEEE_C76}, where RSUs are placed first in locations with higher weights. It was shown in~\cite{IEEE_C76} that taking into account 3D space in rural mountainous areas leads to more accurate results.

In~\cite{IEEE_C27}, a wireless mesh network is used to provide access to a local intranet in a rural area in Bangladesh. A pilot was implemented where users are provided with tablets connected to the intranet, on which they can share local content and discuss local issues related to farming, weather, local environmental issues, etc.  In~\citep{IEEE_C12, IEEE_C38}, a system is proposed to provide local connectivity for rural areas. It is based on having a local ISP, providing connectivity using a local network, based for example on WiFi. The ISP also handles billing, deploys servers containing local content, and provides an application hub to subscribers such that they can use applications relevant to their local context. This network can operate without internet connectivity, and can use internet access whenever a suitable backhaul link is provided. For example, as described in~\citep{IEEE_C12}, if connectivity is intermittent, popular internet content can be downloaded to the local cache, where it can be used on the local network. In addition, if some requests to the internet cannot be serviced, they can be placed in a DTN bundle and pushed to be routed through DTNs. Transportation networks can then carry the requests, or users that can regularly travel to areas with internet connectivity can have their devices act as data mules. The approach of~\citep{IEEE_C12} suggests providing incentives to these users so that they perform this crowd-sourced DTN service. In~\cite{IEEE_C7}, \lq\lq Near Cloud" is proposed as a cloud-less platform in rural areas without internet connectivity. It uses the presence of IoT devices to build a wireless mesh network, where local content can be exchanged, stored, and processed in a distributed way. Machine learning algorithms an be implemented, and the system acts as a cloud-like platform without the existence of an actual cloud connected to the internet. Naturally, whenever a gateway connected to the internet is attached to the system, internet connectivity can be instantly provided. In addition, DTN-based internet connectivity can be achieved using vehicles or drones~\cite{IEEE_C7}.

\section{Case Studies of Rural Connectivity}
\label{sec:Case_Studies}
%
\label{subsec:Countries}
This section describe specific implementation scenarios of rural connectivity from the literature. It provides a collection of experiences from several countries (the countries are listed in alphabetical order). A summary of the main references is presented in Table~\ref{tab:Summary_Countries}.

\subsection{Bangladesh}
In~\cite{IEEE_C49}, 802.22 coverage for Bangladesh was investigated. Antenna design was considered using the \lq\lq Radio Mobile" planning tool. Interestingly, due to the relatively large flat areas of rural Bangladesh, a limited number of antennas was needed to provide coverage. Other types of antennas for point-to-point links, thus having shorter communication distances, were also investigated in~\cite{IEEE_C49}.

In~\cite{IEEE_C29}, a server was constructed using off-the-shelf equipment and open source software. It is connected to a WiFi access point to provide local connectivity through a WLAN for a rural village in Bangladesh. A GPRS module can be added to provide connectivity to the internet. The server can be used for eEducation to local schools, for eGovernment services, and to provide information to local farmers. The system can be powered by solar panels. The microgrid model to power rural areas proposed in~\cite{IEEE_C109} is adopted in Bangladesh. By 2012, $320,000$ houses in rural areas have deployed a solar home system to participate in a local microgrid, out of an estimated market of $500,000$. Typically, a microgrid would consist of ten houses, among which six have deployed solar panels to generate electricity and four are only consumers~\cite{IEEE_C109}. Such a microgrid can easily power a local LAN system like the one proposed in~\cite{IEEE_C29}.

\subsection{Cameroon}
In~\cite{nonIEEE_RuralCameroon}, the author discusses the Multipurpose Community Telecenters (MCTs) project launched by the government in Cameroon. MCTs aim at providing internet and telecommunications access to the local community. However, the majority of MCTs are connected via VSAT technology, which increases the cost and makes the technology expensive for most of the targeted rural population~\cite{nonIEEE_RuralCameroon}. A survey showed that most people use the MCTs for education purposes. Furthermore, the survey showed that the most desired services are (in decreasing order of popularity): eEducation, eHealth, eGovernment, and eCommerce~\cite{nonIEEE_RuralCameroon}.  Recommendations were provided in~\cite{nonIEEE_RuralCameroon} to make the system more affordable and attract more users. The main recommendations consist of adopting a mesh/multihop network to provide access instead of using more expensive VSAT technology, providing elementary services to users, providing local content and cache proxies, in addition to appropriate subscription packages (bronze for accessing local content only, silver for additional basic internet connectivity, gold for more expansive internet access)~\cite{nonIEEE_RuralCameroon}.

\subsection{Ecuador}
The telemedicine project in Ecuador was launched in 2002 with the aim of providing diagnosis, prevention, and support to rural areas~\cite{IEEE_C86}. As a slow start due to numerous challenges~\cite{IEEE_C86}, several phases of the project were completed by 2009-2010~\cite{IEEE_C86_Comp1, IEEE_C86_Comp2}. To provide connectivity in rural areas targeted by the project, satellite communications were used whenever ADSL or fiber optics were not available~\cite{IEEE_C86_Comp2}. Due to the importance of this project, telemedicine has become an important part of medical education in Ecuador by 2017-2018, with students performing telemedicine activities as part of their medical training~\cite{IEEE_C86_Comp3} using a specific telemedicine platform designed for this purpose~\cite{IEEE_C86_Comp4}. The possibility of using other technologies to provide broadband connectivity to rural areas in Ecuador are being investigated, e.g., in~\cite{IEEE_C31} where CDMA in the $450$~MHz band is proposed to allow for longer propagation distances.

\subsection{India}
BharatNet is a project that aims to provide broadband connectivity to rural areas in India. It will use fiber optic cables to connect $250,000$ village offices or councils named Gram Panchayats (GPs), then WiFi access points will be provided to each village and will be connected to the fiber backbone at the corresponding GP~\cite{nonIEEE_BharaNet2_pdf}. The project aims to serve around $650,000$ villages~\cite{nonIEEE_BharaNet2_pdf}. In~\cite{IEEE_J18}, TVWS was proposed as a middle-mile solution, linking the WiFi APs in villages to the fiber backhaul at GPs, and a testbed covering seven villages was implemented. In~\cite{IEEE_C83}, the BharatNet network was used in the VIVEKDISHA system, providing tele-medicine and tele-education services to rural areas in India. A kiosk named \lq\lq Kshema" was proposed in~\cite{IEEE_C124} to provide healthcare access to population in rural India. Kshema kiosks can be operated by a trained technician (who does not have to be a doctor). It allows measuring vital signs of the patients, provides a digital microscopy product, and can transmit radiology images. It also allows maintaining electronic records of patients, and allows patients to communicate with doctors using an attached video camera, that can also be used to take and transfer images of injuries. The kiosk can adapt to available connection speeds and can operate in online or offline mode~\cite{IEEE_C124}. A mobility aspect was added in~\cite{IEEE_C53}, where instead of kiosks, Rickshaw vehicles carrying the medical equipment in rural areas in India, and equipped with wireless transmitters powered by solar energy connected to WiMAX base stations, are proposed. A van equipped with medical devices was used in~\cite{IEEE_C25}. It circulates in rural villages according to a predefined schedule. The medical equipment in the van is operated by a junior doctor, and the van communicates with a central hospital through CDMA 2000 1X connectivity. The deployment of kiosks can be useful to issue epidemic alert generation as demonstrated in~\cite{IEEE_C9}, since kiosks are distributed over villages. Thus, the symptoms of patients visiting a given kiosk can be correlated with those of a given disease, and an alert an be issued when the number of affected people crosses a certain threshold. The transmission of the symptoms, along with the patient ID, to a healthcare center can be performed using available technology, e.g., GPRS as in~\citep{IEEE_C9, IEEE_C118}. GPRS was also used in~\cite{IEEE_C44} to test the performance of SmartHTTP, a proposed enhancement to the traditional HTTP protocol in order to enhance the quality of user experience in rural areas with intermittent and poor connectivity, especially for multimedia transmission. It is based on subdividing the content into smaller chunks depending on network conditions, and avoiding the retransmission of already received parts when the connection breaks down. In~\cite{IEEE_C104}, a system using mVAS based on IVR was proposed to support micro entrepreneurs in rural areas in their business activities. To address the issue of paper-based transactions with microfinance groups in rural India, the authors of~\cite{IEEE_C24} proposed the CAM system, where papers are scanned using a mobile phone's camera, and the information is read by the CAM application on the mobile device. In~\cite{IEEE_C77}, banking services were proposed for rural areas in India by using a terminal device that can be used for reading smart cards, and that can be carried by an agent who can visit villages in rural areas according to a certain schedule. Users can then perform financial transactions (deposit, withdrawal, loan payments, etc.) without the need to visit the nearest bank branch (which would be distant and requires long travel time). The terminal communicates via GPRS with a backend server connected to the bank's server.

In~\cite{IEEE_C108}, the deployment of LTE networks in rural India is investigated, and a feasibility study is performed. The authors suggest the use of the $800$~MHz band to allow for larger coverage areas per BS. They show that although the deployment can be profitable in the whole country, profitability in rural areas is better guaranteed when some form of government subsidies are provided. One of the important reasons for high costs in~\cite{IEEE_C108} are the spectrum license fees. This problem can be alleviated by resorting to license-exempt systems, like TVWS. An experiment conducted in~\cite{IEEE_C52} showed that in Urban New Delhi, around $85$\% of the TV $470$-$698$ MHz band is unused, whereas in rural areas, $95$\% is unused. The largest contiguous TWVS varies between $51$~MHz and $242$~MHz in rural areas, thus indicating that TV white spaces can be used to provide rural connectivity at affordable prices. Simulation results conducted in~\cite{IEEE_C78} supported the conclusion that TVWS is suitable for wireless broadband in rural India. Similar conclusions where reached for TV white spaces in rural Bangalore in~\cite{IEEE_C196}, where WiFi was suggested as a solution for providing connectivity in the VHF and UHF bands, using the IEEE 802.11af known as Super WiFi dedicated for use in TVWS, instead of the IEEE 802.22 WRAN standard~\cite{IEEE_C196}. An 802.11af prototype using a geo-location database at Bengaluru was presented in~\citep{IEEE_C48, IEEE_C105}. In~\cite{IEEE_C121}, unlicensed spectrum was used with LoRa technology to transmit IoT measurement data for monitoring water quality in the tanks and distribution network in Mori village near the Bay of Bengal. IoT was also used in~\cite{IEEE_C192} for monitoring biogas plants in rural India. The measurements were sent by SMS to the user's phone where an Android application pushed them to a remote database whenever internet connectivity was available.

All these activities help moving towards the \lq\lq Digital India" vision, leading to eGovernance and internet connectivity, along with promoting population awareness throughout India. It will require collaboration between several ministries, private sector entities, advisory groups, innovative entrepreneurships, and local governments to reach its objectives~\cite{IEEE_C143}. Future prospects seem promising as far as rural connectivity is concerned~\citep{IEEE_J26}.

\subsection{Malaysia}
In~\cite{IEEE_C34}, internet access was provided to a group of remote villages in Malaysia. The central village, called Bario, was connected to the network via a VSAT terminal. The objective was to connect nearby villages within a $10$~km radius with difficult transportation in the jungle. The solution was to resort to multihop long range WiFi connectivity using directive antennas. Within each small village, these solar-powered long-range WiFi stations would provide access to the villagers using omnidirectional coverage.
In~\cite{IEEE_C39}, a feasibility study for deploying IEEE 802.22 WRAN in rural Malaysia was performed, and simulation results indicated the possibility of covering large cells with long transmission distances in rural areas.
In~\citep{IEEE_C10}, a method is proposed to transform echocardiography videos into text that can be transmitted by SMS over 2G connections. Then, at the destination, the text can be transformed into video that can be analyzed by a physician. The objective of~\citep{IEEE_C10} is to provide appropriate diagnosis while overcoming the problem of limited or non-existent internet in rural Malaysia.

\subsection{New Zealand}
Conversely to most of the countries described in this section, New Zealand is a developed economy. However, it has vast rural regions that represent a challenge for providing ubiquitous connectivity. The government of New Zealand had launched the Ultra-Fast Broadband (UFB) initiative to provide fiber-based broadband access to $75$\% of the population~\cite{IEEE_C75}. For rural areas, the Rural Broadband Initiative (RBI) was launched and allowed providing access to $90$\% of New Zealanders by 2018, with connected rural households having a broadband connection of $5$~Mbps. In Despite the efforts done, a survey conducted between February-July 2018~\cite{nonIEEE_SurveyNZ1} showed that around $28$\% of the respondents are not satisfied with the internet speed and reliability in rural areas. The survey showed that the vast majority of users use the internet for email, reading the news, entertainment, and social media. Very few rural users used the internet for business purposes~\cite{nonIEEE_SurveyNZ1}. For example, farmers have shown some dissatisfaction due to their inability to use broadband connectivity efficiently for their business purposes (using applications to support smart agriculture for example)~\cite{nonIEEE_FarmersNZ1}. However, rural connectivity could provide strong support for e-Education in rural New Zealand, especially that schools have access to fiber broadband under the RBI~\cite{nonIEEE_SurveyNZ1}. In fact, in~\cite{IEEE_C150}, a study was performed on students from schools urban, provincial, and rural areas in New Zealand. Although students from all backgrounds enjoyed learning using computers and benefited most when learning using computers, those from rural and provincial areas provided higher satisfaction scores than their urban counterparts. The RBI approach uses FTTN, complemented by wireless access to the residents. The government of New Zealand, as part of the RBI, aims to provide connectivity for the remaining $10$\% by 2025~\cite{IEEE_C75}. A collaborative approach between the government, network operators, and the rural population is proposed in~\cite{IEEE_C75} in order to reach this objective sooner. It is based on the joint implementation of several technologies discussed in Sections~\ref{sec:Backhaul} and~\ref{sec:Fronthaul}, namely: 5G with D2D to provide better accessibility in rural areas, the use of a mesh network with collaboration between the local nodes in rural zones, the use of TV white space, and resorting to drones, balloons or satellites to connect hard to reach areas.

\subsection{South Africa}
Internet connectivity was provided to the Dwesa-Cwebe rural area in South Africa via VSAT technology, with WiMAX used to distribute the signal to the subscribers~\citep{IEEE_C65, IEEE_C60, IEEE_C6}. This allowed citizens to be more informed about announcements posted on government websites that were related to their daily lives, e.g., see~\cite{IEEE_C95}. IEEE 802.22 WRAN was tested in South Africa by providing access to several secondary schools, and results comparable to WiMAX were obtained~\cite{IEEE_C171}. In~\cite{IEEE_C42}, WiFi mesh was proposed for providing coverage in a rural area of South Africa. In the approach of~\cite{IEEE_C42}, the WiFi access points can communicate via multiple hops, with WiMAX providing backhaul connectivity to the internet, whereas in~\cite{IEEE_C92}, WiFi mesh was used for access and VSAT was used for backhaul in another South African rural area. In~\cite{IEEE_C188}, an emulation system is presented, with the aim of deploying a DTN between the rural area of Kwaggafontein and the city of Pretoria in South Africa. The buses of the public transport system are used as data mules carrying traffic between the rural area and the city, where the access points at the city's central station provide backhaul connectivity to the internet. In~\cite{IEEE_C71}, a system for eProcurement was designed over low end smartphones for small scale retailers in the Kgautswane rural area of South Africa. This allowed them to perform stock replenishment without having to be displaced to an urban center $70$~km away. Furthermore, it allowed the providers to schedule bulk delivery to specific delivery points closer to the retailers, where the payments can be made in cash upon delivery.

\subsection{Sweden}
According to~\cite{nonIEEE_LuleaTechRep_2016}, the Swedish government aims to provide $90$\% of households and companies by at least $100$~Mbit/s. However, the EU commission requires that all EU citizens have at leas a $30$~Mbit/s connectivity by 2020. To reach this goal for the remaining $10$\% of the Swedish population, who mainly live in rural areas, the authors of~\cite{nonIEEE_LuleaTechRep_2016} proposed the use of TV transmission towers to provide backhaul connectivity, as each covers a radius of $100$~km and are naturally connected to the power grid. However, the challenge is in powering the 5G base stations that will be providing access to the rural areas, since many of them might not have access to the power grid. Therefore, it was proposed in~\cite{nonIEEE_LuleaTechRep_2016} to resort to solar panels. However, due to the weather in Sweden, around three months with no solar coverage necessitate the use of large batteries, at a prohibitively high cost. A possible solution is to fill the gap by using wind energy during this period, and in reducing the power consumption of the BSs by resorting to massive MIMO beamforming and ultra-lean design~\citep{nonIEEE_LuleaTechRep_2016, IEEE_C214}. On another topic related to rural connectivity in Sweden, in~\cite{IEEE_C115}, it was noted that DTNs can be used for reindeer tracking by herders in arctic areas of Sweden, with the possibility of using helicopters as data mules.

\subsection{Other Countries}
In~\cite{IEEE_C184}, the use of mobile phones was described as the main method of communication in Liberia after its fixed line infrastructure was completely destroyed during the civil war. Rural users expressed the importance of the phone as a useful tool, especially in emergency situations. In~\citep{IEEE_C73}, SMS was proposed as a method for ePayment in the rural areas of Indonesia. Registered users can top up their accounts via vouchers and use SMS for their transactions.
In~\cite{IEEE_C3}, a study on the deployment of 3G CDMA deployment over the $900$~MHz frequency to provide connectivity in Tanzania showed that the deployment is economically viable. The use of $900$~MHz spectrum allows for longer propagation distances to serve rural areas, similarly to the proposition of using CDMA in the $450$~MHz in rural Ecuador in~\cite{IEEE_C31}. Multichannel Multipoint Distribution Service was proposed in~\cite{IEEE_C5} to leverage the fiber connectivity reaching the Nelson Mandela African Institute of Science and Technology in Tanzania, in order to provide connectivity to the surrounding rural area within a $50$~km radius.

In~\cite{IEEE_C11}, IEEE 802.22 was found as a suitable solution for providing connectivity to rural areas in Zimbabwe. Similar conclusions were reached in~\cite{nonIEEE_Congo_Africom18} concerning the use of TVWS for the Democratic Republic of Congo. In addition, tests on the feasibility of IEEE 802.22 for rural connectivity have been performed in several African countries~\cite{IEEE_C171}, including Malawi~\cite{IEEE_C178}. In fact, in~\cite{IEEE_J29}, schools in a rural area in Malawi were provided with broadband internet for the first time through the use of TVWS. In~\cite{nonIEEE_Nigeria_AfricanJrnl2016}, the challenges facing rural connectivity in Nigeria are analyzed. They mainly include deployment cost, financial sustainability, security, and regulatory challenges. In~\cite{IEEE_C189}, TVWS measurements using software defined radio were performed in the Philippines.

In a survey conducted in the UK, farmers have indicated limited connectivity and poor coverage in their farms, which affects the productivity, despite the high penetration rate of mobile phones~\cite{nonIEEE_FarmersUK1}. A possible solution to this problem could be to resort to TVWS in rural areas, as indicated in~\cite{nonIEEE_BroadwayPartners1}, where relatively high speeds were achieved over woody and hilly terrain.

A public-private partnership to provide connectivity to rural parts of Ontario is described in~\cite{IEEE_C51}. A fiber based backbone was deployed, providing $10$~Gbit ethernet connections to $160$~villages. From there, access was provided via fixed wireless communications or via DSL at speeds of $10$~Mbps. In very sparsely populated areas (less than three houses per square km), access was provided via satellite after reducing the initial high prices~\cite{IEEE_C51}.

Botswana, a cattle farming country, used RFID technology to identify cattle animals~\citep{IEEE_C64}. The data is stored on a mobile extension officer personal computer, and then transferred to a central database containing all the information on cattle and their owners. In case good connectivity is available (wired or wireless), data can be transferred online to the database. Otherwise, it was transferred offline every three weeks to update the database~\citep{IEEE_C64}. Sunflower farming was initiated in the Macha rural area in Zambia after the introduction of internet to the village and its surroundings~\citep{IEEE_C50}. The network is based on WiFi mesh for local access, with two VSAT terminals providing backhaul connectivity, although at high costs. The network was also used for eLearning, enhancing the procedures of the local health system, and for other personal uses (email, chatting, etc.) by the inhabitants~\citep{IEEE_C50}. In the rural area of Maseru, Lesotho, \lq\lq eKiosks" were deployed to provide basic internet connectivity, with the WiBACK system (based on WiFi mesh networking) relaying the traffic to reach the backhaul network~\citep{IEEE_C35, IEEE_C81}.

Several telemedicine projects in Pakistan are described in~\citep{IEEE_C93}, fueled by the large development of the telecommunications industry. Telemedicine in rural areas of North Carolina in the United States is discussed in~\citep{IEEE_C111}. In~\cite{IEEE_C46}, a system for performing ultrasound imaging in remote rural areas by of Peru was tested. Trained no-physician personnel conduct the tests using a portable ultrasound device be connected to a WiFi router, and the results are transferred to a cloud system where the physicians could perform diagnosis in an urban hospital.

In~\cite{IEEE_C100}, connectivity was provided to rural clinics in Crete island (Greece), to support the use of electronic health records (EHR). Local clinics in villages, called community offices (COs), are provided with WiFi connectivity and connected to the nearest primary health center (PHC), usually $5$-$20$~km away, via multihop WiFi whenever possible, otherwise VSAT was used. Similarly, PHCs were connected to the wired backbone network whenever possible, or through a VSAT link in the opposite case. The approach of~\cite{IEEE_C100} helped increase the adoption of EHR and ensure a better streamline of the healthcare process between COs and PHCs. In~\cite{IEEE_C106}, an EHR system was piloted in a rural area in Sri Lanka. It allowed patients to access an eClinic where remote consultations can be performed with specialist doctors located in distant urban hospitals, thus saving patients transportation costs and travel time, while allowing the medical records to be organized and stored in a central database. The healthcare centers in Sri Lanka have ADSL access, even in the rural areas. Hence, the challenge in~\cite{IEEE_C106} was in developing the EHR and the eClinic system. In~\cite{IEEE_C21}, WiMAX was used to provide access to Guangshan County, a rural area in Henan Province in China, mainly for eHealth services. This area had no public transport, no electricity, poor cellular coverage, and harsh topographical conditions that made it hard to deploy wired networks. Therefore, VSAT was used for backhaul connectivity, allowing the data from local health centers to be sent to the nearest urban hospitals.

In~\cite{IEEE_C43}, VoIP telephony was provided to a rural area in Cyprus by using a WiFi access point with high antenna gain to provide access to CPEs connected to VoIP phones inside homes or businesses, an open source implementation of a SIP server, and a VSAT terminal to provide backhaul access to the internet. The system can be used to serve local calls within the area without resorting to satellite connectivity. In~\cite{IEEE_C132}, connectivity was provided to the unconnected Verrua Savoia rural area in Italy. Four IEEE 802.11n access points working in the 5Ghz band were used, equipped with directive antennas along with polarization diversity. They were positioned in strategic positions on the hills surrounding the village, and connected to a gateway located $35$~Km via a microwave radio link. In Cambodia, project \lq\lq iReach" described in~\cite{IEEE_J27} allowed sharing of local information on health and agriculture. It also allowed local capacity building and distance learning. The project consists of ten WiFi hotspots, each connected to a central office via WiMAX. The central office is connected to the internet via a satellite link~\cite{IEEE_J27}.

In Thailand, the internet of educational things is proposed in~\cite{IEEE_C67} for underprivileged rural areas. Weather sensors were connected to a Raspberry Pi device, and used by Grade~1 students equipped with tablet computers, accessing the readings via a local WiFi connection~\cite{IEEE_C67}. In~\cite{IEEE_C16}, the use of solar panels to provide electricity to off-grid schools in rural areas for the purpose of distance education is discussed for several countries in Latin America.

In Kenya, around $43$ Million mobile phones are used, versus $70,000$ landlines. However, a large part of the population lives in rural areas. To provide cellular coverage, balloons are being deployed, flown over from Puerto Rico and traversing the Atlantic ocean before being positioned above their coverage areas in Kenya~\cite{IEEE_Spectrum_Loon2}.

In Tonga, a country consisting of $171$ islands, although a submarine fiber optic cable links the country to Fiji, different technologies are needed to provide internet connectivity for the various islands. Therefore, a multi-tenancy approach is proposed in~\cite{IEEE_C213}, where 5G, TVWS, DTNs, UAVs, and balloons can be used jointly in different areas in order to provide ubiquitous broadband connectivity.

\begin{table}[h!]
\begin{tiny}
\caption{Summary of Key References for the Different Services Provided to Rural Areas and the Corresponding Countries}
\label{tab:Summary_Countries}
\begin{center}
\vspace{-1.0cm}
\begin{tabular}{|p{2cm}|p{3cm}|p{5cm}|p{5cm}|}
  \hline
   {\bf Ref.} & {\bf Country} & {\bf Service/Application}  & {\bf Technology}\\
  \hline
   \citep{IEEE_C29}  & Bangladesh & eEducation, eGovernment, Farming & WiFi/GPRS\\
  \hline
   \citep{IEEE_C64}  & Botswana & Farming/Cattle & RFID\\
  \hline
   \citep{nonIEEE_RuralCameroon}  & Cameroon & Basic internet connectivity & VSAT\\
  \hline
   \citep{IEEE_C51}  & Canada (Ontario) & Internet connectivity & Fiber backbone and fixed wireless access\\
  \hline
   \citep{IEEE_C21}  & China & eHealth & WiMAX/VSAT\\
  \hline
   \citep{IEEE_C43}  & Cyprus & VoIP & WiFi/VSAT\\
  \hline
   \citep{IEEE_C86, IEEE_C86_Comp1, IEEE_C86_Comp2, IEEE_C86_Comp3, IEEE_C86_Comp4}  & Ecuador & Telemedicine & Satellite\\
  \hline
   \citep{IEEE_C100}  & Greece (Crete) & Healthcare & WiFi/VSAT\\
  \hline
   \citep{nonIEEE_BharaNet2_pdf, nonIEEE_BharaNet1_ppt}  & India & Basic internet connectivity & Fiber/WiFi\\
  \hline
   \citep{IEEE_C124, IEEE_C83, IEEE_C25, IEEE_C9}  & India & Healthcare & Broadband / Wireless Connectivity\\
  \hline
   \citep{IEEE_C83}  & India & Education & Broadband / Wireless Connectivity (WiMAX)\\
  \hline
   \citep{IEEE_C104}  & India & Trading Services & 2G/3G Cellular\\
  \hline
   \citep{IEEE_C77}  & India & Financial Services & GPRS\\
  \hline
   \citep{IEEE_C121}  & India & Water Quality Monitoring & LoRa\\
  \hline
   \citep{IEEE_C192}  & India & Biogas Monitoring & GSM\\
  \hline
   \citep{IEEE_C73}  & Indonesia & ePayment & SMS/2G GSM\\
  \hline
   \citep{IEEE_Spectrum_Loon2}  & Kenya & Cellular connectivity & Balloons\\
  \hline
   \citep{IEEE_C35, IEEE_C81}  & Lesotho & Basic internet connectivity & WiBACK (WiFi mesh)\\
  \hline
   \citep{IEEE_C184}  & Liberia & Mobile connectivity & 2G/3G Cellular\\
  \hline
   \citep{IEEE_J29}  & Malawi & Education & TVWS\\
  \hline
   \citep{IEEE_C34}  & Malaysia & Basic internet connectivity & Long range WiFi/Multihop\\
  \hline
   \citep{IEEE_C10}  & Malaysia & Healthcare & SMS/2G Cellular\\
  \hline
   \citep{IEEE_C75, nonIEEE_SurveyNZ1, nonIEEE_FarmersNZ1}  & New Zealand & Broadband connectivity & Fiber/Wireless\\
  \hline
   \citep{IEEE_C65, IEEE_C60, IEEE_C6}  & South Africa & Basic internet connectivity & VSAT/WiMAX\\
  \hline
   \citep{IEEE_C171}  & South Africa & Basic internet connectivity & IEEE 802.22\\
  \hline
   \citep{IEEE_C188}  & South Africa & Basic internet connectivity & DTN (using WiFi)\\
  \hline
   \citep{IEEE_C71}  & South Africa & e-Procurement & GPRS/3G\\
  \hline
   \citep{IEEE_C106}  & Sri Lanka & eHealth & ADSL\\
  \hline
   \citep{IEEE_C67}  & Thailand & Internet of Educational Things (IoET) & WiFi\\
  \hline
   \citep{IEEE_C50}  & Zambia & Basic internet connectivity, Farming, eLearning & WiFi mesh/VSAT\\
  \hline
\end{tabular}
\end{center}
\end{tiny}
\end{table}

\section{Foundations/Initiatives}
\label{subsec:Foundations}

\begin{table}[h!]
\begin{footnotesize}
\caption{Summary of Key Initiatives Aiming to Provide Connectivity to Rural Areas}
\label{tab:Summary_Initiatives}
\begin{center}
\begin{tabular}{|p{2cm}|p{3cm}|p{3cm}|p{7cm}|}
  \hline
   {\bf Ref.} & {\bf Initiative} & {\bf Technology}  & {\bf Description}\\
  \hline
   \citep{OneWeb}  & One Web & Satellite & Broadband access from orbit with 2,000 LEO satellites\\
  \hline
    \citep{SpaceX} & Space X - Starlink & Satellite & Broadband access from orbit with 12,000 LEO satellites forming the Starlink constellation\\
  \hline
   \citep{Sat_Amazon_Kuiper}  & Amazon - Project Kuiper & Satellite & Broadband access from orbit with 3,236 LEO satellites\\
  \hline
   \citep{Loon}  & Alphabet - Project Loon & Balloons & Broadband access from balloons navigating the stratosphere. Initially launched by Google before its restructuring to become a subsidiary of Alphabet\\
  \hline
   \citep{Terragraph1, Terragraph2, Terragraph3, Facebook_Connectivity}  & Facebook Connectivity & Fiber, WiFi, mmWave/ Terragraph & Ambitious project that includes partnerships with different industry players and different initiatives in various countries, mostly in Africa and South America. It also included an initiative related to providing backhaul connectivity through solar powered UAVs, the Aquila drone project, that was discontinued~\citep{Facebook_Aquila1, Facebook_Aquila2}.\\
  \hline
\end{tabular}
\end{center}
\end{footnotesize}
\end{table}

In addition to the various country initiatives, this paper has discussed several efforts made by foundations, companies, and non-governmental organizations for providing connectivity to rural and under-privileged areas. This section is a sort of recapitulation that aims to provide a condensed summary of the main foundations working on this topic. They are listed in Table~\ref{tab:Summary_Initiatives}.
The main categories are:
\begin{itemize}
  \item Companies aiming to provide global connectivity through large satellite constellations, like One Web~\citep{OneWeb}, Space X~\citep{SpaceX}, and Amazon~\citep{Sat_Amazon_Kuiper}. The efforts of these companies were discussed mainly in Section~\ref{subsubsec:Backhaul_Sat}.
  \item Companies using balloons through the stratosphere, mostly represented by the Loon project~\citep{Loon}. Examples related to this approach were discussed in Section~\ref{subsubsec:Backhaul_UAV}. The Loon project was initially launched by Google before its restructuring to become a subsidiary of Alphabet. A small number of balloons can relay a signal over long distances $1,000$~km~\cite{IEEE_Spectrum_Loon2}. They are being used in actual projects to provide backhaul connectivity to cellular communications in rural areas~\cite{IEEE_Spectrum_Loon2}.
  \item Facebook connectivity, which includes various initiatives using different technologies. For example, the Terragraph technology described in Section~\ref{subsubsec:Fronthaul_mmWave} is part of this project. Another part was the use of solar powered UAVs traveling for long distances, serving similar purposes as the balloons discussed above. The drone project was named \lq\lq Aquila\rq\rq, and it was discontinued by Facebook since other more established companies in the airplane industry started efforts on similar areas~\citep{Facebook_Aquila1, Facebook_Aquila2}.
  \item Other initiatives such as the \lq\lq basic internet\rq\rq initiative~\cite{BasicInternet_Website} and the Global Access to the Internet for All (GAIA~\cite{GAIA_IETF1}. They are not listed in Table~\ref{tab:Summary_Initiatives} since they are not related to adopting a specific technology, but they also aim to provide global internet coverage. The \lq\lq basic internet\rq\rq initiative supports the argument that internet access should be provided free of charge when only static content (text and images) is requested, since this kind of traffic amounts to around 2-3\% of the bandwidth~\cite{IEEE_J30}, whereas those requesting dynamic content can be charged.
\end{itemize}

\section{Future Directions/Trends}
\label{sec:Future_Directions}
This section presents a discussion of future directions for providing ubiquitous connectivity based on the current trends and achievements in rural connectivity.

\subsection{Current Situation: Putting it All Together}
\label{subsec:Putting_Together}
To provide sustainable connectivity to rural areas, the different parts or components surveyed in this paper need to be successfully integrated: fronthaul technologies, backhaul technologies, innovative methods for providing electricity, user awareness that creates local services and drives local demand, which in turn leads to more advanced connectivity, with the whole governed by suitable government policies and wise governance.

In addition, it should be noted that there is no single technology that is best suited to provide connectivity to rural areas. Each technology can be the best fit for certain scenarios while not being convenient for other scenarios. Furthermore, no technologies should be excluded from this process. Technologies that are considered dead can be adopted in certain rural areas as they might be the best fit: For example, WiMAX can be deployed in certain areas, be it for simple nomadic or fixed access to CPEs, with the CPE connected to WiFi through a wired connection to provide local access and limited mobility (especially that WiMAX integrates smoothly with WiFi, both being IEEE 802 standards).

Thus, different solutions consisting of various fronthaul/backhaul combinations might coexist in different rural areas, while eventually converging to a common core or national backbone network. Combining the different rural connectivity technologies while providing the flexibly to evolve to 5G/5G+/6G as the demand increases and the infrastructure is gradually provided will eventually help achieve global broadband coverage.

\subsection{Next Steps: Where to Go From Here}
\label{subsec:Next_Steps}
Internet has become a commodity or a merit good that users should access to on the basis of need~\cite{IEEE_C36}. To provide this access regardless of costs and business aspects, Government intervention might be necessary in several places. Although the technology is different, this is similar to what happened historically for railroads, postal mail, and the fixed telephone networks. Thus, one can learn from history to develop a policy for the future, as suggested in~\cite{IEEE_C36}. Policies for providing broadband access need to be reached and implemented through collaboration between various stakeholders, including Government, policy makers (e.g. regulatory authority), business players such as equipment manufacturers and telecom operators, service and content providers, and citizens~\citep{Stakeholders_Analysis, Policies_Connect5B}. In~\cite{Stakeholders_Analysis}, guidelines were noted to help stakeholders bridge the broadband divide, e.g., (i) establishing an independent national telecommunications regulatory authority, (ii) sharing the investment in the physical infrastructure and human capital, and (iii) competition between telecommunication service providers to reduce costs. Once basic infrastructure is established and the demand over broadband services starts to grow in rural areas, the wired fiber backbone will expand towards these areas since the business case becomes economically viable and gradually profitable~\citep{Policies_Connect5B, Nature_Photonics_Africa}.

\begin{figure}[t!]
 \begin{center}
  \includegraphics[scale=0.7]{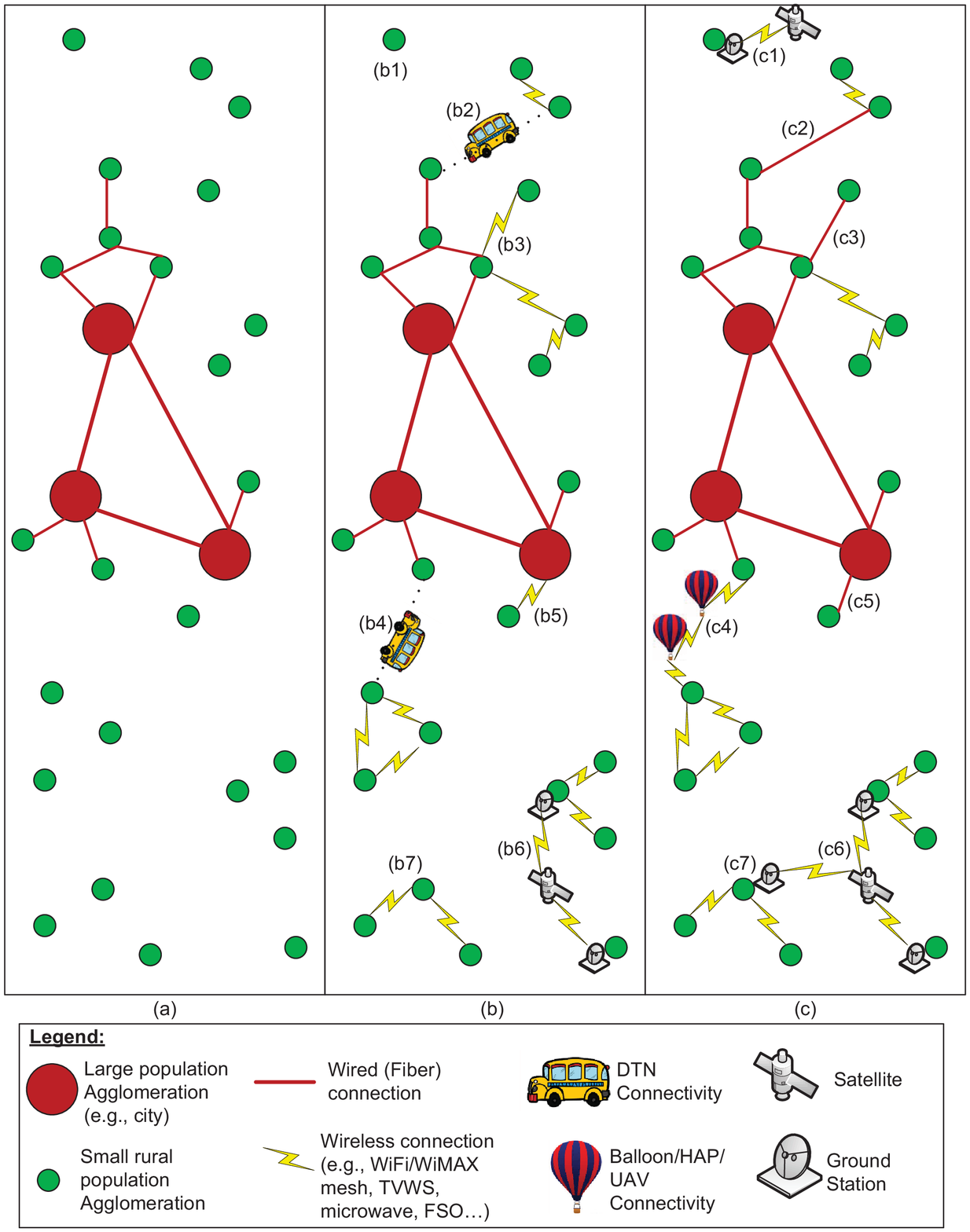}
 \end{center}
  \vspace{-1.0cm}
\caption{Illustrative example showing gradual expansion of broadband connectivity from urban to rural areas.}
\label{fig:Connectivity_Progress}
\end{figure}
An illustrative example in this regard is shown in Fig.~\ref{fig:Connectivity_Progress}, where Fig.~\ref{fig:Connectivity_Progress}(a) shows a geographical area where urban centers are connected by fiber, along with some connectivity to neighboring rural villages, with most rural areas being unconnected. Fig.~\ref{fig:Connectivity_Progress}(b) shows an enhanced scenario with some basic rural connectivity in most areas: Although some areas are still unconnected (b1), some other remote areas have local networks that are not yet connected to the backbone (b7), whereas some others have local connectivity with intermittent connectivity to the internet, e.g., through DTN connectivity (b2, b4). The remote connectivity is mainly wireless (b3, b5), which could include WiFi long distance, WiMAX, multihop/mesh networks (typically using WiFi and/or WiMAX), cellular broadband, or TVWS. Some of these wireless connections are connected to the internet via satellite links (b6), as they might be too far to reach a fiber point of presence. As demand grows in rural areas, the infrastructure expands to meet the increasing demand, and we reach the situation shown in Fig.~\ref{fig:Connectivity_Progress}(c): Fiber has expanded to new areas that are the closest to the previously connected zones (c2, c3, c5), more efficient and permanent connections using advanced technologies have replaced previous intermittent ones (c4), and no areas are isolated as satellite backhaul was provided to the most remote spots (c1, c7).

Hence, suitable policies can contribute to bridging the digital divide and preparing the rural areas to enter the 5G era~\cite{ICT_Handbook1}, in the hope that user awareness and the market applications will allow them to catch up with the beyond 5G next generation ICT technologies such as machine learning, artificial intelligence, augmented reality, autonomous driving, blockchain and cloud solutions~\cite{ICT_Handbook2}.

\subsection{Ultimate Target: Reaching Smart Living}
\label{subsec:Smart_Living}
As discussed in Section~\ref{subsec:Next_Steps}, rural connectivity will hopefully evolve to reach a level comparable to that in urban hyper connected areas. Reaching this level will help achieve the concept of \lq\lq smart living\rq\rq everywhere, as opposed to having only smart cities.

In fact, smart city advocates aim to deal with the problem of having half the world population already living in cities and another couple billion being on their way. As stated in~\citep{Deloitte_MiddleEast}: \lq\lq {\em The world is becoming far more urbanized, and mega cities with populations greater than ten to twenty million people are emerging, there is a greater need for large-scale operations and management for cities to effectively serve its inhabitants.}\rq\rq

However, with the evolution of rural connectivity to reach levels comparable to those of urban areas, we can have not only smart cities, but also smart villages, smart towns, smart suburbs, etc. In such a scenario, one can enjoy quality healthcare, quality education, and several jobs can tolerate employees working remotely by benefiting from the huge advancements in communications technologies. All this can take place in a less crowded, less polluted rural environment. Thus, quality living can be enjoyed without having to move to a big city.

Hence, while the technological progress has made the existence of smart cities with huge populations possible, that same progress, when rural connectivity evolves sufficiently, can also make it possible for people to enjoy quality living in their rural areas. This \lq\lq smart everywhere\rq\rq concept could lead to a more balanced population deployment between cities and rural areas, while allowing all citizens to enjoy quality living.

Finally, we mention a final note supporting another side of the story, where limited connectivity is recommended even when advanced infrastructure is available: In~\cite{nonIEEE_AltIntModels_ITUjrnl18}, a discussion was provided about the benefits of adopting a hybrid model of connectivity, where people have internet access but are not connected all the time, such that some of the side effects of the internet are reduced. In other words, even if people have full connectivity in cities, it might be good to willingly adopt a model similar to certain rural areas, in order to preserve their mental health and social relationships.

\section{Conclusions}
\label{sec:Conclusions}
This paper surveyed the literature related to providing connectivity to rural areas. The problem of providing connectivity to around half of the World population living in rural or underprivileged areas is indeed a major challenge. Thus, in this paper, fronthaul and backhaul technologies used for connecting rural areas were presented and analyzed. The long term CAPEX and OPEX costs of backhaul solutions for rural areas were discussed based on inputs from the relevant literature. In addition, fronthaul specific challenges were listed and analyzed, while focusing on major issues like electricity provision, spectrum allocation, user awareness and acceptability, and gradual deployment from simple to complex networks in order to guarantee sustainability. Typical application scenarios in rural areas were presented, and several country-specific use cases were surveyed ana analyzed. In addition, the future trends in the evolution of rural connectivity were outlined, in the hope of reaching ubiquitous global connectivity, a goal hopefully achievable by 5G+/6G networks.

\bibliographystyle{ieeetr}
\end{document}